\documentclass[twocolumn]{aastex63}

\newcommand\s{$\sim$}
\newcommand{\e}{\'{e}}

\usepackage{xcolor}
\usepackage[normalem]{ulem}
\usepackage{blindtext}
\shorttitle{Major merger rate for AGNs with the highest Eddington ratios at $z<0.2$}
\shortauthors{Marian et al.}

\begin{document}

\title{
A significant excess in major merger rate for AGNs with the highest Eddington ratios at $z<0.2$
}

\correspondingauthor{Victor Marian}
\email{marian@mpia.de}

\author[0000-0003-1733-9281]{Victor Marian}
\affiliation{Max-Planck-Institut f\"ur Astronomie,
K\"onigstuhl 17,
69117 Heidelberg, Germany}
\affiliation{International Max Planck Research School for Astronomy \& Cosmic Physics at the University of Heidelberg}

\author[0000-0003-3804-2137]{Knud Jahnke}
\affiliation{Max-Planck-Institut f\"ur Astronomie,
K\"onigstuhl 17,
69117 Heidelberg, Germany}

\author[0000-0001-6102-9526]{Irham Andika}
\affiliation{Max-Planck-Institut f\"ur Astronomie,
K\"onigstuhl 17,
69117 Heidelberg, Germany}
\affiliation{International Max Planck Research School for Astronomy \& Cosmic Physics at the University of Heidelberg}

\author[0000-0002-2931-7824]{Eduardo Ba\~{n}ados}
\affiliation{Max-Planck-Institut f\"ur Astronomie,
K\"onigstuhl 17,
69117 Heidelberg, Germany}

\author[0000-0003-2064-0518]{Vardha N. Bennert}
\affiliation{Department of Physics, 
California Polytechnic State University, 
San Luis Obispo, 
CA 93407, USA}

\author[0000-0003-3329-1337]{Seth Cohen}
\affiliation{School of Earth and Space Exploration, Arizona State University, P.O. Box 871404, Tempe, AZ 85287-1404, USA}

\author[0000-0003-2901-6842]{Bernd Husemann}
\affiliation{Max-Planck-Institut f\"ur Astronomie,
K\"onigstuhl 17,
69117 Heidelberg, Germany}

\author[0000-0002-1173-2579]{Melanie Kaasinen}
\affiliation{Max-Planck-Institut f\"ur Astronomie,
K\"onigstuhl 17,
69117 Heidelberg, Germany}
\affiliation{Universit\"{a}t Heidelberg, Zentrum f\"{u}r Astronomie, Institut f\"{u}r Theoretische Astrophysik, Albert-Ueberle-Stra\ss e 2, D-69120 Heidelberg, Germany}
\affiliation{International Max Planck Research School for Astronomy \& Cosmic Physics at the University of Heidelberg}

\author[0000-0002-6610-2048]{Anton M. Koekemoer}
\affiliation{Space Telescope Science Institute, 3700 San Martin Drive, Baltimore, MD 21218, USA}

\author[0000-0001-6462-6190]{Mira Mechtley}
\affiliation{School of Earth and Space Exploration, Arizona State University, P.O. Box 871404, Tempe, AZ 85287-1404, USA}

\author[0000-0003-2984-6803]{Masafusa Onoue}
\affiliation{Max-Planck-Institut f\"ur Astronomie,
K\"onigstuhl 17,
69117 Heidelberg, Germany}

\author[0000-0002-4544-8242]{Jan-Torge Schindler}
\affiliation{Max-Planck-Institut f\"ur Astronomie,
K\"onigstuhl 17,
69117 Heidelberg, Germany}

\author[0000-0001-7825-0075]{Malte Schramm}
\affiliation{Graduate school of Science and Engineering, Saitama Univ.,
255 Shimo-Okubo, Sakura-ku, Saitama City, Saitama 338-8570, Japan}

\author[0000-0002-6660-6131]{Andreas Schulze}
\affiliation{National Astronomical Observatory of Japan, 
Mitaka, Tokyo 181-8588, Japan}

\author[0000-0002-0000-6977]{John D. Silverman}
\affiliation{Kavli Institute for the Physics and Mathematics of the Universe, The
University of Tokyo, Kashiwa, Japan 277-8583 (Kavli IPMU, WPI)
}
\affiliation{Department of Astronomy, School of Science, The University of
Tokyo, 7-3-1 Hongo, Bunkyo, Tokyo 113-0033, Japan
}

\author[0000-0002-2260-3043]{Irina Smirnova-Pinchukova}
\affiliation{Max-Planck-Institut f\"ur Astronomie,
K\"onigstuhl 17,
69117 Heidelberg, Germany}
\affiliation{International Max Planck Research School for Astronomy \& Cosmic Physics at the University of Heidelberg}

\author[0000-0002-5027-0135]{Arjen van der Wel}
\affiliation{Sterrenkundig Observatorium, Universiteit Gent, Krijgslaan 281 S9, B-9000 Gent, Belgium}

\author[0000-0002-8956-6654]{Carolin Villforth}
\affiliation{University of Bath, Department of Physics, Claverton Down, BA2 7AY, Bath, United Kingdom}

\author[0000-0001-8156-6281]{Rogier A. Windhorst}
\affiliation{School of Earth and Space Exploration, Arizona State University, P.O. Box 871404, Tempe, AZ 85287-1404, USA}

%\received{}
%\revised{}
%\accepted{}

\begin{abstract}

Observational studies are increasingly finding evidence against major mergers being the dominant mechanism responsible for triggering AGN. After studying the connection between major mergers and AGN with the highest Eddington ratios at $z=2$, we here expand our analysis to $z<0.2$, exploring the same AGN parameter space. Using ESO VLT/FORS2 $B-$, $V-$ and color images, we examine the morphologies of 17 galaxies hosting AGNs with Eddington ratios $\lambda_{\mathrm{edd}} > 0.3$, and 25 mass- and redshift-matched control galaxies. To match the appearance of the two samples, we add synthetic point sources to the inactive comparison galaxies. The combined sample of AGN and inactive galaxies was independently ranked by 19 experts with respect to the degree of morphological distortion. We combine the resulting individual rankings into multiple overall rankings, from which we derive the respective major merger fractions of the two samples. With a best estimate of $f_\mathrm{m,agn}$ = 0.41 $\pm$ 0.12 for the AGN host galaxies and $f_\mathrm{m,ina}$ = 0.08 $\pm$ 0.06 for the inactive galaxies our results imply that our AGN host galaxies have a significantly higher merger rate, regardless of the observed wavelength or applied methodology. We conclude that although major mergers are an essential mechanism to trigger local high Eddington ratio AGNs at $z<0.2$, the origin of $\gtrsim50\%$ of this specific AGN subpopulation still remains unclear.

\end{abstract}


\section{Introduction} \label{sec:intro}

An ever-growing number of empirical studies are finding that the properties of the black holes (BH) at the center of galaxies are closely correlated with the properties of the host galaxy, i.e. BH mass, bulge velocity dispersion and mass, stellar host mass, velocity dispersion or luminosity \citep[e.g.][]{marconi_relation_2003, haring_black_2004, jahnke_massive_2009, bennert_cosmic_2010, bennert_relation_2011, beifiori_correlations_2012, graham_m_2013, mcconnell_revisiting_2013, davis_black_2018, davis_black_2019, de_nicola_fundamental_2019, sahu_revealing_2019, shankar_probing_2019, ding_mass_2020}. These findings are complemented by state-of-the-art cosmological hydrodynamical simulations \citep{habouzit_linking_2019, li_correlations_2019, terrazas_relationship_2019} that attempt to capture the physics behind these relations.
Combined with the widely-accepted assumption that every major galaxy hosts a supermassive BH in its center \citep{kormendy_coevolution_2013}, this strongly indicates that hierarchical structure formation applies to black holes in the same way as it does to galaxies as a whole \citep{jahnke_non-causal_2011}.

The potential feedback of the emitted radiation, winds, jets, or a combination thereof, when a BH becomes active, (i.e. starts accreting matter) may have a broad range of effects on the host galaxy, depending on the physical nature, geometry and/or size of those different outflow mechanisms \citep{silk_quasars_1998, harrison_agn_2018}. These range from the total quenching to the enhancement of star formation due to various processes affecting the interstellar and circumgalactic medium \citep{husemann_reality_2018, weinberger_supermassive_2018, davies_quenching_2019, nelson_first_2019, truong_x-ray_2019, valentini_impact_2019, oppenheimer_feedback_2020}, although the impact may also be negligible \citep{schulze_no_2019, oleary_emerge_2020}. In addition, individual AGN feedback processes could even have an impact on larger scales by affecting satellite galaxies and the surrounding intracluster or intragroup medium \citep{blanton_active_2010, chowdhury_cosmological_2019, dashyan_agn-driven_2019, li_direct_2019, martin-navarro_quantifying_2019}.

Considering this interplay between galaxies and their central BH in its active phase, it is imperative to understand the mechanisms responsible for triggering the period of significant black hole accretion.
For decades it has been assumed that galaxies follow an evolutionary path that includes at least one merging event with another galaxy of a similar mass (i.e.\ a major merger). This gravitational encounter would strip part of the gas of its angular momentum, funneling it into the most central regions where the BH(s) reside \citep{barnes_dynamics_1992, sanders_luminous_1996}. Such an incident would ultimately lead to the active galactic nucleus (AGN) phase, in which the coalescing galaxy hosts at least one active BH in the center. This theoretical scenario was comprehensively presented in the seminal work of \citet{sanders_ultraluminous_1988}, and further studied with numerous simulations \citep{springel_modelling_2005, hopkins_unified_2006, hopkins_cosmological_2008, somerville_semi-analytic_2008, mcalpine_rapid_2018, mcalpine_galaxy_2020, weigel_fraction_2018} and observations \citep[e.g.][]{yue_quasars_2019, gao_mergers_2020}.
%that confirmed this sequence of events. 
These causal connections, between major mergers and the presence of an active BH, have been found especially for particular AGN populations at low redshift \citep{koss_merging_2010, cotini_merger_2013, sabater_effect_2013, hong_correlation_2015, ellison_definitive_2019}, and high-luminosity AGNs at different cosmic epochs \citep{urrutia_evidence_2008, schawinski_heavily_2012, treister_major_2012, glikman_major_2015, fan_most_2016, donley_evidence_2018, goulding_galaxy_2018, urbano-mayorgas_host_2019}. %However, in the last years an abundance of studies have refuted the theory that major mergers are the dominant trigger of AGNs, in terms of a major merger incidence for AGN host galaxies $>50\%$.

In recent years, however, a number of studies have found that the fraction of major mergers amongst AGN hosts is $<50\%$, implying that major mergers are not the dominant trigger of AGNs.
For example, no predominant connection between major mergers and AGNs could be found for both the general population of X-ray-detected and optically-observed AGNs at various redshifts \citep{gabor_active_2009, georgakakis_host_2009, cisternas_bulk_2011}. Likewise, studies that investigated luminosity-selected AGNs with low or moderate X-ray luminosities, with an upper limit of $L_\mathrm{X}\leq10^{43}$erg s$^{-1}$ \citep{grogin_agn_2005, allevato_xmm-newton_2011, schawinski_hst_2011, kocevski_candels_2012, bohm_agn_2013} or high X-ray luminosities with $L_\mathrm{X}\geq10^{43}$erg s$^{-1}$ \citep{karouzos_mergers_2014, villforth_morphologies_2014, villforth_host_2017} found no significant connection. 
Studies examining more specific samples of AGNs have obtained similar results: neither sources that possess the highest BH masses \citep{mechtley_most_2016} nor heavily obscured AGNs \citep{schawinski_heavily_2012, zhao_role_2019} appear to be triggered predominantly by major mergers. Even AGNs assumed to be in an early evolutionary stage \citep{villforth_host_2019}, or those exhibiting the highest Eddington ratios \citep{marian_major_2019} show no signs of an enhanced merger fraction. %Additional studies detected in fact slight enhancements in merger rate for AGNs at different luminosities and redshifts, however the vast majority of AGNs were still not of a major merger induced origin \citep{silverman_impact_2011, rosario_host_2015,hewlett_redshift_2017}. 
Additional studies detected slight enhancements in the merger rate for AGNs at different luminosities and redshifts; however, the vast majority of AGNs were still not major merger induced \citep{silverman_impact_2011, rosario_host_2015,hewlett_redshift_2017}. In contrast, recent work examining secularly powered outflows \citep{smethurst_secularly_2019} and the dependence of local AGNs on environment \citep{man_dependence_2019} suggest that secular processes are the dominant mechanisms to trigger AGN activity.
%These studies, which have examined AGNs with a wide range of different redshifts, luminosities and masses, have come to the unanimous conclusion that mergers should be viewed as only one of multiple possible mechanisms to initiate black hole growth, making it necessary to consider alternative processes and/or differences in the lifetimes of merger features and AGNs.
These studies, in which AGNs with a variety of different redshifts, brightnesses and masses have been examined, have come to the unanimous conclusion that mergers should only be considered as one of several possible mechanisms for initiating black hole growth. Therefore, it is necessary to consider alternative processes and/or differences in the lifetime of merger features and AGNs.

Large-scale galactic bars \citep{cheung_galaxy_2015, cisternas_role_2015, goulding_galaxy-scale_2017} and a time delay between a major merger event and the onset of an AGN \citep{cisternas_bulk_2011, mechtley_most_2016, marian_major_2019} appear to be an inadequate explanation for these contrary results regarding the relevance of large-scale mergers for triggering AGNs. Instead, \citet{goulding_galaxy_2018} propose an intriguing alternative, which may ease this tension: although AGNs are indeed triggered by major mergers, their activity and therefore luminosity during the merging process depend on the merger stage and thus can vary heavily. At larger separations between the two galaxies the arising torques are not sufficient to provide enough gas to trigger an AGN phase or feed the black hole(s). However, at close passages the torques as well as the gas inflow increase, boosting the AGN activity, as long as the distance between the two galaxies is sufficiently small. Before coalescence, this would result in a periodic AGN variability, while the morphological features, like tidal tails, shells or asymmetries of this encounter would be continuously visible, explaining the lack of observed AGNs in merging systems. 

%\subsection{Fueling  black holes with the highest specific accretion rates}

In this study, we investigate the possibility that the AGNs with the highest Eddington ratios $\lambda_\mathrm{edd}=L/L_\mathrm{edd}$, i.e. the highest specific accretion rates at $z<0.2$ are predominantly triggered by major mergers. We also expand on the work presented in \citet{marian_major_2019}, in which we studied comparable BHs %with the highest Eddington ratios $\lambda_\mathrm{edd}=L/L_\mathrm{edd}$, i.e. the highest specific accretion rates, 
at $z\sim2$. Contrary to $z\sim2$, which marks the peak of cosmic black hole activity \citep{boyle_2df_2000, aird_evolution_2015} and star formation rate \citep{madau_cosmic_2014}, the comparable population of local AGN host galaxies at $z<0.2$ exhibit up to $\sim$ 10 times lower 
%-- a time when both the 
black hole activity and star formation rates %are $\sim$ 10 times lower than at $z\sim2$ 
\citep{aird_evolution_2015}. Moreover, only a small fraction ($\lesssim 10\%$) of today's massive galaxies ($\mathrm{log(}M_*\mathrm{/M_{\odot})} > 10$) may have undergone one or more major merger events since $z\sim$1, with the majority of such galaxies being undisturbed for the past $\sim$ 7\,Gyr \citep{lopez-sanjuan_galaxy_2009, lotz_major_2011, xu_galaxy_2012}. 
In addition, the mean black hole accretion rate \citep{delvecchio_mapping_2015,aird_x-rays_2019}, as well as the cold gas fraction \citep[e.g.][]{santini_evolution_2014,popping_inferred_2015} of a galaxy are substantially lower at $z<0.2$ than at $z\sim2$. 
Hence, we may expect different physical processes to be dominant at such a low redshift, which makes it necessary to also examine the role of major mergers with respect to triggering AGNs at such a cosmic time. Despite the expected small overall merger rates at low redshifts, especially for the particular population of AGNs showing the highest Eddington ratio, major mergers may still be the only viable option to deliver enough gas to the BH for it to reach such high specific accretion rates. 
% this paragraph is really long and I'm not sure what the main point is. Can you still shorten it? 

\begin{figure*}[t]
\centering
\includegraphics[width = 18cm]{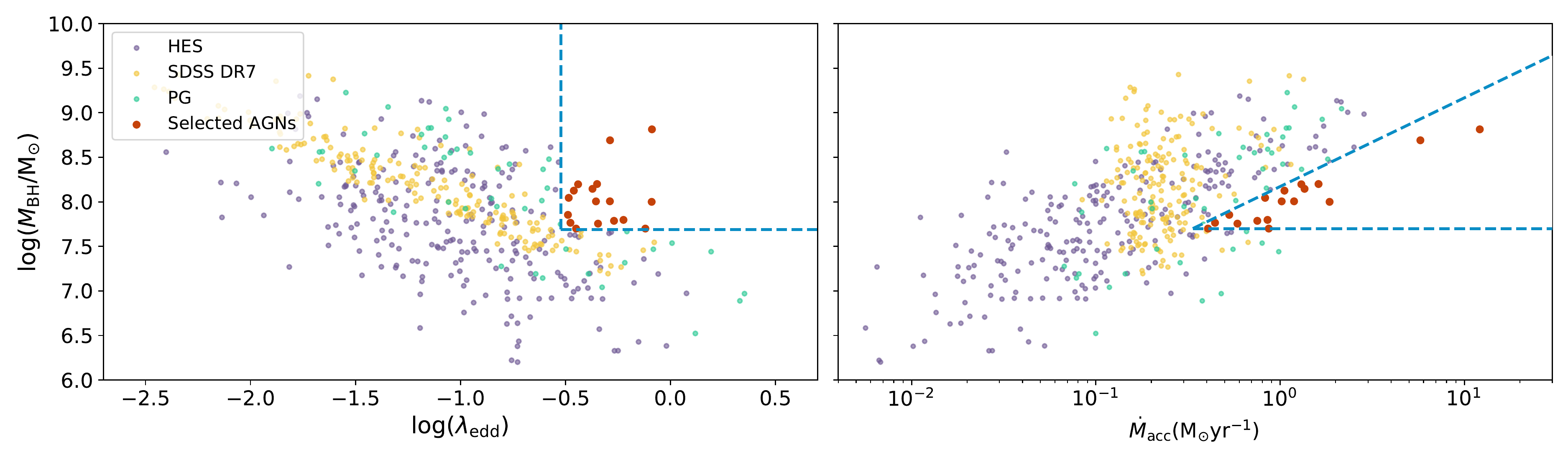}
\caption{\textit{Left}: Eddington ratio $\lambda_\mathrm{edd} = L/L_\mathrm{edd}$ vs.\ black hole mass for the parent sample of AGNs at redshift $z<0.2$. Overplotted are our selection limits in black hole mass and Eddington ratio (blue box) and our final selection of AGNs (red dots). \textit{Right}: Black hole mass accretion rate vs.\ black hole mass for the same sample indicating that our final selection consists of AGNs possessing the highest specific accretion rates.}
\label{fig:AGN_sample}
\end{figure*}

Like in almost all the aforementioned studies that reject major mergers as the dominant triggering mechanism of AGNs, we compare a  specific sample of AGN host galaxies to a sample of inactive comparison galaxies, matched in redshift, stellar mass, observed wavelength, depth, and signal-to-noise ratio (S/N). We examine 17 galaxies hosting AGNs with $\lambda_\mathrm{edd}>0.3$ at $z<0.2$ and 25 inactive control galaxies and compare the relative difference of the respective merger fractions in order to conclude whether major mergers play a dominant role. We derive the merger fractions by having experts visually classify and rank a joint blinded and randomized sample with respect to the appearance of distinct (major) merger features, such as tidal tails, shells, or asymmetries, which serve as proxies for an ongoing or recent past merger event. We then create a `consensus ranking' and subsequently split the sample again into AGN hosts and inactive galaxies in order to determine the separate fraction of distorted sources as the basis for discussion.

%This paper is structured as follows. In Section \ref{sec:data} we describe the data of our two samples, as well as their reduction and preparation for the subsequent analysis, which is presented in Section \ref{sec:analysis}. In Section \ref{sec:discussion} we discuss our results and their implications and compare them to previous studies, while we summarize our findings in Section \ref{sec:summary}. 
All magnitudes are given in the AB system and we adopt a concordance cosmology, with $\Omega_{\Lambda} = 0.7, \Omega_0 = 0.3$ and $h = 0.7$. At our sample's median redshift of $z\sim0.15$ $B$- and $V$- approximately correspond to rest-frame $U$- and $B$-band.

\section{Data} \label{sec:data}

We base the sizes of our two samples on the goal to identify a potential predominant presence of major merger signatures in AGN host galaxies with respect to a matched sample of inactive galaxies. 
As a fiducial initial condition, we assume  a merger fraction for our control sample of inactive sources of $f_{\mathrm{m,ina}} = 0.15$ with the goal to be able to detect for an AGN host galaxy merger fraction of $f_{\mathrm{m,agn}} \geq 0.5$ a significance difference between these two fractions with $\sim$99\% confidence. As the confidence of a detected difference in merger fractions can only increase for smaller values of $f_{\mathrm{m,ina}}$ and to ensure we achieve this desired level of confidence, we use
this, when compared to literature results \citep[e.g.][]{lotz_major_2011, man_resolving_2016, mundy_consistent_2017}, rather large value for $f_{\mathrm{m,ina}}$. We expect this fiducial fraction to be an upper limit of the real merger rate for inactive galaxies in our mass and redshift range.

Since the number of available AGNs with high Eddington ratios at $z<0.2$ is limited, we first create our sample of AGN host galaxies and then derive the number of inactive galaxies required to satisfy our conditions. With our final sample sizes we can then conclude whether or not AGN host galaxies show a significant enhancement in merger rates, indicating a causal dependence of our population of AGNs on major mergers.  

\subsection{AGN Host Galaxies} \label{subsec:AGN_sample}

We construct our parent AGN sample by making use of the catalogs provided by the Hamburg/ESO survey \citep[HES,][]{schulze_low_2010}, the Palomar Green Survey \citep[PG, ][]{vestergaard_determining_2006}, and the SDSS DR7 \citep{shen_catalog_2011}. We constrain our selection of potential targets to sources with a redshift of $z<0.2$. Since we require an estimate of the central BH mass and are interested in the AGNs with the highest specific accretion rates, we only select unobscured broad-line AGNs with an Eddington ratio $\lambda_{\mathrm{edd}} = L/L_{\mathrm{edd}} > 0.3$. To derive $\lambda_{\mathrm{edd}}$, we use the BH mass determinations based on single epoch H$\beta$ measurements and the bolometric luminosities, which, in turn, are based on the luminosities at 5100 \AA\ multiplied by a bolometric correction factor of $k_{\mathrm{bol}} = 9$ \citep{schulze_low_2010, netzer_bolometric_2019}. Both the BH masses and luminosities at 5100 \AA\ are taken from the respective catalogs.

We apply a minimum BH mass threshold of $\mathrm{log}(M_{\mathrm{BH}}/\mathrm{M_{\odot}}) = 7.7$, which results in a median BH mass for our AGN sample of $\mathrm{log}(M_{\mathrm{BH}}/\mathrm{M_{\odot}}) \sim 8.0$. Using the $\mathrm{M_{BH}}-\mathrm{M_{bulge}}$ scaling relation of \citet{kormendy_coevolution_2013} as a proxy to predict stellar host galaxy masses, the corresponding median stellar mass for our AGN host galaxies yields $\mathrm{log(}M_*\mathrm{/M_{\odot})} \sim 11$. This mass selection results in feasible exposure times for our inactive galaxies, which are required to be of equal stellar mass, and enables us to compare the results presented in this work with the findings of \citep{marian_major_2019}%for AGNs and matched comparison galaxies at $z\sim2$
, which are based on similar stellar host masses. Furthermore, we
%We apply a minimum BH mass threshold of $\mathrm{log}(M_{\mathrm{BH}}/\mathrm{M_{\odot}}) = 7.7$ and 
only select targets with a declination of $\mathrm{dec} < +15^{\circ}$ for better visibility with the \textsf{VLT}. All of these constraints yield a total number of 19 suitable AGN host galaxies, of which we observe 17 with \textsf{VLT FORS2} in $V$- and $B$-band (ESO programs 091.B-0672(A), 095.B-0773(A), and 098.A-0241(A), PI: Knud Jahnke). The median redshift of these 17 sources lies at $z=0.15$. 

The left panel of Fig.~\ref{fig:AGN_sample} summarizes our selection process. The smaller colored points show the respective parent catalogs (with HES in violet, PG in green and SDSS in  yellow) whereas the blue box shows the limits of our parameter space. Our final target selection is indicated by the red points. Since our AGNs show high Eddington ratios ($\lambda_{\mathrm{edd}} > 0.1$), we do not have to consider a potential trend of decreasing radiative efficiency $\eta$ with low accretion rates \citep{churazov_supermassive_2005, weinberger_supermassive_2018, nelson_first_2019} and can calculate the BH mass accretion rates $\dot{M}_\mathrm{acc}$ (Fig.~\ref{fig:AGN_sample}, right panel) as \begin{equation}
    \dot{M}_\mathrm{acc} = L/\eta c^2,
\end{equation} where we define $L$ as the derived bolometric luminosities and assume an efficiency parameter $\eta = 0.1$. The right panel of Fig.~\ref{fig:AGN_sample} %shows the resulting BH mass accretion rates versus the corresponding BH masses and
highlights that we target the AGNs with the highest specific accretion rates, i.e. those with the highest absolute mass accretion rates relative to their BH masses.

\begin{deluxetable*}{lhccccccDc}
\tablenum{1}
\tablecaption{AGN sample properties\label{tab:qso_prop}}
\tablewidth{0pt}
\tablehead{
\multicolumn2l{AGN designation} & \colhead{$z$} &  \colhead{$m_\mathrm{I}$} & \colhead{$L_{5100}$} & \colhead{$L_\mathrm{bol}$} & \colhead{FWHM} & \colhead{$M_{\mathrm{BH}}$} & \multicolumn2c{$\lambda_{\mathrm{edd}}$} & \colhead{$\dot{M}_\mathrm{acc}$}  \\
\multicolumn2l{\ } &  & \colhead{mag} & \colhead{$\textrm{erg\,s}^{-1}$}  & \colhead{log($\mathrm{L_{\odot}}$)} & \colhead{H$\beta$ (km s$^{-1}$)} & \colhead{log($\mathrm{M_{\odot}}$)} & \nocolhead{} & \nocolhead{} & \colhead{$\mathrm{M_{\odot} yr^{-1}}$} 
}
\decimalcolnumbers
\startdata
HE0119--2836	&&	0.12	&	14.8	&	44.92	&	12.29	&	3363.00	&	8.2	&	0.36	&	1.3	\\
HE0132--0441	&&	0.15	&	15.8	&	44.81	&	12.18	&	1719.00	&	8.0	&	0.44	&	1.0	\\
HE0157+0009	&&	0.16	&	16.1	&	44.73	&	12.10	&	2369.00	&	7.8	&	0.60	&	0.9	\\
HE0444--3449	&&	0.18	&	16.0	&	44.83	&	12.20	&	1714.00	&	8.1	&	0.35	&	1.1	\\
HE0558--5026	&&	0.14	&	15.5	&	44.88	&	12.25	&	1583.40	&	8.0	&	0.51	&	1.2	\\
HE1201--2408	&&	0.14	&	16.8	&	44.45	&	11.82	&	1820.86	&	7.8	&	0.33	&	0.4	\\
HE1226+0219	&&	0.16	&	13.2	&	45.89	&	13.26	&	3835.03	&	8.8	&	0.82	&	12.1	\\
HE1228+0131	&&	0.12	&	14.4	&	44.93	&	12.31	&	1866.19	&	8.1	&	0.43	&	1.4	\\
HE2011--6103	&&	0.12	&	16.3	&	44.53	&	11.90	&	2862.51	&	7.9	&	0.32	&	0.5	\\
HE2152--0936	&&	0.19	&	14.2	&	45.56	&	12.93	&	2183.42	&	8.7	&	0.52	&	5.8	\\
HE2258--5524	&&	0.14	&	15.9	&	44.68	&	12.05	&	2419.42	&	7.8	&	0.54	&	0.8	\\
PG1001+054	&&	0.16	&	16.3	&	44.74	&	12.11	&	1700.00	&	7.7	&	0.76	&	0.9	\\
PG1012+008	&&	0.19	&	16.2	&	45.01	&	12.38	&	2615.00	&	8.2	&	0.45	&	1.6	\\
PG1211+143	&&	0.09	&	14.3	&	45.07	&	12.44	&	1817.00	&	8.0	&	0.81	&	1.9	\\
SDSS-J032213.89+005513.4	&&	0.18	&	16.1	&	44.72	&	12.09	&	2440.00	&	8.0	&	0.33	&	0.8	\\
SDSS-J105007.75+113228.6	&&	0.13	&	15.7	&	44.57	&	11.94	&	1906.00	&	7.8	&	0.45	&	0.6	\\
SDSS-J124341.77+091707.1	&&	0.19	&	16.8	&	44.41	&	11.78	&	1979.00	&	7.7	&	0.36	&	0.4	\\
\enddata
\tablecomments{Properties of the AGNs in our sample: Columns 1--3, 6, and 7 are taken from the respective catalogs \citep{vestergaard_determining_2006, schulze_low_2010, shen_catalog_2011}. The bolometric luminosities $L_{\mathrm{bol}}$ in column 5 are calculated by applying a bolometric correction factor of 9 to $L_{5100}$\citep{schulze_low_2010, netzer_bolometric_2019}. Column 6 presents the FWHM of the broad component of H$\beta$. We calculate the Eddington ratios $\lambda_{edd}$ and black hole mass accretion rates $\dot{M}_\mathrm{acc}$ in column 8 and 9 by using the bolometric luminosities $L_{\mathrm{bol}}$, the respective BH masses $M_{\mathrm{BH}}$, and a radiative efficiency parameter of $\eta=0.1$. 
}
\end{deluxetable*}

Each target has been observed for at least three long exposures, to detect large-scale distortion features down to $B$ and $V\sim 23.4\ \mathrm{mag}/\mathrm{arcsec}^2$, and three short exposures, for an unsaturated image of the bright central region. The actual individual exposure times amount to 430\,s and 14\,s for $B$ and 150\,s and 8\,s for $V$, respectively. In Table \ref{tab:qso_prop}, we summarize the properties of our AGN sample. We cite the corresponding catalog designations, redshifts, apparent $I$-band magnitudes, as well as the luminosities at 5100~\AA, $L_{\mathrm{5100}}$, and the bolometric luminosities, derived by applying a correction factor of 9 to $L_{\mathrm{5100}}$ \citep{schulze_low_2010, netzer_bolometric_2019}. In addition, we state the catalog values for the FWHM of the single-epoch measurements of the (broad) $H\beta$ line, the respective BH masses $M_{\mathrm{BH}}$, along with the calculated Eddington ratios $\lambda_{\mathrm{edd}}$ and mass accretion rates $\dot{M}_\mathrm{acc}$.

\subsection{Inactive comparison sample} \label{subsec:comp_sample}

Given the size of the AGN sample and our assumptions for the merger fractions for our AGN and control sample ($f_{\mathrm{m,agn}} \geq 0.5$ and $f_{\mathrm{m,ina}} = 0.15$) we need to observe at least 25 inactive galaxies to meet our criterion of detecting a difference in those merger fractions with $\sim99\%$ confidence.
The comparison galaxies are randomly chosen from a parent sample of $\sim$2900 galaxies, which are part of the SDSS MPA/JHU catalog \citep{kauffmann_stellar_2003, brinchmann_physical_2004}. We perform this initial selection by constraining the declination to $\mathrm{dec}<10^{\circ}$ and the redshift to $z<0.2$, resulting in a median redshift of $z\sim0.13$ for our control sample. Furthermore, we only choose sources that possess comparable stellar masses to our AGN host galaxies. 

As described in Section \ref{subsec:AGN_sample}, we adopt
%To that end we use 
the $\mathrm{M_{BH}}-\mathrm{M_{bulge}}$ scaling relation of \citet{kormendy_coevolution_2013}
to derive the median stellar host mass 
%as a proxy to predict stellar host galaxy masses 
for the AGN sample from the inferred BH masses. We restrict the inactive galaxies to a small range around the median derived stellar mass of the AGN host galaxies, $\mathrm{log(}M_*\mathrm{/M_{\odot})} = 11 \pm 0.01$. Finally, we vet all potential sources against hard X-ray AGN signatures \citep{baumgartner_70_2013}, to remove any galaxies with a hidden, obscured AGN. In Table~\ref{tab:comp_gal_prop} we provide the coordinates, redshifts, $k$-corrected and dereddened $I$-band magnitudes, and median stellar masses from the MPA-JHU catalog for our comparison galaxies. 

\begin{deluxetable*}{lDDccc}
\tablenum{2}
\tablecaption{Comparison galaxy sample properties\label{tab:comp_gal_prop}}
\tablewidth{0pt}
\tablehead{
\colhead{Galaxy designation} & \multicolumn2c{$\alpha$(J2000)} &  \multicolumn2c{$\delta$(J2000)} & \colhead{$z$} & \colhead{$m_\mathrm{I}$} & \colhead{$M_\mathrm{*}$}  \\
\colhead{\ } & \multicolumn2c{deg} & \multicolumn2c{deg} &  & \colhead{mag} & \colhead{log($\mathrm{M_{\odot}}$)}}
\decimalcolnumbers
\startdata
Gal000232	&	0.164	&	-0.013	&	0.08	&	16.9	&	11.0 \\
Gal003114	&	2.083	&	-0.772	&	0.16	&	17.8	&	11.0 \\
Gal030481	&	19.605	&	-9.962	&	0.11	&	17.8	&	11.0 \\
Gal050873	&	34.151	&	-8.233	&	0.18	&	18.1	&	11.0 \\
Gal079769	&	50.365	&	-6.309	&	0.16	&	18.0	&	11.0 \\
Gal095873	&	58.093	&	-6.748	&	0.09	&	16.6	&	11.0 \\
Gal176221	&	132.158	&	7.598	&	0.13	&	17.9	&	11.0 \\
Gal185580	&	133.941	&	3.320	&	0.12	&	17.2	&	11.0 \\
Gal204260	&	137.351	&	9.810	&	0.16	&	18.0	&	11.0 \\
Gal210148	&	138.539	&	4.123	&	0.14	&	17.3	&	11.0 \\
Gal221730	&	140.921	&	-0.891	&	0.14	&	18.5	&	11.0 \\
Gal270096	&	150.303	&	-0.089	&	0.10	&	17.5	&	11.0 \\
Gal286443	&	153.515	&	7.057	&	0.10	&	17.2	&	11.0 \\
Gal347112	&	164.300	&	6.874	&	0.14	&	17.7	&	11.0 \\
Gal391560	&	171.878	&	-2.142	&	0.10	&	17.1	&	11.0 \\
Gal419090	&	176.075	&	-1.720	&	0.11	&	17.1	&	11.0 \\
Gal458007	&	181.927	&	1.421	&	0.11	&	17.5	&	11.0 \\
Gal498251	&	188.551	&	-1.446	&	0.16	&	17.7	&	11.0 \\
Gal510223	&	190.692	&	0.540	&	0.08	&	17.0	&	11.0 \\
Gal510224	&	190.692	&	0.540	&	0.08	&	17.0	&	11.0 \\
Gal534882	&	195.327	&	-0.937	&	0.19	&	18.2	&	11.0 \\
Gal557614	&	199.167	&	9.361	&	0.17	&	17.8	&	11.0 \\
Gal656010	&	215.724	&	8.849	&	0.14	&	17.3	&	11.0 \\
Gal676011	&	218.892	&	0.672	&	0.11	&	17.2	&	11.0 \\
Gal698144	&	222.606	&	6.647	&	0.16	&	18.0	&	11.0 \\
Gal782980	&	236.689	&	-0.860	&	0.07	&	16.1	&	11.0 \\
\enddata
\tablecomments{Our designations (column 1), coordinates (columns 2 and 3), redshifts (column 4) $k$-corrected and dereddened $I$-band magnitudes (column 5), and photometric median stellar masses for the inactive galaxies in our comparison sample taken from the MPA-JHU catalog \citep{kauffmann_stellar_2003, brinchmann_physical_2004}.}
\end{deluxetable*}

With the exception of one source\footnote{Due to weather losses one target was only observed in $V$-band}, all of the 25 galaxies in our final sample were observed in the $B$- and $V$-band with a comparable observational setup as for our AGN host galaxies. Each target has been observed with at least three individual, 470\,s and 180\,s long, exposures in $B$ and $V$, respectively. This selection and observational approach enables us to analyze two distinct samples of AGN host galaxies and inactive comparison galaxies, which are nonetheless matched in redshift, stellar (host) mass, depth, spatial resolution, filter band and S/N. Thus, we can directly compare potential relative differences in the merger fractions of both populations.

\vspace{1cm}

\subsection{Data reduction and preparation}

We require a seeing of 1\arcsec\ or better to diagnose large-scale merger signatures at a minimum required spatial resolution of $\sim$2.5\,kpc at our sample's median redshift. Hence, prior to reducing the raw images, we automatically determine the average seeing for each exposure by measuring the FWHM of 100 local peaks, using the \texttt{Astropy} package \texttt{photutils} \citep{larry_bradley_astropyphotutils_2019}, and calculating the corresponding median FWHM of all sources. We visually check and re-measure every single exposure with a median FWHM $> 1$\arcsec\ and discard individual exposures with a median FWHM above this threshold. Out of a total of $\sim$450 individual frames, we reject 22 from the subsequent reduction process and analysis. Despite the exclusion of these images, we end up with at least three individual exposures per band for every object.

To execute all the initial data reduction steps, i.e. the bias and flat-field correction, sky background subtractions, astrometry and aligning, and combination of individual exposures, we use the data processing pipeline \textsf{THELI}\footnote{https://www.astro.uni-bonn.de/theli/gui/index.html \\ https://github.com/schirmermischa/THELI} \citep{erben_gabods_2005, schirmer_theli_2013}. The resulting pixel scale of 0\farcs252 corresponds to $\sim$0.6\,kpc at our median redshift. We combine the respective $B$- and $V$-band observations to create color images using \textsf{MultiColorFits}\footnote{https://multicolorfits.readthedocs.io} \citep{cigan_multicolorfits_2019}.

To ensure that the samples are directly comparable, we mimic the appearance of the AGN host galaxies in the images of the inactive galaxies  by adding a synthetic point source on top of the respective flux centers. To this end, we first detect the 15 brightest, unsaturated stars within the central image regions around each inactive galaxy with the help of the \texttt{DAOStarFinder} algorithm within the \texttt{photutils} package. For each galaxy, we then visually select and cut out one of the detected stars, and upscale the brightness correspondingly, such that they possess a central brightness comparable to HE2152--0936, our second brightest AGN source. In the course of this procedure we also downscale noise in the outer parts. Since an upscaling with a constant factor would lead to a noticeable discrepancy in flux between the galaxy and the edge of the artificially enhanced point source, % we utilize radius-dependent scaling factors. More specifically,
we fit the original point sources with a 2D Gaussian and determine a circular region centered around the brightest pixel with a radius of 5$\sigma$. We divide this region into five bins and upscale the pixel values depending upon which bin they lie in. For the innermost region, i.e.\ within 1$\sigma$, we upscale with the total scaling factor, whereas for the outermost region, i.e. between 4 and 5$\sigma$ we apply a scaling factor lower by $5 \times 10^{-4}$. For the intermediate bins we choose a multiple of the scaling factor such that the distribution of the scaling factor with radius follows a Gaussian function.    
Using this approach we create point sources that resemble the central regions of our AGN host galaxies, but also blend in unrecognizably and smoothly into the respective galaxies. We add these point sources randomly at the centroid of each inactive galaxy, mimicking % eventually 
the appearance of our AGN host galaxies. Our point sources have a similar size to the upper limit of $\sim$1\arcsec\  set on the seeing, whereas the typical diameter of our sample galaxies, both AGN and inactive, is of the order of 5--6\arcsec. %Hence, a modeling and eventual point-source subtraction would not add significant information.
Thus, in contrast to our study of highly-accreting AGNs at $z\sim$2 \citep{marian_major_2019} %and after doing preliminary test runs with our low-redshift sample, 
%the angular sizes of the central point sources compared to the total extent of the galaxies are negligible. Thus, 
there was no necessity to model and subtract point sources for the samples here.
%modeling our sources and an eventual subtraction of the point-source model. Given the image quality and the redshifts of our sources, the angular sizes of the central point sources compared to the total extent of the galaxies are negligible.

\begin{figure*}
\centering
\includegraphics[width = 18cm]{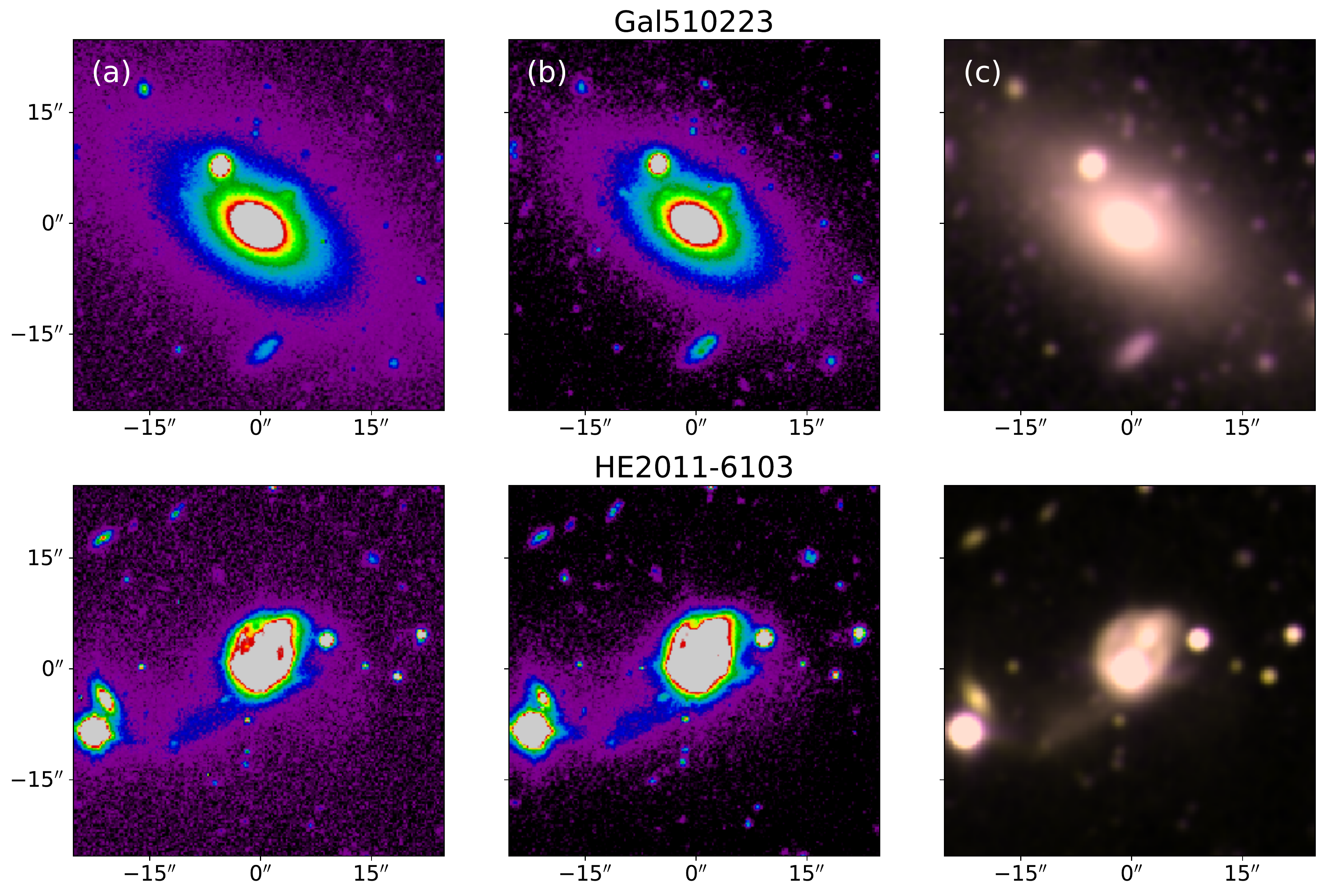}
\caption{Two sources representative for our targets. On the top row we show one of the comparison galaxies and on the lower row one of our AGNs is displayed. From left to right we present a postage stamp in (a) V-band, (b) B-band and (c) color, respectively. Note: In order to enhance the visibility the images are not shown with the same cuts and color map parameters. %25" cutout region
}
\label{fig:showcases}
\end{figure*}

An example of an AGN host galaxy and an inactive comparison galaxy are shown in Fig.~\ref{fig:showcases}. The left (a) and middle column (b) depict the $V$- and $B$-band images, respectively, whereas the right column (c) shows the color images. To optimize the visibility of large-scale structures and possible merger signatures, while blending out the brightest inner regions, we chose different parameters for the color cuts and color map for the single band images as well as the color images. However, within one set, i.e.\ $V$-, $B$-band or color images, the parameters are constant. In addition, we adopted a Gaussian 2-pixel smoothing for the color images only. %in order to %detect
Due to the different visualization of the sources, we can test for any systematic differences in the subsequent distortion rankings or the resulting merger fractions 
(see Section \ref{sec:discussion}). 

\section{Morphological analysis \& merger fractions} \label{sec:analysis}

\begin{figure*}[t]
\centering
\includegraphics[width = 18cm]{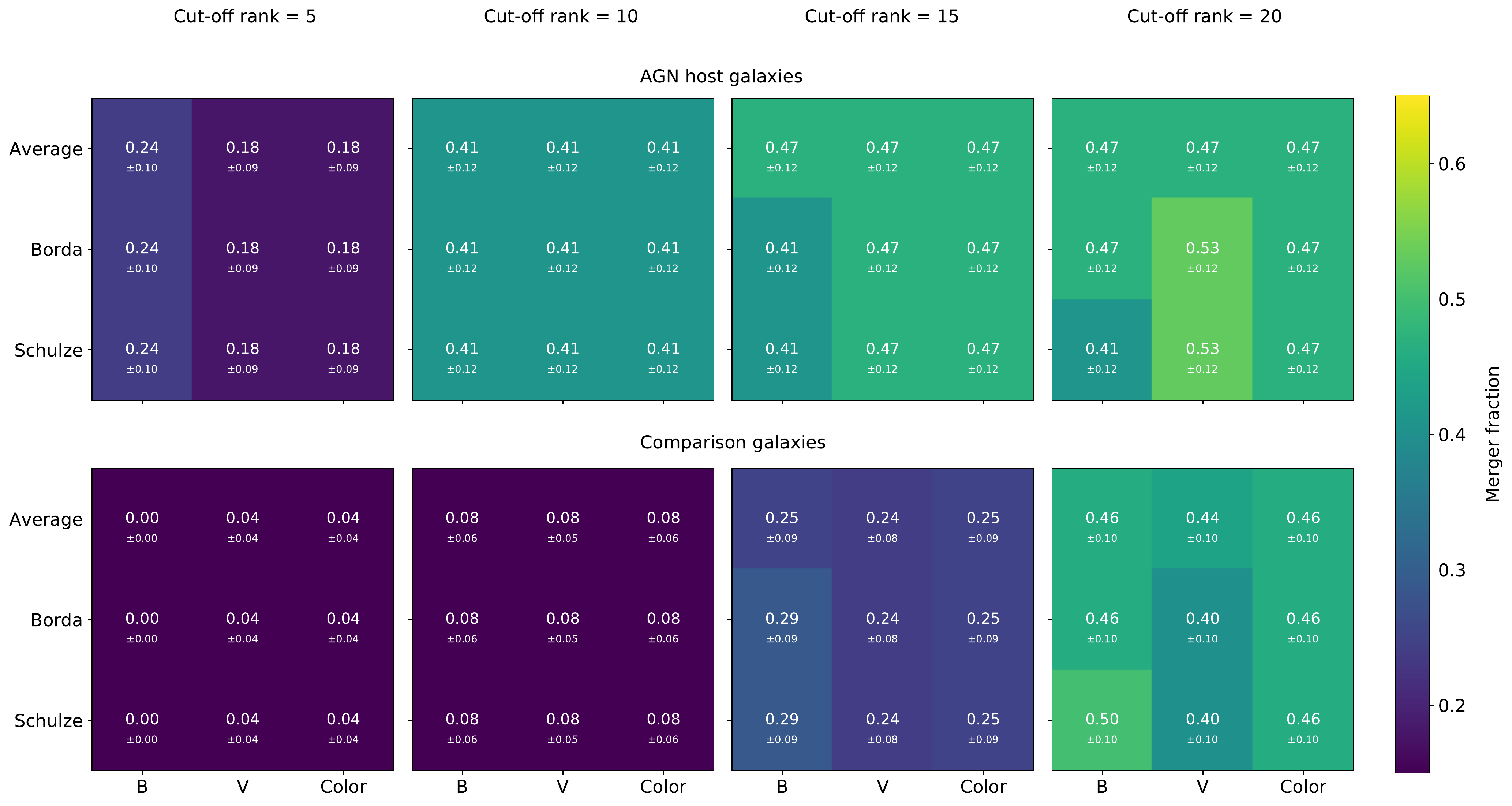}
\caption{The merger fractions for every set ($B-$, $V-$band, and color images) and ranking combination method (Average, Borda, Schulze) for four distinct cut-off ranks. In the top row we show the corresponding fractions of disturbed AGN host galaxies, the bottom row depicts analogously the inactive comparison galaxies. The smaller numbers below the actual merger fraction values give the standard deviations (i.e. 1$\sigma$) of the corresponding beta distributions.}
\label{fig:Merger_Fractions}
\end{figure*}

% After the preparation process  all objects are indistinguishable with respect to the presence of an AGN, i.e. it is not visually recognizable whether the respective source is an AGN host galaxy or an inactive galaxy. 

We join both processed samples (for which the galaxies can no longer be visually separated as AGN or not)  resulting in a final sample of 42 sources in $V$- and 41 in $B$-band and color, respectively. To derive the merger fractions, 19 experts\footnote{The rankings were done by the coauthors Andika, Ba\~{n}ados, Bennert, Cohen, Husemann, Jahnke, Kaasinen, Koekemoer, Marian, Onoue, Schindler, Schramm, Schulze, Silverman, Smirnova-Pinchukova, van der Wel, Villforth and Windhorst}, proficient in working with imaging data of galaxies, perform a visual assessment of the targets, ranking them from most to least distorted with respect to the appearance of large scale distortion features. These features are indicative of  ongoing or recent major merger events. Each set of $V$-, $B$-band and color images  is ranked independently by each expert. We note that there are an increasing number of machine learning algorithms that can classify galaxies, based on their morphologies and possibly merger state \citep[e.g.][]{bottrell_deep_2019, cheng_optimising_2019, snyder_automated_2019}. However, we rely on the human interpretation and judgment due to the manageable sample size and the extensive time and logistic requirement to teach an automatic classification routine with a  matching `external' training set. % to %setup and
Since the sources in the joint sample are indistinguishable with respect to whether or not they are active, every expert's individual bias regarding the classification of a major/minor merger applies equally to AGN host galaxies and comparison galaxies. Thus, in our subsequent analysis any personal subjectivity in classification will have the same impact on either of the two subsamples. To further reduce any systematic bias, the dataset provided to each of the 19 ranking experts is randomized. As an additional task we request every classifier to choose a ``cut-off'' rank below which they deem all sources to be in a merging state, or, to at least show signs of a recent gravitational disturbance, like asymmetries, tidal tails or double nuclei. Every galaxy with a rank higher than the cut-off is interpreted to be completely free of major disturbances stemming from interactions. In our ensuing analysis we will use this property to determine the merger fractions of our two samples and also discuss the dependence of those fractions on different cut-off ranks (see Sect. \ref{sec:discussion}). 

We combine the 57 individual rankings (19 experts times three sets) into three consensus sequences for each respective set.
We apply the same methods as in \citet{marian_major_2019} to combine the individual rankings and repeat this task for each set, i.e.\ separately for $B$-, $V$-band, and color images. For our first approach we calculate and weigh the average ranks of each galaxy, whereas for our second and third approach we use the Borda count \citep{emerson_original_2013} and Schulze algorithm \citep{schulze_new_2011, schulze_schulze_2018}, respectively. More information on the different methods and on how we implement them are provided in Appendix~\ref{appendix:details_methods}. Ultimately, by applying all three methods to all three sets we obtain nine overall rankings. 

We select various cut-off ranks and
%We 
split the combined rankings back into AGN host and comparison galaxies. Subsequently, we derive the merger fractions for each chosen cut-off rank by counting how many active and inactive galaxies are above and below this threshold.
The merger fraction is then simply defined as, 
\begin{equation}\label{mean}
    f_m = \frac{a}{a+b} ,
\end{equation}
where $a$ represents the number of merging galaxies, whereas $b$ counts the sources that are undisturbed. 
However, since we only examine samples of limited size, we need to quantify the probability densities and uncertainties introduced by the shot noise for our resulting merger fractions. Based on those two parameters $a$ and $b$ we can quantify the probability densities for a continuous range of merger fractions in the feasible interval $[0,1]$ by using the
beta distribution \citep[see also][]{mechtley_most_2016, marian_major_2019}, 
\begin{equation}
f(x) = \frac{(a+b+1)!}{a!\,b!} x^{a-1} (1-x)^{b-1} \textrm{.}
\end{equation}
The respective standard deviations and means of the associated merger fraction probability distributions are then derived by,
\begin{equation}
    \sigma(x) = \sqrt{\frac{ab}{(a+b)^2(a+b+1)}} ,
\end{equation}
and Eq.~\ref{mean}, respectively.
%The probability distributions of the merger fractions is described 
%represents the number of disturbed galaxies above the cut-off rank, whereas $b$ counts the sources that are below this limit. 

In Figure~\ref{fig:Merger_Fractions}, we present the corresponding means and standard deviations of the various probability distributions for every combination of method and set for four distinct cut-off ranks at 5, 10, 15, and 20. The merger fractions increase with cut-off rank, because a higher cut-off rank means that more galaxies are below this limit and are thereby considered to exhibit merger features. We find no evidence that the choice of combination method or the choice of $B$-, $V$-, or color image-set affect the resulting merger fractions. For all combinations the results for a given sample and cut-off rank are well within the errors of each other or even equal. However, it is also evident that for cut-off ranks $\lesssim$15 the merger fractions for the AGN host galaxies (Fig.~\ref{fig:Merger_Fractions}, upper row) are significantly larger then the fraction of disturbed inactive galaxies (Fig.~\ref{fig:Merger_Fractions}, bottom row). This is not the case for larger cut-off ranks. 
% Thus, in the next section we discuss which choices of cut-off ranks are valid and its 
We discuss the implications of the chosen cut-off ranks on our recovered merger fractions and the potential causal connection between major mergers and the triggering of AGNs in the following section.

%are of our thus recovered merger fractions, and a potential causal connection between major mergers and the triggering of AGNs. 

\subsection{Constraining the absolute merger fractions}\label{subsec:absolute_merger_fracs}

We have calculated the merger fractions for two samples of 17 AGN host galaxies and 25 inactive comparison galaxies. As mentioned in the preceding section, the final merger fractions depend on the choice of cut-off rank. In Appendix~\ref{appendix:frac_m_rank_dependence} we present the continuous evolution of merger fractions with cut-off rank for all combinations of set and method, while in this section we describe the two approaches we used to analyze and interpret our results. Firstly, we base the cut-off rank on our experts' opinions, and secondly, we construct this limit so that the resulting merger fraction of our inactive control sample is consistent with the merger rates presented in the literature. To obtain a valid first estimate, we calculated the means of the individual cut-off ranks chosen by each classifying expert for each set. The average cut-off ranks are 21$\pm$8, 22$\pm$9, and 18$\pm$8 for the $B$, $V$, and color sets, respectively.

We suspect that the reason for such high cut-off ranks, which are almost bisecting our joint samples, lies in the visual determinations of our experts. Since our galaxies are well-resolved, any minor asymmetries  (which do not need to stem from a recent major merger event, but can be of a minor merger or secular origin) can be easily identified. This leads our experts to put those particular sources into the `merger bin', i.e. below the cut-off rank, increasing %obviously 
the percentage of galaxies classified as merging. 
With a corresponding cut-off rank = 20,  the merger fractions range between $f_\mathrm{m,agn}$ = 0.41 $\pm$ 0.12 and $f_\mathrm{m,agn}$ = 0.53 $\pm$ 0.12 for the AGN sample and $f_\mathrm{m,ina}$ = 0.40 $\pm$ 0.10 and $f_\mathrm{m,ina}$ = 0.50 $\pm$ 0.10 for the inactive sample. Therefore, the fractions of disturbed sources in both samples are not significantly different, which would indicate a negligible contribution of mergers of any strength to the triggering of AGNs.

However, our primary goal is to determine the distinct impact of \textit{major} mergers on the formation of AGNs, without considering the effects of minor gravitational encounters or other processes shaping the morphology of a galaxy. Thus, we have to correct our recovered merger fractions for the contamination by sources with minor asymmetries. Such a high merger rate of $\sim$40--50\%, indicates that approximately half of the population shows signs of a recent or ongoing gravitational encounter of any strength. This significantly exceeds our initial assumption for inactive galaxies (see Sect.~\ref{sec:data}) and also the assessments by previous studies %examining observed and simulated major merger rates per galaxy for sources at different redshifts and stellar masses
\citep{lotz_galaxy_2008, lotz_evolution_2008, lotz_major_2011, bridge_cfhtls-deep_2010, xu_galaxy_2012, casteels_galaxy_2014, man_resolving_2016, ventou_muse_2017, ventou_new_2019, duncan_observational_2019, oleary_emerge_2020}. Based on these studies we adopt a major merger rate per galaxy of $R_\mathrm{m} \sim 0.05~[\mathrm{Galaxy}^{-1}\, \mathrm{Gyr}^{-1}]$. This number represents the number of galaxies currently in a merger state, divided by the timescale of the visibility of merger signatures. In order to obtain an absolute merger fraction, representative of our comparison sample, we need to multiply this rate with the timescale $T_\mathrm{m}$ in which a major merger is observable. This property not only depends strongly on the mass ratio, individual masses, and gas fractions of the two progenitor galaxies, but also on the depth of the observations. Considering our targets' low redshifts and surface brightness limits we choose a comparatively conservative value of $t_\mathrm{m} \sim 1.5$~Gyr, which results in a major merger fraction of $f_\mathrm{m} \sim 0.08$ for galaxies in our mass bin and at our sample's redshift. 

\begin{figure}[t]
\centering
\includegraphics[width = 9cm]{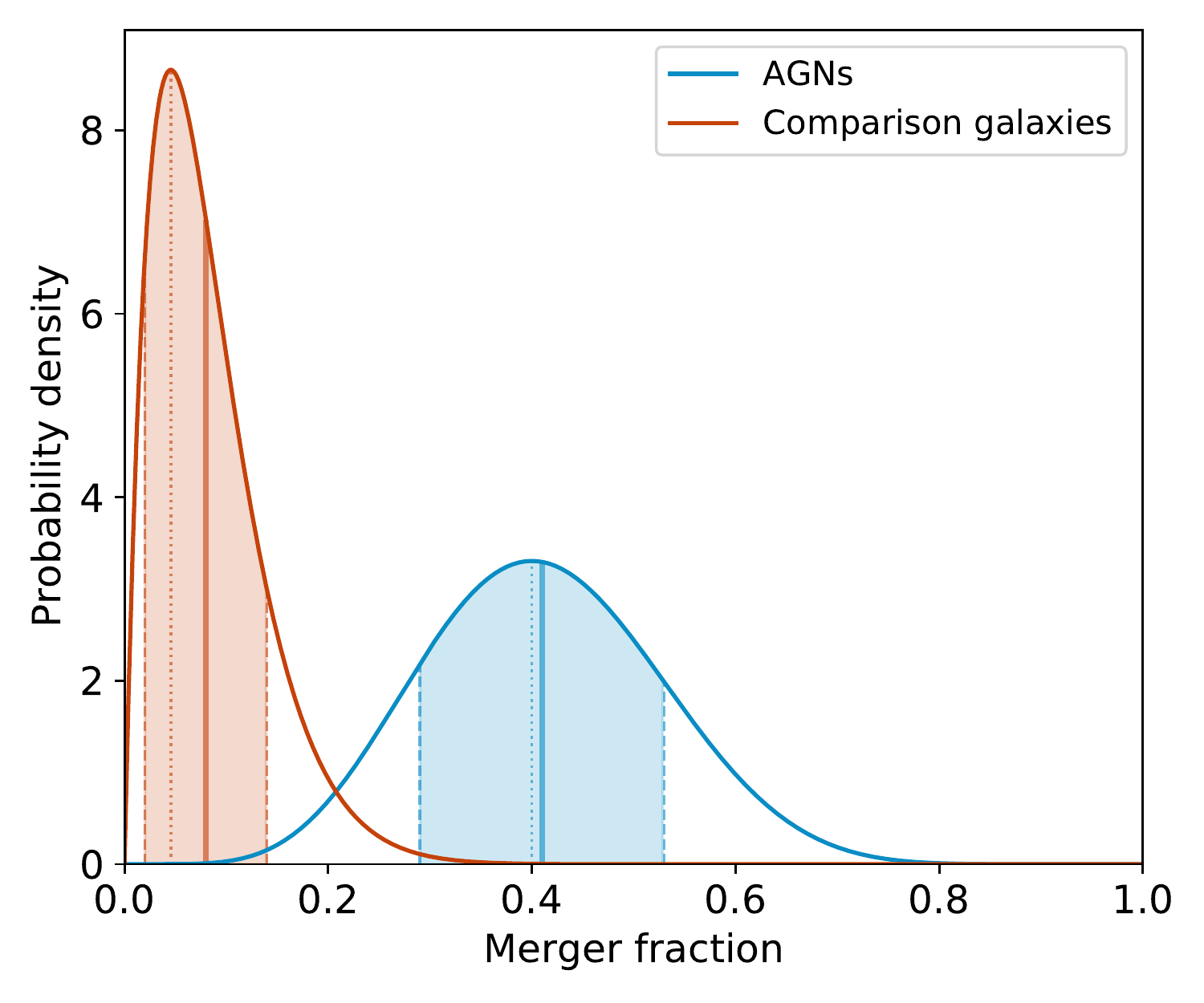}
\caption{Probability distributions for the derived merger fractions of our $z<0.2$, high-accretion AGN host galaxies (blue) and inactive galaxies (red) at a cut-off rank = 10. The solid and dotted lines show the means and modes of the respective merger fractions, while the dashed lines and shaded regions depict the central 68\% confidence intervals. At this particular cut-off rank the respective merger fractions are identical, independent of method and set.}
\label{fig:Merger_Fractions_PDF_showcase}
\end{figure}

Such a value for the merger fraction for our comparison galaxies corresponds to a cut-off rank $= 10$. Coincidentally, at this cut-off rank the respective merger fractions are equal over all sets and methods  for each of the two samples (Fig.~\ref{fig:Merger_Fractions}) and yield $f_\mathrm{m,agn}$ = 0.41 $\pm$ 0.12 for the AGN host galaxies and $f_\mathrm{m,ina}$ = 0.08 $\pm$ 0.06 for the comparison galaxies. This value of $f_\mathrm{m,ina}$ is not only in excellent agreement with the major merger rates found in the 3DHST survey by \citet{man_resolving_2016} for all five fields (AEGIS, COSMOS, GOODS-N, GOODS-S, UDS) in CANDELS \citep{grogin_candels_2011, koekemoer_candels_2011}, but also for the major merger fractions recovered by MUSE deep observations \citep{ventou_muse_2017, ventou_new_2019} as well as studies by \citet{duncan_observational_2019} in CANDELS, and in GAMA by \citet{mundy_consistent_2017}.

The two corresponding probability distributions are shown in Figure~\ref{fig:Merger_Fractions_PDF_showcase}, with blue and red denoting the probability distributions for the AGN sample and the comparison sample, respectively. The shaded regions represent the 1$\sigma$ intervals and the solid and dotted lines depict the corresponding means and the modes. Due to the low number of merging comparison galaxies the associated probability distribution appears considerably skewed with the corresponding mean not coinciding with the peak position. Thus, we also report the merger fraction associated with the mode of the distribution, which yields $f_\mathrm{m,ina}\sim$ 0.04 and is still well within the error of the mean.

\section{Discussion} \label{sec:discussion}

\subsection{Robustness of results}\label{subsec:robustness_of_fracs}

For a cut-off at rank 10, the resulting merger fractions translate to, 

\begin{itemize}
\renewcommand\labelitemi{--}
    \item 7/17 AGN host galaxies showing merger features, and,
    \item 2/25 inactive galaxies showing merger features.
\end{itemize}

The order and appearance of the sources in the various consensus rankings do not have to be congruent, e.g. the 7 as merger classified AGN host galaxies could vary in the different consensus rankings. However, we find that, despite a difference in order, the first eight positions of every combined ranking feature the same targets, with seven of them being the same AGN host galaxies. Out of these seven targets, five stem from the HES sample, while one each is listed initially in the SDSS and PG catalogs, respectively. Since we selected a total of 11 AGNs from the HES catalog and in each case three from the SDSS and PG catalogs, we conclude that the parent catalogs from which the AGN host galaxies are drawn from are not introducing any bias with respect to morphological classification. A repeated visual inspection also confirms that a distinction at exactly this cut-off rank into merging and non-disturbed systems reveals a noticeable separation into sources with clearly obvious large-scale merger features like tidal tails and shells and galaxies with explicitly less asymmetries.

Eventually, we created one singular overall ranking by re-applying the Schulze method on the nine consensus rankings (see Appendices $\ref{appendix:meta_ranking}$ and $\ref{appendix:meta_ranking_tab}$). The same sources that occupy the first eight ranks in the nine initial consensus sequences, populate the highest positions in this final ranking as well. Therefore, we obtain an unchanged result for both merger fractions after again applying a cut-off at rank 10. 

Considering the appearance of seven AGN host galaxies among the eight highest-ranked sources and the clear excess in merger fractions for the AGN host galaxies with respect to the inactive sample with a significant difference of $>2.5\sigma$, we conclude that major mergers are an essential triggering mechanism for AGNs with the highest Eddington ratios at $z<0.2$. However, based on the mean of our recovered probability distribution for the AGN merger fraction we only find a $\sim$22\% probability that the merger fraction is above the threshold of $f_\mathrm{m,agn}$ = $0.5$. %, which would imply a majority of AGNs being of a major merger triggered origin. 
This means that although major mergers are indeed a non-negligible mechanism in triggering our specific population of AGNs, more than half of the BHs must be activated by different means, like secular processes or minor mergers. We discuss the role of the latter in triggering AGNs with the highest specific accretion rates at low redshifts in more detail in Sect.~\ref{subsec:minor_mergers}.

\subsection{Comparison to previous studies}

Our result, which shows an excess in AGN merger fraction compared to a matched control sample, stands in contrast to recent simulations \citep{steinborn_cosmological_2018, ricarte_tracing_2019} and several previous empirical studies examining the potential causal connection between major mergers and the triggering of different populations of AGNs. 
\citet{villforth_morphologies_2014} found no increase in merger signatures with luminosity and also reported consistent disturbance fractions between the AGNs and comparison galaxies for their sample of observed low- and moderate-luminosity AGNs ($41 \lesssim L_X \mathrm{[erg\ s^{-1}]} \lesssim 44.5$) at $0.5 \lesssim z \lesssim 0.8$. %A study by
In contrast, \citet{silverman_impact_2011} found an enhanced merger rate for AGNs of moderate X-ray luminosities in spectroscopic pairs at $z < 1$. However, their rate of $17.8^{+8.4}_{-7.4}\%$ is still significantly lower than what we find here. 

AGNs and host galaxies at comparable redshifts and luminosities as our sample were explored by \citet{bohm_agn_2013} and \citet{grogin_agn_2005}. They assessed the neighboring counts, asymmetries, and various morphological indices (concentration, Gini coefficient and $M_\mathrm{20}$ index) to characterize the respective host galaxies, but found no significant causality between major mergers and AGNs. Likewise, \citet{allevato_xmm-newton_2011}, \citet{ schawinski_hst_2011}, and \citet{rosario_host_2015} detected no redshift evolution of morphological properties for similar AGNs up to $z\sim$~2.5 and \citet{kocevski_candels_2012} found that only $16.7^{+5.3}_{-3.5}$\% of comparable AGNs at $z\sim$~2 are highly disturbed. X-ray-selected and optically observed AGNs with higher luminosities ($43 \lesssim L_X \mathrm{[erg\ s^{-1}]} \lesssim 46$) at $0.5 \lesssim z \lesssim 2.2$ also appear to show no causal link to major mergers \citep{cisternas_bulk_2011,hewlett_redshift_2017,villforth_host_2017}.  
% find no evidence to suggest that galaxies hosting such AGNs to be systematically more disturbed when compared to a matched inactive sample. 
Instead they all reported consistent merger fractions of $\sim$~15--20\%. Regarding more specific populations at $z\sim$~2 \citet{schawinski_heavily_2012} presented a major merger fraction between 4\% and 11\% for their analyzed sample of 28 dust-obscured AGNs, while \citet{mechtley_most_2016} has found consistent merger fractions for 19 galaxies hosting the most massive supermassive BH ($M_\mathrm{BH} = 10^9 - 10^{10} \mathrm{M_{\odot}}$) and a sample of 84 matched inactive galaxies. Similarly, in our previous work \citep{marian_major_2019} in which we examined 21 AGNs with the highest Eddington ratios ($\lambda_\mathrm{edd} > 0.7$) at $z\sim$2 and compared them to 92 matched inactive galaxies we found no dominant connection between major mergers and the occurrence of AGNs.

Similar to the results presented in this work other studies have found considerably enhanced merger rates for particular populations of AGNs. For their sample of hard X-ray detected, moderate luminous AGNs at $z<0.05$ \citet{koss_merging_2010} reported an enhanced merger fraction of 18\% when compared to a matched control sample, in which only 1\% of the sources display merger features. However, they speculated that their AGNs may not be classified correctly via means of optical diagnostics due to superimposing features of ongoing star formation and optical extinction. In fact it appears that, independent of redshift, obscured and luminous AGNs are more likely to be connected to major merger events. Albeit, it should be noted that this is expected since by focusing on obscured sources a bias towards merging systems is most likely introduced as that obscuration may be due to dust within a merging (U)LIRG-like host. With this caveat in mind, \citet{glikman_major_2015}, \citet{fan_most_2016}, and \citet{donley_evidence_2018} detected merger fractions $>50\%$ for such reddened or obscured AGNs sources at $z\sim$~2, $z\sim$~3, and $0<z<5$, respectively. Also at low redshifts ($z\lesssim0.2$) \citet{koss_population_2018} and \citet{ellison_definitive_2019} presented comparable results. In addition, in the latter study the authors described an increase of merger fraction with AGN luminosity, with the most luminous AGNs exhibiting the highest merger incidence. Corresponding findings have also been reported by \citet{treister_major_2012}, \citet{hong_correlation_2015} and \citet{goulding_galaxy_2018}, who have analyzed luminous AGNs ($\mathrm{log}(L_\mathrm{bol}\mathrm{[erg~s^{-1}]})>45 $) at various redshifts. Especially with a merger fraction of $\sim$~44\% for luminous AGNs at $z<0.3$ the results published in \citet{hong_correlation_2015}  are very consistent with the distortion rate we find for our sample of AGNs of comparable bolometric luminosity. Similar results are also reported by \citet{gao_mergers_2020} for their sample of AGNs at $0<z<0.6$, who detected a merger fraction of $\sim$40\% and a general increase of distortion incidence with stellar mass. Finally, \citet{mcalpine_rapid_2018, mcalpine_galaxy_2020} reported for the \texttt{EAGLE} simulation that major mergers -- while of no great importance at high redshifts -- play a significant role at low redshifts and present a consistent major merger fraction of $\sim$~40\% for BHs growing rapidly at $z\sim$~0.

\vspace{1cm}

\subsection{Physical interpretation and comparison to AGN counterparts at z\texorpdfstring{$\sim$}{\~{}}2}

%Since we have chosen the AGNs with the highest Eddington ratios, i.e.\ the sources with the highest accretion rates relative to their BH mass, we have also not necessarily selected the total most luminous AGNs from our parent samples, but rather the most luminous with respect to the corresponding BH masses. Except for the two aforementioned AGNs -- HE1226+0219 and HE2152--0936, which possess bolometric luminosities of $\mathrm{log}(L_\mathrm{bol}\mathrm{[erg~s^{-1}]})>46.5 $ -- all our remaining sample AGNs have luminosities of $45.3 \lesssim \mathrm{log}(L_\mathrm{bol} \mathrm{[erg~s^{-1}]}) \lesssim 46$, but feature the smallest BH masses in that luminosity bin ($7.7 < \mathrm{log}(M_\mathrm{BH}/ \mathrm{M_{\odot}}) < 8.2$).

In light of our previous work at $z \sim$~2 \citep{marian_major_2019}, which also focuses especially on AGNs with the highest Eddington ratios, but yields an opposite result, we need to consider the different epochs of the studied AGNs. To make a comparison in absolute terms between the AGN major merger fractions, which we recover for the respective two samples at $z\sim2$ and $z<0.2$, we have to factor in the impact of surface brightness dimming on detecting possible faint morphological distortion features. With a drop in surface brightness of $\sim$5mag/arcsec$^2$ at z\s2, we miss at this redshift most definitely merger features we otherwise would see at $z\sim0.2$. 
This effect can be enhanced by the fact that galaxies at $z\sim2$ are on average more compact than at $z\sim0$ \citep[e.g.][]{van_der_wel_3d-hstcandels_2014}.
If the triggering of an AGN follows immediately after a starburst caused by a galaxy merger, the resulting potential extensive amount of dust can obscure the starburst at $z\sim2$ more easily than at $z\sim0.2$. In the latter case the starburst may happen as much in the galaxy's outer spiral arms and tidal streams, whereas the starburst in a galaxy at $z\sim2$ is  much more confined to the central region due to its more compact nature. Hence, in addition to the difference in surface brightness dimming between $z\sim2$ and $z\sim0.2$, a more complex situation is possible where the visibility of an AGN host galaxy at $z\sim2$ is not only reduced by surface brightness dimming, but also obscuring dust.
Thus, the AGN merger fraction at $z\sim2$ could be significantly underestimated with $f_\mathrm{m,agn}$ = 0.24 $\pm$ 0.09 for the AGN sample at $z\sim2$ and $f_\mathrm{m,agn}$ = 0.41 $\pm$ 0.12 for the AGNs presented in this study (see Sect.~\ref{subsec:absolute_merger_fracs}). Therefore, this effect could explain the discrepancy in the derived AGN major merger rates and would lead us to the conclusion that a substantial part of AGNs with the highest Eddington ratios at $z\sim2$ is actually triggered by major mergers as well. However, in \citet{marian_major_2019} as well as in this study we draw our main conclusions by comparing the respective AGN samples to two matched control samples of inactive galaxies at both redshifts and determining primarily the relative differences between the respective merger fractions. The corresponding merger fractions for the inactive galaxies are $f_\mathrm{m,ina}$ = 0.19 $\pm$ 0.04 and $f_\mathrm{m,ina}$ = 0.08 $\pm$ 0.06 for the sources at $z\sim2$ and $z\sim0.2$ (see Sect.~\ref{subsec:absolute_merger_fracs}), respectively. 

We assume now that the actual merging process is independent of the presence of a potential future AGN and consider the mechanisms causing the detectable morphological features to be identical between the respective AGN host galaxies and their corresponding inactive counterparts. 
As a result the merger fractions at $z\sim2$ are affected equally by surface brightness dimming and we actually do not have to consider this effect.
Similarly, a merger-driven starburst creating an abundant amount of obscuring dust can happen equally in both an inactive galaxy or a system that will host an AGN triggered by this merger event. Hence, dust would only impact the findings described in \citet{marian_major_2019} if the dust were to obscure the actual AGNs, which would lead to a misclassification of those particular sources as inactive galaxies. In this earlier study, however, we investigated the importance of hidden and intermittent AGNs at $z\sim2$, which would implicitly include such sources, but found no significant effect on the resulting merger rates. In addition, just as in this work, we deliberately have only selected type-1 AGNs, minimizing the probability of dust obscured sources influencing the reported result. We expect the number of such sources with a dust content low enough to be not classified as type-2 AGNs, but sufficiently high to actually hide a potential AGN or morphological merger features to be relatively low. %that could have influenced the findings in \citet{marian_major_2019} 
%However, if AGNs are predominantly triggered by major mergers and the respective host galaxies have therefore a higher distortion rate, it is also more probable for those systems to be obscured by dust \textbf{compared to the overall galaxy population}. 
Therefore, similar to the surface brightness dimming, we can neglect the effect of obscuring dust when considering the relative difference in merger fractions at $z\sim2$.
%Comparable to the surface brightness dimming affecting inactive and AGN host galaxies similarly, we consider the obscuration by dust to be independent of the galaxy hosting an AGN or being inactive, 
%if the dust hides only the morphological signs of a recent major merger. % and the responsible starbursts are not more prevalent in one of the two samples. 
%Still, it would impact the findings described in \citet{marian_major_2019} if \textbf{the dust were to obscure the actual} AGNs, which would lead to a misclassification of those particular sources as inactive galaxies. In this earlier study, however, we investigated the importance of hidden and intermittent AGNs at $z\sim2$, which would implicitly include such sources, but we found no significant effect on the resulting merger rates.  
%In addition, just as in this work, we deliberately have only selected type-1 AGNs, minimizing the probability of dust obscured sources influencing the reported result. Hence, only sources with a dust content low enough to be not classified as type-2 AGNs, but sufficiently high to actually hide a potential AGN or morphological merger features could have influenced the findings in \citet{marian_major_2019}.
%We expect the number of such sources to be relatively low and thus consider the overall impact of obscuring dust on the result to be negligible. 
A rigorous analysis would require a larger sample and data at longer wavelengths, as the James Webb Space Telescope (JWST) will be able to provide at z$\sim$2, enabling spatial modeling of dust in more detail.

Since we only compare the relative differences in merger fractions at both redshifts with no significant distinction of merger fractions at $z\sim2$, but a clear excess of the AGN merger fraction when compared to inactive galaxies at $z<0.2$ we still conclude that major mergers play an essential role for AGNs with high Eddington ratios at low redshift. In addition, the major merger fractions for both samples of inactive galaxies are consistent with previous findings of major merger rates for galaxies at comparable redshifts and masses \citet{man_resolving_2016,snyder_automated_2019,steinborn_cosmological_2018,ventou_muse_2017, ventou_new_2019}. This agreement corroborates the findings presented in \citet{marian_major_2019} and indicates that surface brightness dimming or dust is actually not impacting the merger fractions at $z\sim2$ considerably. Hence, we have to consider an alternative explanation for this excess of merger fraction in our subpopulation of AGNs at low redshift.

Besides the mean BH accretion rate/bolometric luminosity and Eddington ratio of an AGN \citep{schulze_cosmic_2015}, especially the cold gas fraction of a galaxy at $z<$~0.2 is considerably lower than for a counterpart at $z\sim$~2 \citep[e.g.][]{santini_evolution_2014,popping_inferred_2015}. Hence, with the AGNs in both redshift samples having comparable Eddington ratios, but the sources at lower redshifts a significant smaller intrinsic gas reservoir it is reasonable to assume that while at $z\sim$~2 a sufficient amount of gas is still left to fuel the central supermassive BH via other mechanisms than major mergers, at $z<$~0.2 this process is essential to trigger AGNs with the highest specific accretion rates.  
This scenario is completely consistent with the results of the \texttt{EAGLE} simulations, which sees major mergers in a negligible role for triggering AGNs at high redshifts, but shows that such galaxy encounters play a substantial role at low redshifts, %with respect to the rapid growth of BHs 
yielding comparable major merger fractions \citep{mcalpine_rapid_2018, mcalpine_galaxy_2020}. However, it should be noted that despite the excess in major merger fraction for our AGN host galaxies, $\gtrsim 50\%$ of our sample appear not to be not triggered by such an event, requiring an alternative explanation for the existence of such AGNs.

\subsection{AGN merger fraction and luminosity}

\begin{figure}[t]
\centering
\includegraphics[width = 9cm]{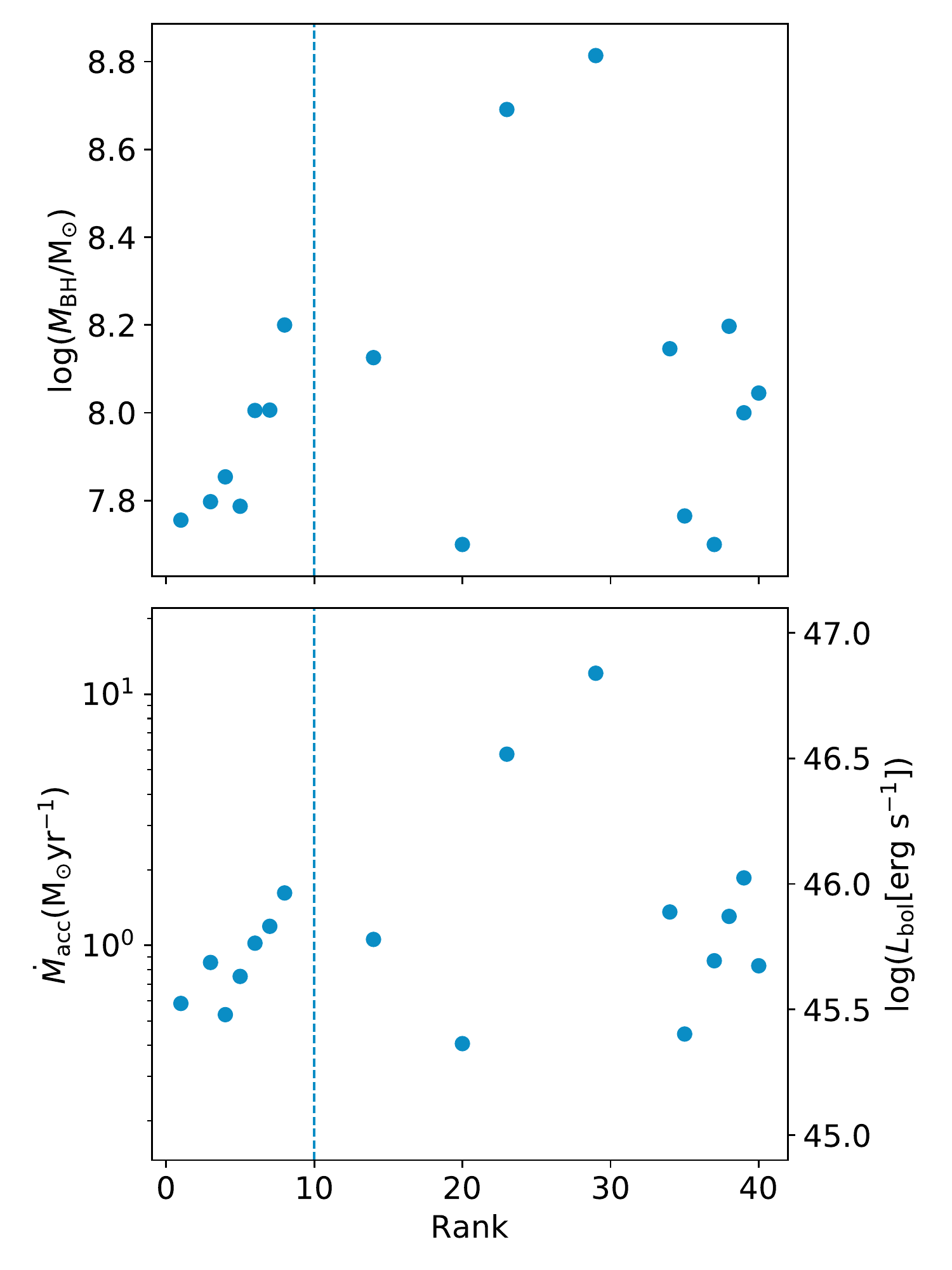}
\caption{Overall consensus rank vs. black hole mass (top) and black hole mass accretion rate and bolometric luminosity (bottom) for our sample of AGNs. The vertical dashed line visualizes a cut-off at rank 10, which was used in our discussion.}
\label{fig:rank_acc_rate_l_bol}
\end{figure}

Although our AGN sources can be considered luminous for sources at $z<$0.2, we emphasize that we have not selected our AGNs on absolute luminosity (see Sect.~\ref{subsec:AGN_sample} for our sample selection). Rather, we have chosen the AGNs with the highest Eddington ratios, i.e.\ the sources with the highest accretion rates and luminosities relative to their BH masses. Except for the two AGNs -- HE1226+0219 and HE2152--0936, which possess bolometric luminosities of $\mathrm{log}(L_\mathrm{bol}\mathrm{[erg~s^{-1}]})>46.5 $ -- all our remaining sample AGNs have luminosities of $45.3 \lesssim \mathrm{log}(L_\mathrm{bol} \mathrm{[erg~s^{-1}]}) \lesssim 46$, but feature the smallest BH masses in that luminosity bin ($7.7 < \mathrm{log}(M_\mathrm{BH}/ \mathrm{M_{\odot}}) < 8.2$).
In fact, $\sim10$ more luminous AGNs in our three initial parent catalogs would have been selectable. Unlike other studies, which detect an enhanced merger rate for luminous AGNs we see no trend of the strength of the merger features -- i.e. rank -- with either BH mass or BH mass accretion rate/luminosity within our AGN sample (Fig.~\ref{fig:rank_acc_rate_l_bol}). In fact HE1226+0219 and HE2152--0936, %the two most luminous AGNs 
both with distinctly higher absolute mass accretion rates with respect to our other sample AGNs, % -- HE1226+0219 and HE2152--0936 -- 
only occupy the ranks $\sim$~30 and $\sim$~25 in all the consensus rankings and show clearly no significant merger features. However, due to our selection of AGNs being based on a combination of BH mass and Eddington ratio, we note that apart from the two aforementioned most luminous AGNs our sources sample a relatively narrow luminosity range. Still, because of the lack of an obvious correlation of merger fraction with AGN luminosity, our results require an alternative explanation -- %than a direct correlation of merger fraction with AGN luminosity, 
especially considering that the existence of such a trend is still inconclusive. Despite some studies have found evidence of such a link between merger rate and luminosity \citep{treister_major_2012, fan_most_2016, goulding_galaxy_2018} others did not \citep{villforth_morphologies_2014, villforth_host_2017, hewlett_redshift_2017}.

\subsection{The (un)importance of minor mergers}\label{subsec:minor_mergers}

In Sect.~\ref{subsec:absolute_merger_fracs} we argue that the initial high merger fraction of our sample of control galaxies, is the result of our experts including galaxies in the merger category, which show features that are only the consequence of minor merger events. \citet{lotz_major_2011} state that the minor merger rate is $\sim$3 times the major merger rate (with a minor merger being in a mass ratio range of 1:4 $< M_\mathrm{sat}/M_\mathrm{primary} \lesssim$ 1:10). Considering our major merger fraction for those galaxies to be correct we end up with a \emph{total} merger fraction of $f_\mathrm{m,ina}$ = 0.33 $\pm$ 0.09 for our inactive galaxies. This would correspond to a cut-off at a rank of 17 and in turn in a total merger fraction of $f_\mathrm{m,agn}$ = 0.47 $\pm$ 0.12 for our AGN host galaxies. Obviously the difference between these two distortion rates is significantly decreased and indeed for our singular overall ranking we find 8 inactive galaxies and 8 AGNs below our cut at rank 17. However, while all experts can easily agree on the most distorted galaxies, it should be noted that sources with such small asymmetries are more difficult to classify. Hence, the rank of a particular galaxy with such features may differ strongly in the individual expert's rankings, which in turn could also influence to some extent the resulting overall rank and hence also the actual number of sources being considered merging in our final ranking. Nevertheless, we do not expect this scatter to be substantial. With only one AGN host galaxy, but eight inactive galaxies added into the merger category, it appears that only a small fraction of AGNs are seen to be in this interval. Hence, we are confident that the number of AGN host galaxies showing weak distortion features is still significantly less than compared to our comparison galaxies. This points to the conclusion that minor merging is comparably unimportant and most of the rest of AGNs require a different triggering mechanism. 

\subsection{Considering AGN and merger timescales}

With major mergers only triggering at most $\sim$50\% of our AGNs and minor mergers playing a subdominant role the question still remains which process(es) are responsible for triggering high Eddington rate AGNs at $z<0.2$. 
With that question in mind and a diminishing number of alternative mechanisms we consider a possible impact of the different timescales. Previous studies, which have found no enhancement in distortion fractions between AGNs and a matched sample of control galaxies, analyzed a potential disparity in AGN and merger lifetimes to be an explanation for their results \citep{cisternas_bulk_2011, mechtley_most_2016, marian_major_2019}. 
The unanimous conclusion is that the difference in life cycles is not sufficient to explain the lack of excess in merger rates, since the timescale of merger features being observable is much longer than the lifetime of the respective AGNs. 

%Building up on that 
We consider a scenario in which some of the galaxies that host no visible AGN and feature only minor distortions are actually the result of a major merger event which also lead to a past phase of active black hole growth. However, since the lifetime of AGNs can be significantly shorter when compared to that of major merger features, the only detectable remains of such a gravitational encounter would be in the form of minor asymmetries. This implies that if we utilize the \emph{total} merger fractions we derived in the previous subsection, a part of the $33\pm9\%$ inactive galaxies that show distortions of various strength have actually hosted a major merger triggered AGN in the past. As a result the AGN major merger fraction with $f_\mathrm{m,agn}$ = 0.41 $\pm$ 0.12 would increase, indicating that major mergers are not only an essential, but indeed the dominant mechanism to trigger high Eddington rate AGNs at $z<0.2$. %Simultaneously, f

Following the scenario outlined by \citet{goulding_galaxy_2018} we also assess the number of AGNs in an ongoing merger after first passage that are currently not visible due to an insufficient gas inflow. Those particular black holes will eventually become active again when the distance between the two galaxies decreases again resulting in growing torques and hence gas inflow. As in \citet{marian_major_2019}, we refer to such AGNs in the following as intermittent AGN. We cannot distinguish between such AGNs or past AGNs that will not be ignited again. However, since we are only interested in the eventual increase of the AGN merger fraction, the origin of this increase is irrelevant. 

We try to constrain the fraction of distorted inactive galaxies, which hosted an AGN in the recent past or currently an intermittent AGN, $f_\mathrm{m,ina\,\&\,agn}$ by adopting the formula presented in \citet{marian_major_2019}:

\begin{equation}
f_\mathrm{m,ina\,\&\,agn} = f_\mathrm{agn} \times f_\mathrm{m,agn} \times \frac{t_\mathrm{m}}{t_\mathrm{agn}} \text{.}
\end{equation}

\begin{figure*}[t]
\centering
\includegraphics[width = 18cm]{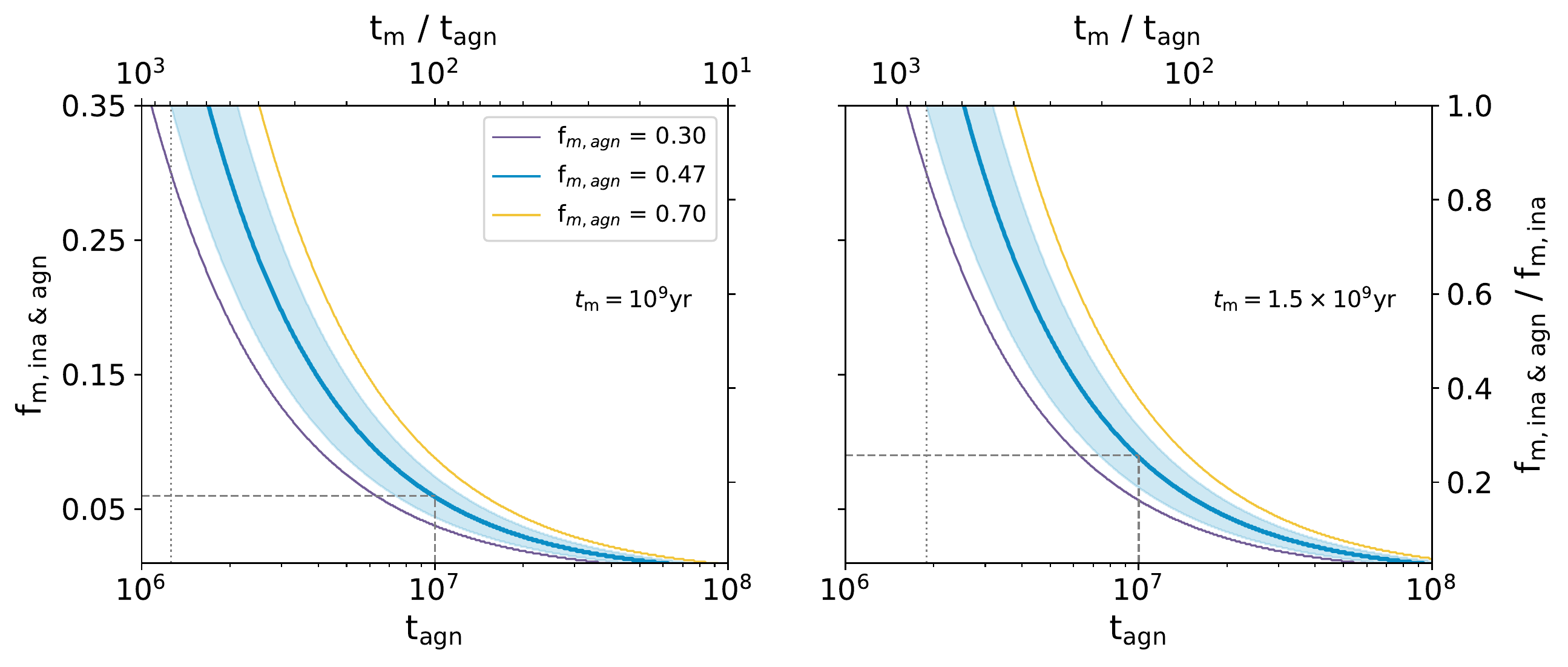}
\caption{Total fraction of merging inactive galaxies that hosted an AGN in the recent past $f_\mathrm{m,ina\,\&\,agn}$ in dependence of the AGN life time $t_\mathrm{agn}$ for a merger timescale $t_\mathrm{m}$ of 10$^9$yr (left) and $t_\mathrm{m} = $1.5$\times 10^9$yr (right). The blue line including the shaded region represents our result of the AGN total merger fraction of $f_\mathrm{m,agn}$ = 0.47 $\pm$ 0.12. The violet and yellow lines correspond to $f_\mathrm{m,agn}$ = 0.30 and 0.70, respectively. The dotted line corresponds to a lower limit of $t_\mathrm{agn}$, the dashed lines display the resulting $f_\mathrm{m,ina\,\&\,agn} \sim 0.09$ for an assumed $t_\mathrm{agn} = 10^7$.}
\label{fig:ina_mer_gal_agn}
\end{figure*}

Here, $f_\mathrm{agn}$ and $t_\mathrm{agn}$ represent the fraction and lifetime of AGNs with an Eddington ratio $>30$\% with respect to the total galaxy population at our redshift and mass bin. The timescale in which the merger features are observable is given by $t_\mathrm{m}$, while $f_\mathrm{m,agn}$ describes the total merger fraction of our specific AGN population. We derive $f_\mathrm{agn}$ by utilizing the number densities provided by stellar mass and quasar bolometric luminosity functions at $z\sim0$ and our stellar mass range and average bolometric AGN luminosity. 
Using the respective median $I$-band magnitudes this yields $\mathrm{log}\Phi \sim -2.9$ Mpc$^{-3}$ mag$^{-1}$ for the total galaxy population \citep{hirschmann_cosmological_2014, henriques_galaxy_2015, furlong_evolution_2015, lacey_unified_2016, pillepich_simulating_2018} and $\mathrm{log}\Phi \sim -5.8$ Mpc$^{-3}$ mag$^{-1}$ for our particular population of AGNs \citep{hopkins_observational_2007, fanidakis_evolution_2012, hirschmann_cosmological_2014, sijacki_illustris_2015}, resulting in $f_\mathrm{agn} \sim 1.3\times10^{-3}$, which is in excellent agreement with the value for the active fraction reported by \citet{schulze_low_2010} for BHs at a redshift $z<0.3$ and a mass of $\mathrm{log}(M_{\mathrm{BH}}/\mathrm{M_{\odot}}) \sim 8$. For $f_\mathrm{m,agn}$ we use our reported value of $f_\mathrm{m,agn} = 0.47 \pm 0.12$, but also repeat our calculations for $f_\mathrm{m,agn} = 0.30$ and $0.70$. Besides our initial estimate of $t_\mathrm{m} = 1.5\times 10^9$yr (see Sect.~\ref{subsec:absolute_merger_fracs}), in addition, we use $t_\mathrm{m}=10^9$yr for comparison. Finally, in accordance to previous studies we constrain our AGN lifetime $t_\mathrm{agn}$ to a range between $10^6$ and $10^8$yr \citep{martini_qso_2004, hopkins_physical_2005, shen_clustering_2007, hopkins_quasars_2009, conroy_simple_2013, cen_new_2015}. Since we can not distinguish between inactive merging galaxies that already went through their AGN phase, are yet to host an AGN or are currently hosting an intermittent AGN, it is not necessary for us to consider any time lag \citep{hopkins_relation_2006, wild_timing_2010, mcalpine_galaxy_2020} between the onset of the actual phase of active black hole growth and the beginning/coalescence of the merger. Hence, from the perspective of timescales our result solely depends on the relative difference between the AGN and merger lifetimes and thus we have to consider our fraction of inactive merging galaxies, which have hosted an AGN to be an upper limit. However, a visual re-examination returned only a low number of galaxies with asymmetries actually having a close companion. Therefore, we conclude that most of the distorted galaxies are already in the late stages of their merging process, indicating that, if at all, they already experienced a potential AGN phase with a low chance of an intermittent AGN becoming active again. 

The total merger fraction of our inactive galaxies, which amounts to $f_\mathrm{m,ina} \sim 0.35$, serves as an upper bound for $f_\mathrm{m,ina\,\&\,agn}$. Both parameters being equal would imply that all distorted, inactive galaxies have hosted (or will host) an AGN. Conversely, $f_\mathrm{m,ina\,\&\,agn} = 0$ would correspond to no such galaxy ever hosting an AGN. 
In Fig.~\ref{fig:ina_mer_gal_agn} we present the results of our computations for different $f_\mathrm{m,agn}$ and $t_\mathrm{m} = 10^9$yr (left) and $1.5\times10^9$yr (right). The blue lines and the shaded regions denote the results for our retrieved AGN merger fraction and the corresponding $1\sigma$ intervals, while the violet and yellow lines display the trend for $f_\mathrm{m,agn} = 0.30$ and $0.70$, respectively. The fraction of merging inactive galaxies hosting an AGN at some point during the merging process increases with shorter AGN lifetimes. In addition, for a given period of AGN activity this share grows with longer merger timescales and larger AGN merger fractions, both due to an enhanced probability to find a distorted galaxy actually hosting an AGN. Depending on the merger timescale and assuming the lower limit of our AGN merger fraction is correct, we can deduce a lower bound for the AGN lifetime by considering every inactive distorted galaxy to host an AGN, i.e. $f_\mathrm{m,ina\,\&\,agn} \equiv f_\mathrm{m,ina}$. The life span of an AGN corresponds then to a minimum of $1.3\times10^6$yr and $1.9\times10^6$yr for merger timescales of $10^9$yr and 1.5$\times$10$^9$yr, respectively (Fig.~\ref{fig:ina_mer_gal_agn}, dotted lines). 

However, based on the best estimates for accretion rate histories we have today \citep{di_matteo_energy_2005, johansson_evolution_2009, johansson_equal-_2009, hopkins_how_2010, jung_multi-scale_2018}, we fix the time period in which an AGN accretes above $\lambda_{\mathrm{edd}} > 0.3$ to $t_\mathrm{agn} = 10^7$yr. 
The inferred fractions of inactive merging galaxies that also host an AGN at any given time yield then $f_\mathrm{m,ina\,\&\,agn} = 0.06_{-0.02}^{+0.01}$ and $0.09_{-0.02}^{+0.02}$ for $t_\mathrm{m}$ = $10^9$yr and 1.5$\times$10$^9$yr, respectively (Fig.~\ref{fig:ina_mer_gal_agn}, dashed lines). So, adding even the upper limit of this fraction onto the AGN major merger rate we derived in Section~\ref{subsec:absolute_merger_fracs} this only results in a revised AGN major merger fraction, which is barely above the threshold of 0.5, which in turn would indicate that the majority of AGNs is triggered by major mergers. This result still leaves $\sim50\%$ of AGNs to be of unknown origin. Only by assuming a significantly lower AGN duty cycle of $t_\mathrm{agn}\sim10^6$yr and thus regarding almost every distorted inactive galaxy hosting an AGN, we can obtain AGN major merger fractions of $\sim80\%$, which would then leave no doubt about the role of major mergers and the triggering of high Eddington rate AGNs at $z<0.2$. Hence, we conclude that neither a difference in AGN and merger timescales nor the potential presence of intermittent AGNs affect significantly our derived AGN merger rate. In order to better constrain our inferred estimates, more detailed simulations predicting especially AGN timescales in dependence of accretion rate are imperative.

\section{Summary \& Conclusions} \label{sec:summary}

We examined a potential direct connection between AGNs specifically exhibiting the highest Eddington ratios and major mergers at $z<$~0.2. We analyzed 17 AGN host galaxies and 25 comparison galaxies, matched in mass, redshift, filter, and the S/N in $V$, $B$ and color images. We adjusted our control galaxies by adding artificial point sources on top of their flux centers, which yielded two indistinguishable samples, that were joined to create a randomized overall sample of 42 targets. This overall sample was ranked according to the presence of merger features (from most to least distorted) by 19 experts. We combined the individual rankings of each set, i.e.\ $V$, $B$ and color, by applying three different methods, resulting in a total number of nine consensus rankings. This allowed us to determine any bias, which might be introduced by visually classifying the galaxies at different wavelengths or the algorithm to combine the individual classifications. Finally, we also created one overall sequence by combining the nine initial consensus rankings. We divided all rankings into; 1) galaxies showing distinct merger features and 2) galaxies showing no signs of a gravitational disturbance, by choosing specific cut-off ranks. As a final step we derived the respective merger fractions by counting the numbers of active and control galaxies above and below these particular limits and applying those quantities to a beta distribution. 

Our findings depend heavily on the choice of distinction between merging and undisturbed systems. To analyze how the selection of the cut-off rank affected our result, we; (1) selected it based on the visual interpretations by the experts and (2) chose it such that the merger rate of our comparison sample was consistent with the overall major merger fraction of galaxies in our mass and redshift range. When we considered the average determinations of the classifiers, approximately half of both populations showed signs of a current or recent merger event, suggesting no causal connection between major mergers and the triggering of this particular population of AGNs.

Since our first approach also considers asymmetries or signatures that stem from processes other than a major merger event, we adjust the major merger fraction of the inactive galaxies to be consistent with recent simulations and observations. As a result we find a substantial excess in the major merger fraction of the AGN sample with respect to the inactive galaxies. Coincidentally, with a separation at the corresponding cut-off rank we also found a clear distinction between strongly-disturbed galaxies and galaxies with either minor or no merger signatures, confirming our classification. 

We summarize our findings as follows.

\begin{itemize}
\renewcommand\labelitemi{--}

\item The merger fractions of the AGN host galaxies and comparison galaxies are $f_\mathrm{m,agn}$ = 0.41 $\pm$ 0.12 and $f_\mathrm{m, ina} = $0.08 $\pm$ 0.06, respectively. 
\item Neither the choice of set nor combination method has impacts the recovered merger fractions.
%\item Within the narrow luminosity range covered by our AGNs, we find no evidence that confirms a correlation of accretion rate or luminosity with the strength of potential merger features. 
\item For our AGNs, with the highest Eddington ratios at $z<0.2$, major mergers are an essential mechanism to trigger black hole growth. %The huge amount of gas needed to attain such high accretion rates at comparable low black hole masses can be easily achieved by a strong gravitational encounter of two gas-rich galaxies of similar mass. 
\item %After adjusting our classification scheme to also take galaxies with minor asymmetries into account and examining the revised distribution of merging inactive and AGN host galaxies 
We rule out that minor mergers play a considerable role in the triggering of our subpopulation of AGNs. 
\item %We also consider a scenario where a fraction of our merging inactive galaxies that only show minor distortions are actually the result of a past major merger, which also led to a phase of active black hole growth that is simply not observable anymore at the present day. This includes also distorted inactive galaxies that may host an intermittent AGN that is currently not visible. With our best estimates the contribution of such a population of galaxies to the AGN merger fraction results in $\sim50\%$ of our AGN population still being of unknown origin. 
Considering AGN and merger lifetimes as well as AGN variability induced by an ongoing merger event, our best estimate results in $\sim50\%$ of our AGN population still being of unknown origin.

\end{itemize}

Extending our study to include IFU-observations and a larger number of sources would enable us to analyze the AGN host galaxies in more detail. By assessing the strength of potential past merger events by examining the kinematics and stellar populations, while larger number provides better statistics we can determine, which processes are responsible for the triggering of the remaining $\sim$~50\% and whether major mergers are indeed the dominant mechanism.  

\acknowledgments
We thank the referee for the constructive feedback, which improved the quality of this work.

We also thank Mischa Schirmer for his helpful guidance in using \texttt{THELI}.

JS is supported by JSPS KAKENHI Grant Number JP18H01251 and the World Premier International Research Center Initiative (WPI), MEXT, Japan.
VNB gratefully acknowledges assistance from a National Science Foundation (NSF) Research at Undergraduate Institutions (RUI) grant (AST-1909297). Note that findings and conclusions do not necessarily represent views of the NSF.
RAW acknowledges support from NASA JWST Interdisciplinary Scientist grants NAG5-12460, NNX14AN10G and 80NSSC18K0200 from GSFC.

Based on observations made with ESO Telescopes at the La Silla Paranal Observatory under program ID 091.B-0672, 095.B-0773 \& 098.A-0241.

Funding for the Sloan Digital Sky Survey (SDSS) has been provided by the Alfred P. Sloan Foundation, the Participating Institutions, the National Aeronautics and Space Administration, the National Science Foundation, the U.S. Department of Energy, the Japanese Monbukagakusho, and the Max Planck Society. The SDSS Web site is http://www.sdss.org/.

The SDSS is managed by the Astrophysical Research Consortium (ARC) for the Participating Institutions. The Participating Institutions are The University of Chicago, Fermilab, the Institute for Advanced Study, the Japan Participation Group, The Johns Hopkins University, Los Alamos National Laboratory, the Max Planck Institute for Astronomy (MPIA), the Max Planck Institute for Astrophysics (MPA), New Mexico State University, University of Pittsburgh, Princeton University, the United States Naval Observatory, and the University of Washington. 

This research has made use of NASA's Astrophysics Data System Bibliographic Services.

 \facility{ESO-VLT(FORS2)}
 \software{
         astropy \citep{astropy_collaboration_astropy_2013,astropy_collaboration_astropy_2018},
         Matplotlib \citep{hunter_matplotlib_2007},
         MultiColorFits \citep{cigan_multicolorfits_2019},
         SAOImageDS9 \citep{joye_new_2003},
        THELI \citep{erben_gabods_2005, schirmer_theli_2013}
 }

\newpage

\appendix

\section{Details on combination methods}\label{appendix:details_methods}

Every method to combine individual votes into a combined consensus sequence violates at least one of three criteria described by Arrow's Impossibility Theorem \citep{arrow_difficulty_1950}. It states that no existing method, which combines two or more individual votes satisfies the following three axioms: (1) non-dictatorship, such that all individual votes are considered to be equal; (2) unanimity or the weak Pareto principle, stating that if all voters agree on $X>Y$, this also holds true for the overall ranking; and (3) the independence of irrelevant alternatives, such that the consensus relation between $X$ and $Y$ only depends on the individual preferences between those two entities and not any additional option(s). As additional conditions we introduce the Condorcet paradox and the Condorcet criterion \citep{condorcet_essai_1785, condorcet_political_1989}. The first one states that an overall sequence can be cyclic -- e.g. $X$ wins over $Y$, which wins over $Z$, which in turn wins over $X$ -- although the individual votes are not. The latter explains that an overall top-ranked candidate wins in every pairwise comparison with every other candidate. 

Below we present the methods we apply to create the overall rankings. As stated in Section~\ref{sec:analysis} we use three different algorithms to construct those combined rankings to determine any potential bias introduced by the method. However, in addition all of our three methods also satisfy or infringe the above mentioned criteria differently, which gives us even more detailed insights in any potential introduction of differences in the merger fractions. 

For our first method to combine the individual expert rankings we adopt the same method applied in \citet{mechtley_most_2016} and \citet{marian_major_2019}. We start with calculating the mean rank for each galaxy from the individual rankings and discard every individual expert classification of each galaxy, if it differs more than 2$\sigma$ from the respective average rank. Out of the 798 individual assessments in $V$-band we reject 25 votes, while out of the total 779 ratings, 17 are discarded for the sets in $B$-band and color, respectively. However, since we weigh individual votes this method obviously violates the non-dictatorship criterion.  

Our second method, the Borda count approach \citep{emerson_original_2013}, satisfies this condition, but violates in exchange the independence of irrelevant alternatives. We adapt the original version of this method in which the first ranked option receives $n$ points, the second one $n-1$ and so on, with $n$ being the total number of candidates, by applying the Dowdall system \citep{reilly_social_2002}. With that approach the candidates receive the reciprocal value of their respective ranks, i.e. the first ranked option is rewarded $1/n = 1$ point, the next one 0.5 points and so on. As low rank galaxies may be ranked more randomly due to a lack of significant merger features, we can decrease the impact those sources might have on our overall ranking by using this variant of the Borda count.

This approach avoids the Condorcet paradox, but only our third method, the Schulze method \citep{schulze_new_2011, schulze_schulze_2018}, also satisfies the Condorcet criterion. With this method all pairwise comparisons between two candidates $X$ and $Y$ for all individual rankings are calculated and put into relation to each other, resulting in an overall ranking, where the top-ranked candidate, wins indeed over all other candidates, being the so-called Condorcet winner. Going to lower ranks within the resulting consensus sequence the second-placed candidate only loses to the first-ranked option and so on \citep[for more details and examples please see][]{schulze_schulze_2018}. 

\section{Dependence of merger fractions on cut-off rank}\label{appendix:frac_m_rank_dependence}

\begin{figure*}[b]
\centering
\includegraphics[width = 15.6cm]{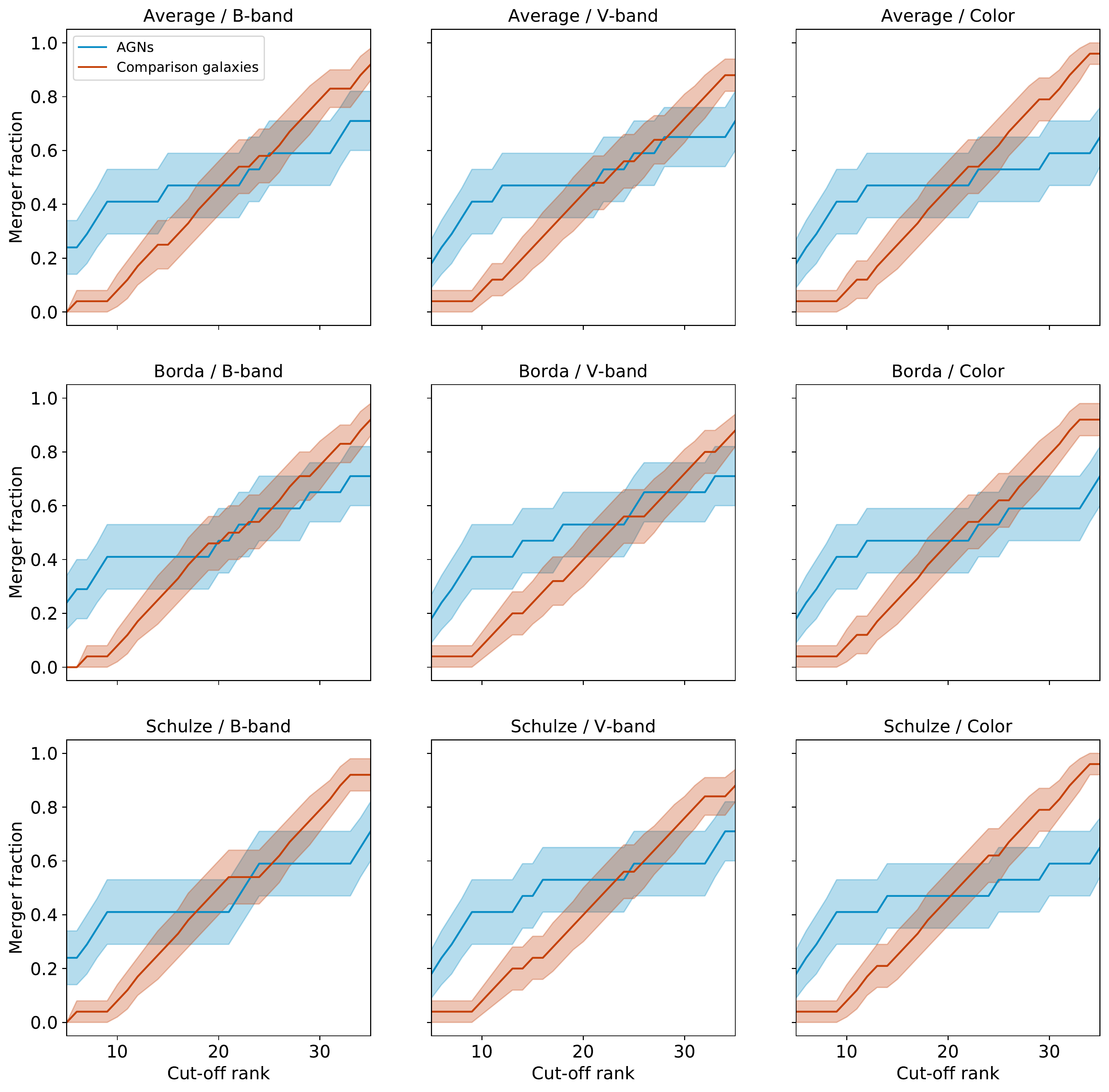}
\caption{Evolution of the merger fractions for the AGN host galaxies (blue) and inactive galaxies (red) in dependence of cut-off rank for each combination of set and method. The shaded regions give the 1$\sigma$ confidence interval.}
\label{fig:Merger_Fractions_Evo}
\end{figure*}

In Sections~\ref{sec:analysis} and \ref{sec:discussion} we describe how the choice of cut-off rank can influence the resulting merger fractions and also present for four selected cut-off ranks the corresponding merger fractions.
In Figure~\ref{fig:Merger_Fractions_Evo} we now present the continuous dependence of merger fractions on cut-off rank for all combinations of method and set. The AGN host galaxies and inactive galaxies are shown in blue and red, respectively. The shaded regions denote the 1$\sigma$ confidence interval from shot- and classification-noise. As already indicated in Figure~\ref{fig:Merger_Fractions} and described in Section~\ref{sec:analysis}, it is also shown in Figure~\ref{fig:Merger_Fractions_Evo} that first, neither the choice of method to combine the individual rankings nor the selection of set has any significant impact on the resulting absolute merger fractions or the relative differences between them. Second, compared to the inactive comparison sample and for cut-off ranks $\lesssim$15 the AGN host galaxies show a clear excess in merger fractions. This clearly indicates that our conclusions rely considerably on the choice of cut-off rank, which is extensively discussed in the main text. 

\clearpage

\section{Visual overall consensus ranking}\label{appendix:meta_ranking}

To have a `meta' singular consensus sequence we apply the Schulze method (see Section \ref{sec:analysis} and Appendix~\ref{appendix:details_methods}) to our final nine overall rankings, which we calculated for each combination of set and method. We show all sources in the resulting order, and include for completeness also the sources already shown in Figure~\ref{fig:showcases}. The respective rank for each object is given in parentheses besides its designation. It should be noted that Gal176221 is only ranked last, because it was only observed in $V$-band and therefore only appears in the three corresponding consensus rankings. In those three respective rankings it is always positioned at rank 14. Clearly visible is the drop-off in strong merger features at a cut-off rank $\gtrsim$10. 

%\newpage

\begin{figure*}[h]
\centering
%\vspace{1.5ex}
\includegraphics[width = 17cm]{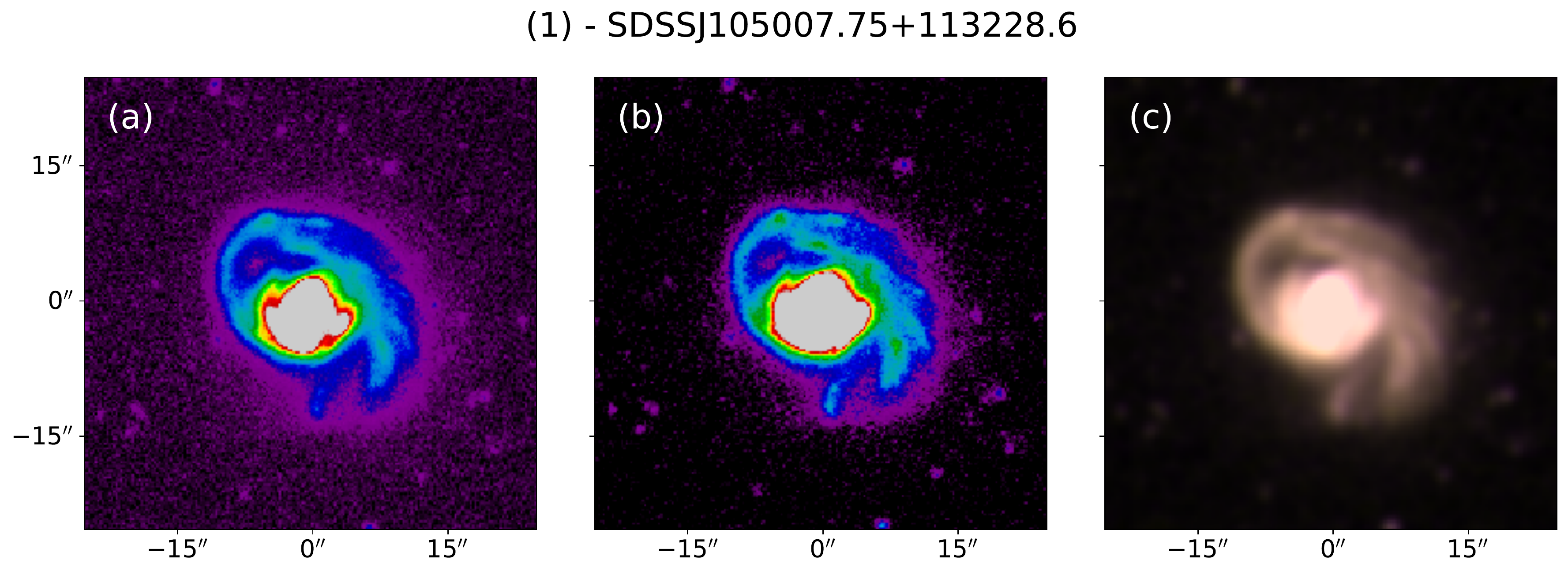}
\label{fig:qso_1}
\end{figure*}

\begin{figure*}[h]
\centering
\includegraphics[width = 17cm]{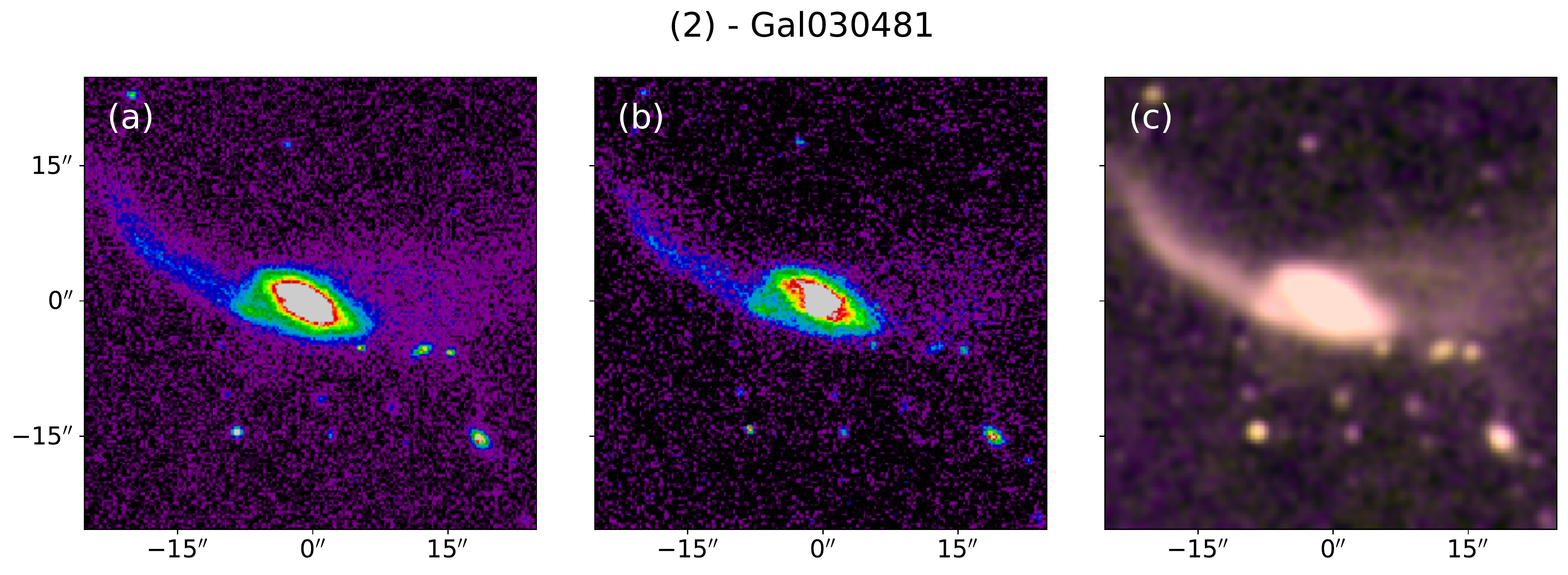}
\caption{From left to right we present a postage stamp in (a) V-band, (b) B-band and (c) color, respectively. Note: In order to enhance the visibility the images are not shown with the same cuts and color map parameters.}
\end{figure*}

\begin{figure*}
\centering
\includegraphics[width = 17cm]{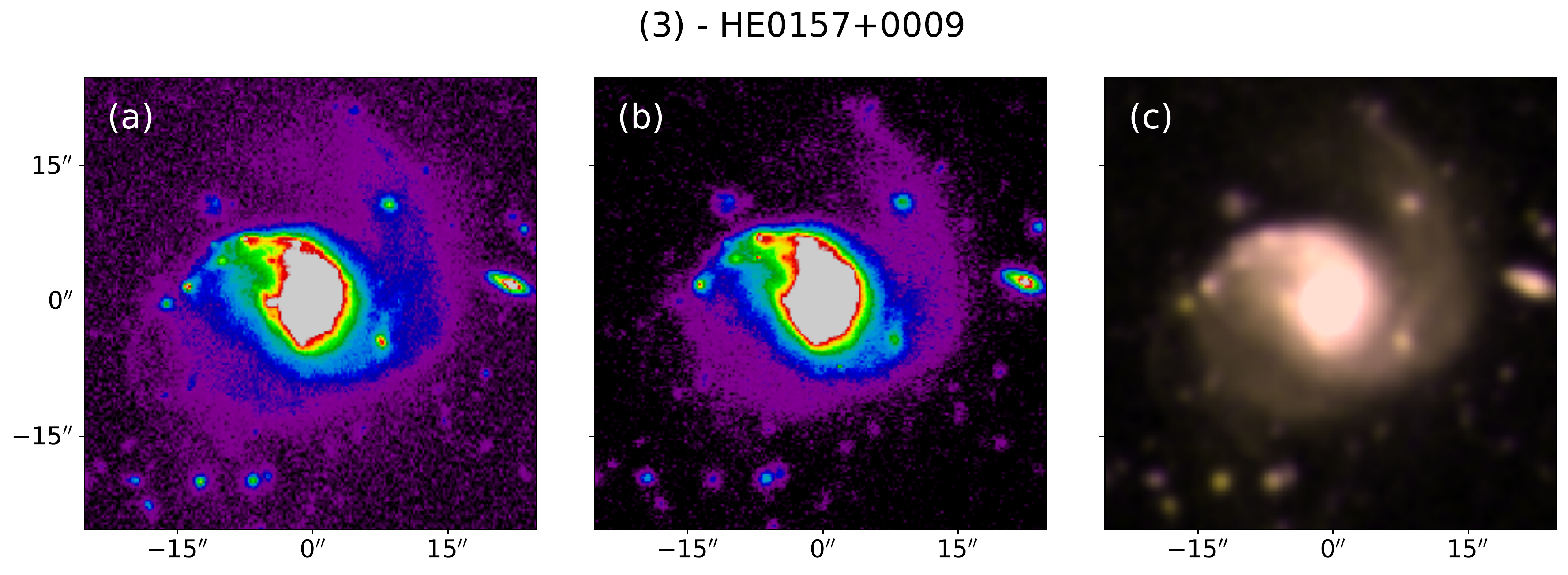}
\end{figure*}

\begin{figure*}
\centering
\includegraphics[width = 17cm]{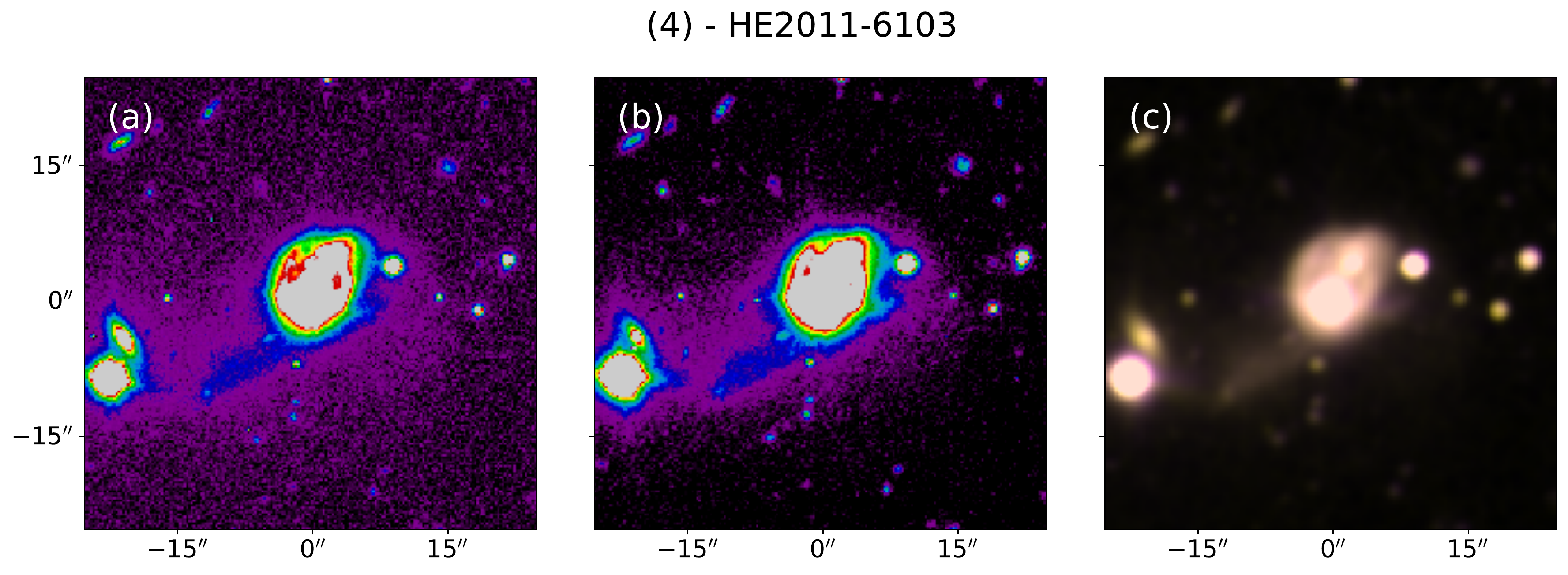}
\end{figure*}

\begin{figure*}
\centering
\includegraphics[width = 17cm]{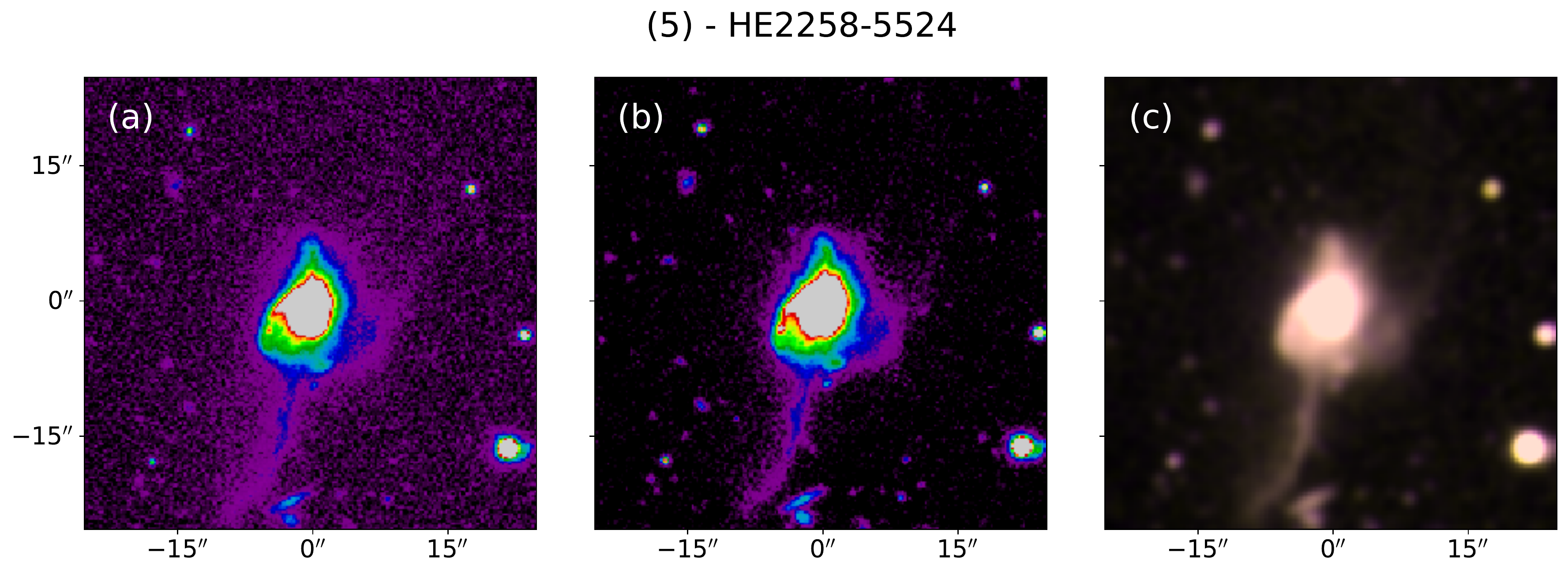}
\figurenum{8}
\caption{(Continued.)}
\end{figure*}

\clearpage

\begin{figure*}
\centering
\includegraphics[width = 17cm]{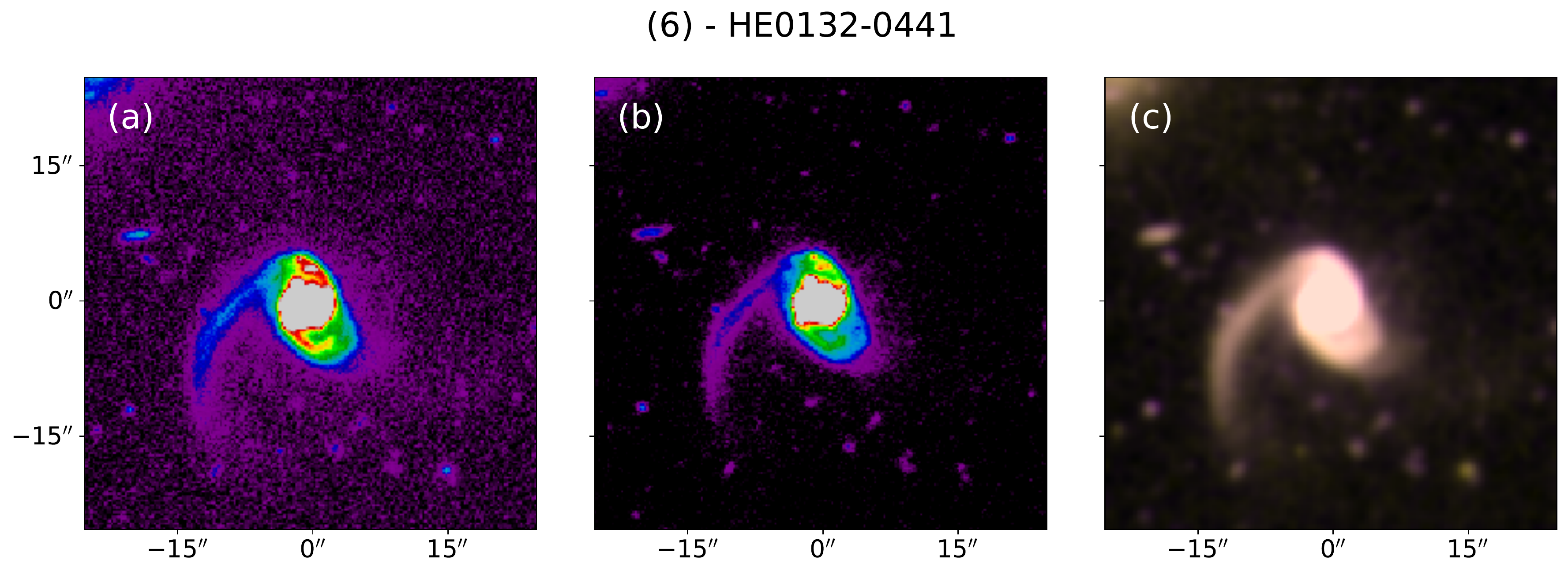}
\end{figure*}

\begin{figure*}
\centering
\includegraphics[width = 17cm]{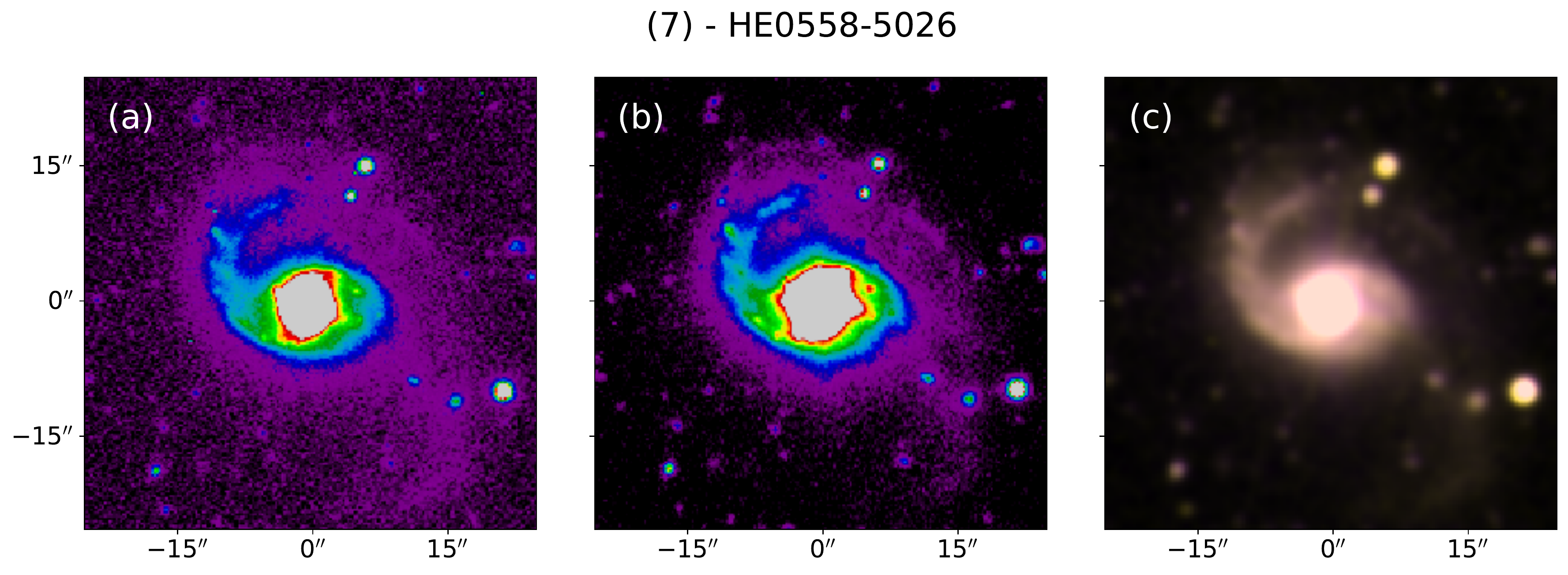}
\end{figure*}

\begin{figure*}
\centering
\includegraphics[width = 17cm]{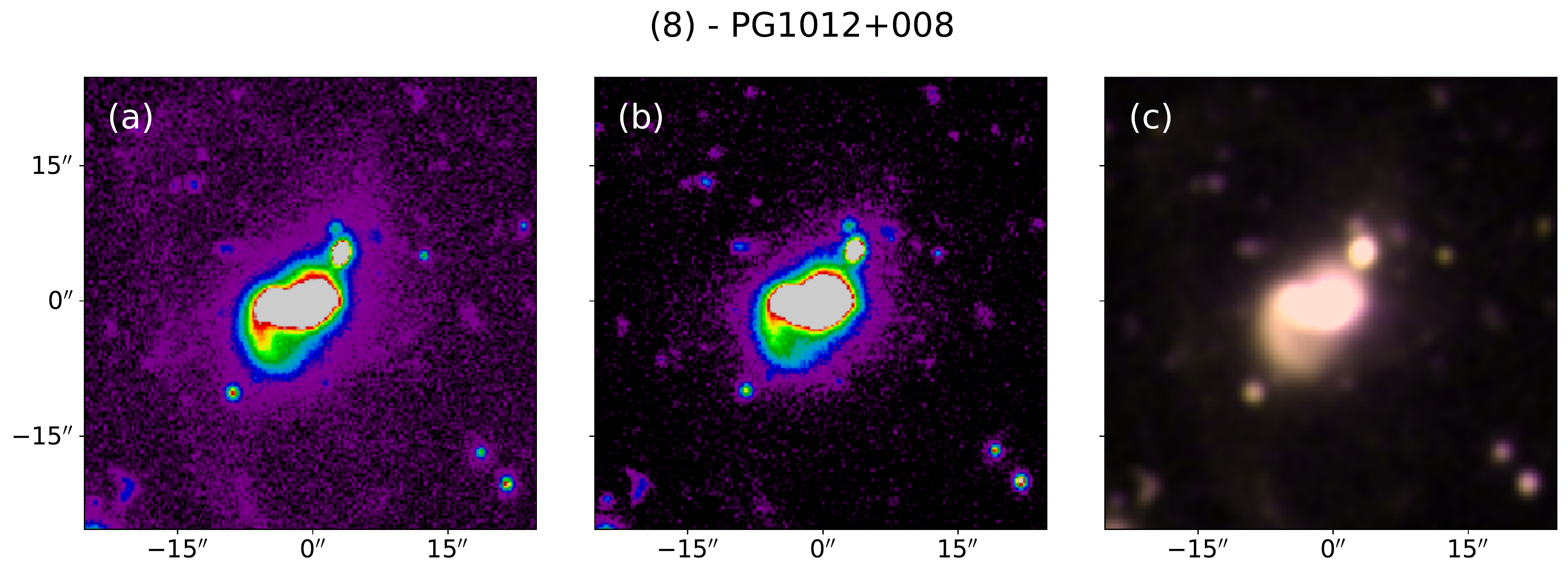}
\figurenum{8}
\caption{(Continued.)}
\end{figure*}

\clearpage

\begin{figure*}
\centering
\includegraphics[width = 17cm]{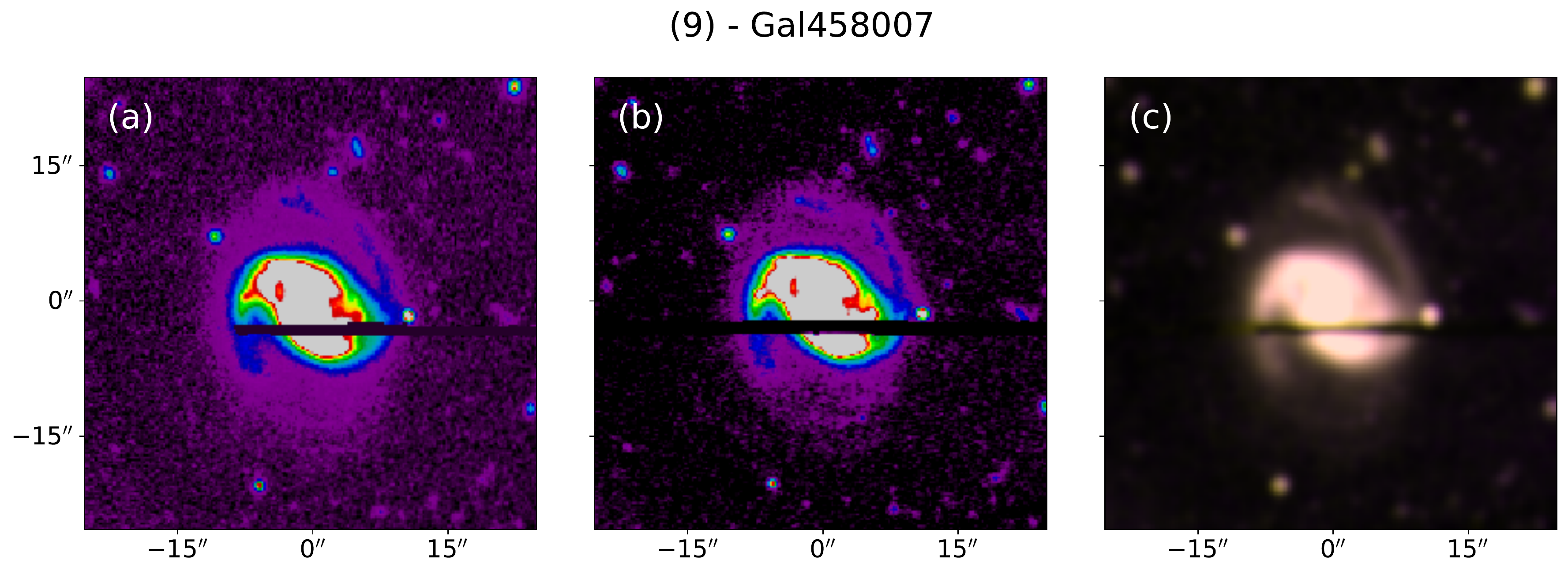}
\end{figure*}

\begin{figure*}
\centering
\includegraphics[width = 17cm]{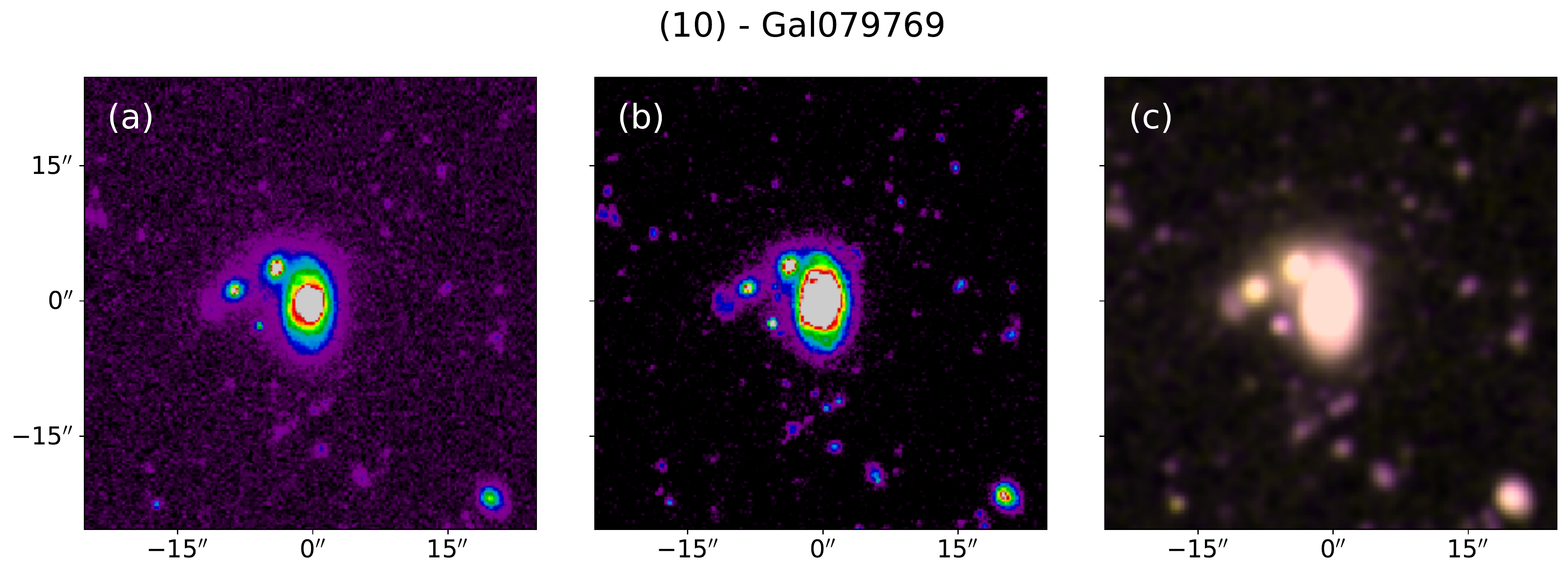}
\end{figure*}

\begin{figure*}
\centering
\includegraphics[width = 17cm]{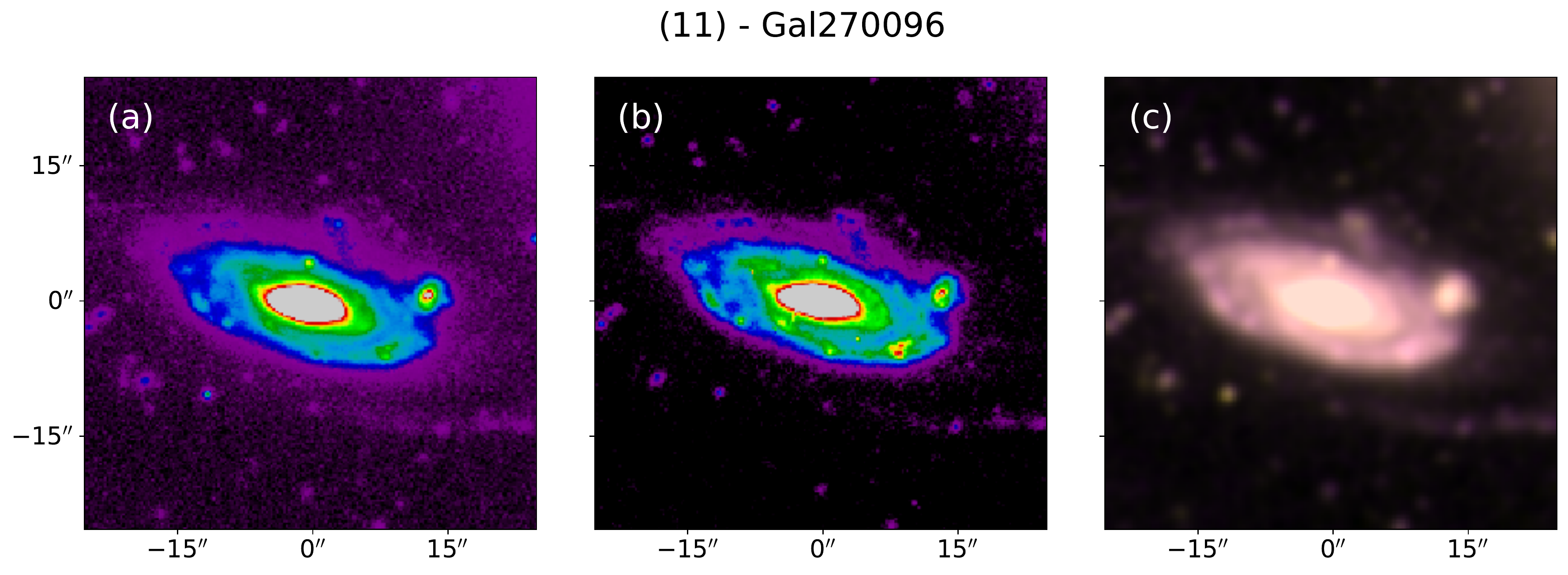}
\figurenum{8}
\caption{(Continued.)}
\end{figure*}

\clearpage

\begin{figure*}
\centering
\includegraphics[width = 17cm]{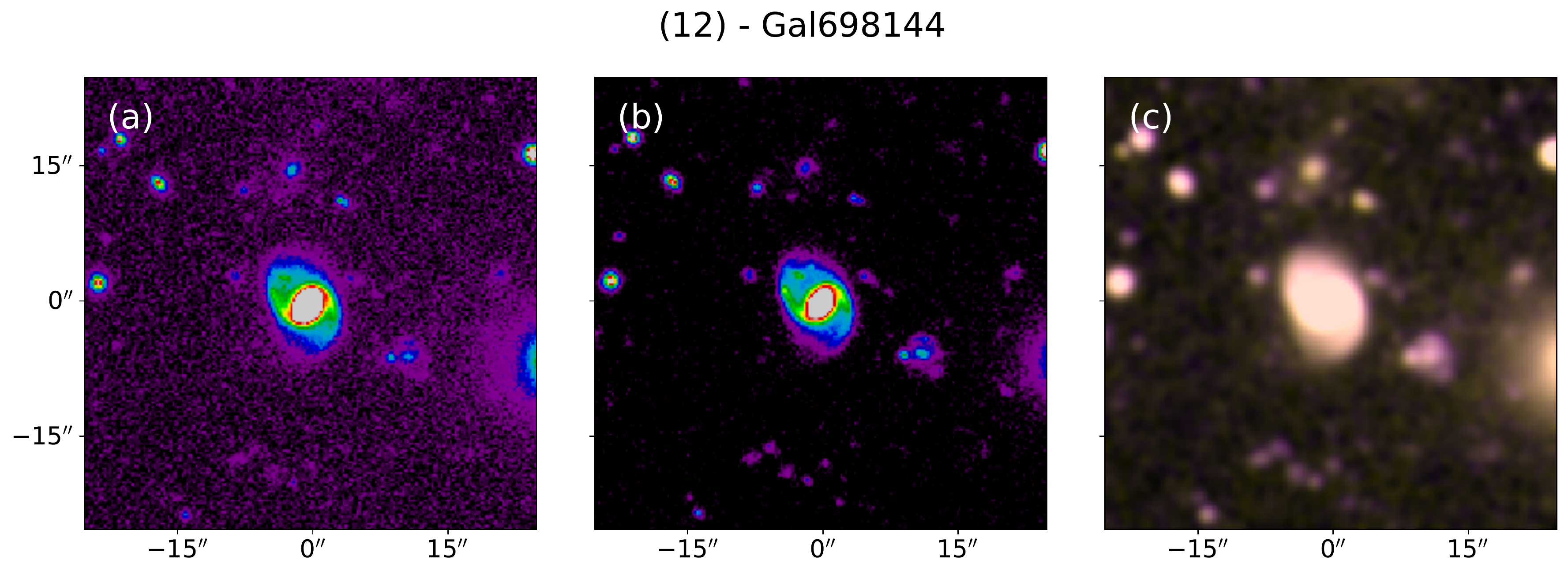}
\end{figure*}

\begin{figure*}
\centering
\includegraphics[width = 17cm]{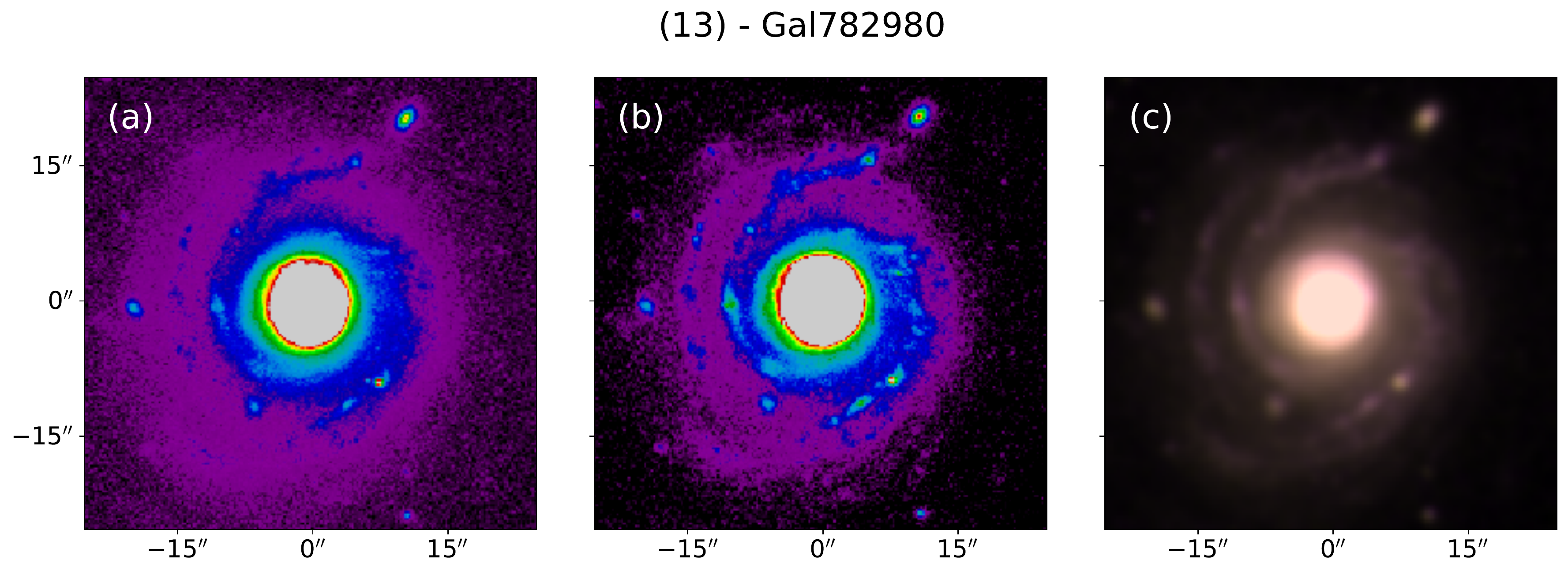}
\end{figure*}

\begin{figure*}
\centering
\includegraphics[width = 17cm]{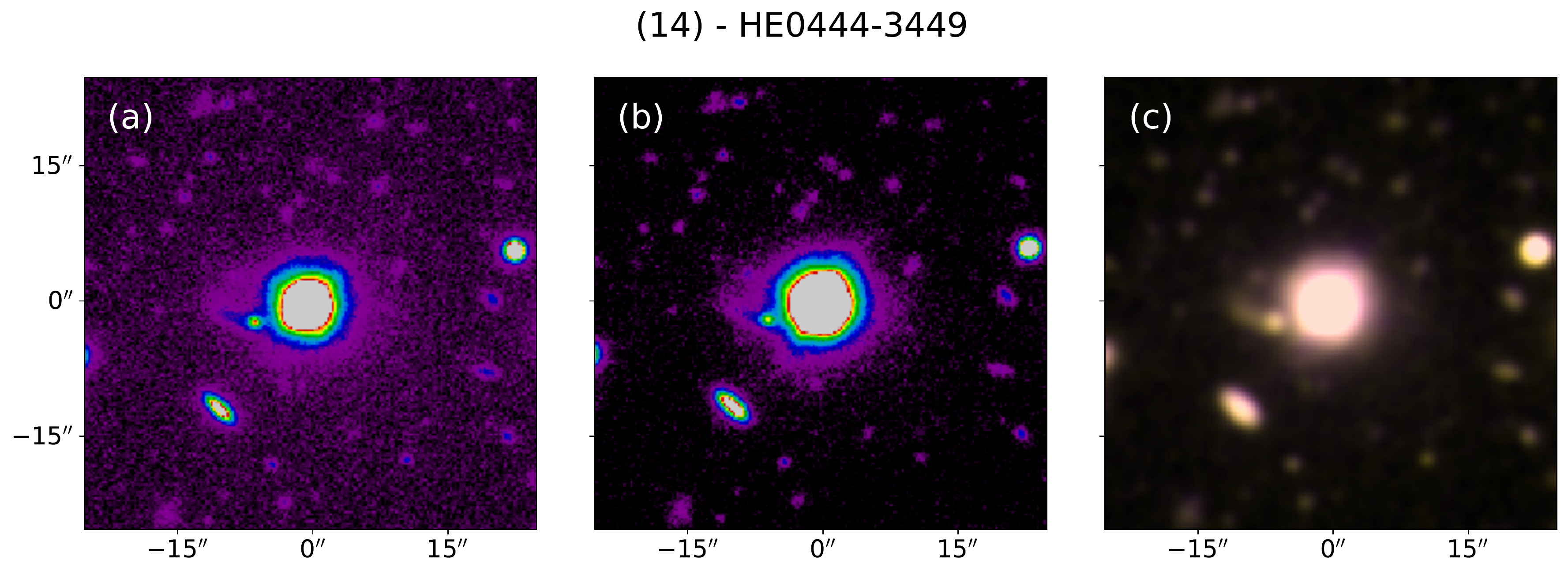}
\figurenum{8}
\caption{(Continued.)}
\end{figure*}

\clearpage

\begin{figure*}
\centering
\includegraphics[width = 17cm]{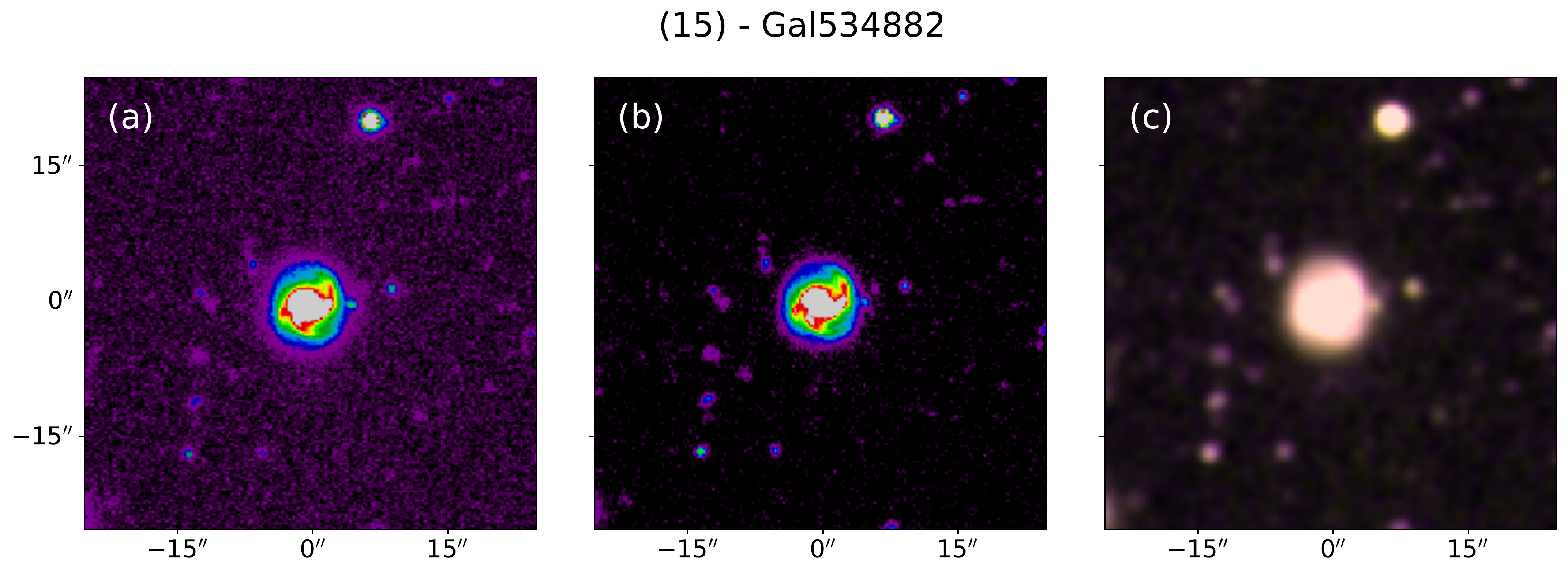}
\end{figure*}

\begin{figure*}
\centering
\includegraphics[width = 17cm]{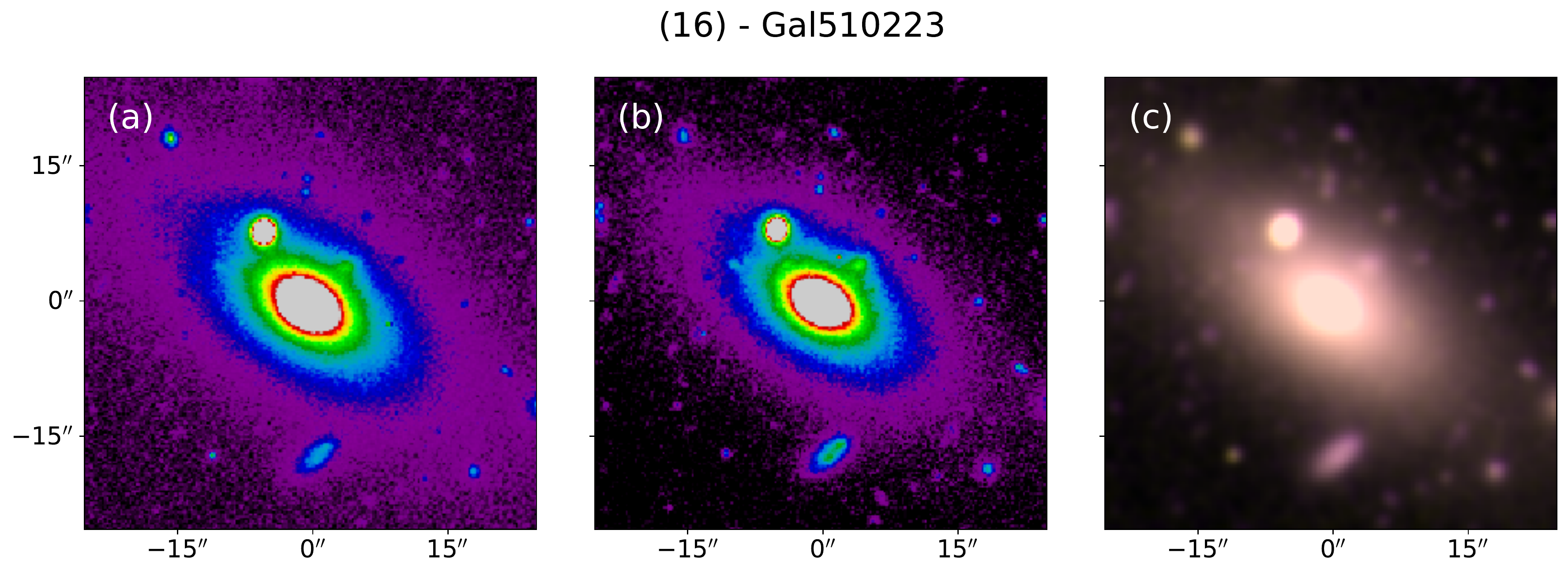}
\end{figure*}

\begin{figure*}
\centering
\includegraphics[width = 17cm]{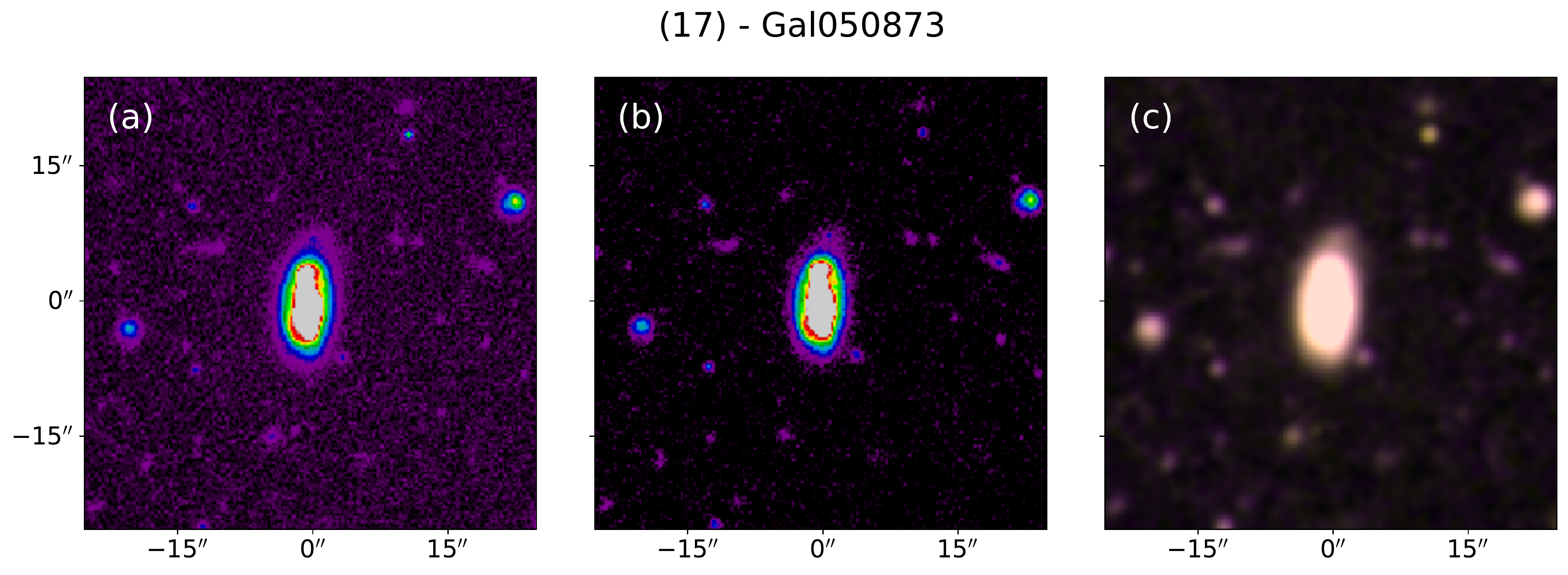}
\figurenum{8}
\caption{(Continued.)}
\end{figure*}

\clearpage

\begin{figure*}
\centering
\includegraphics[width = 17cm]{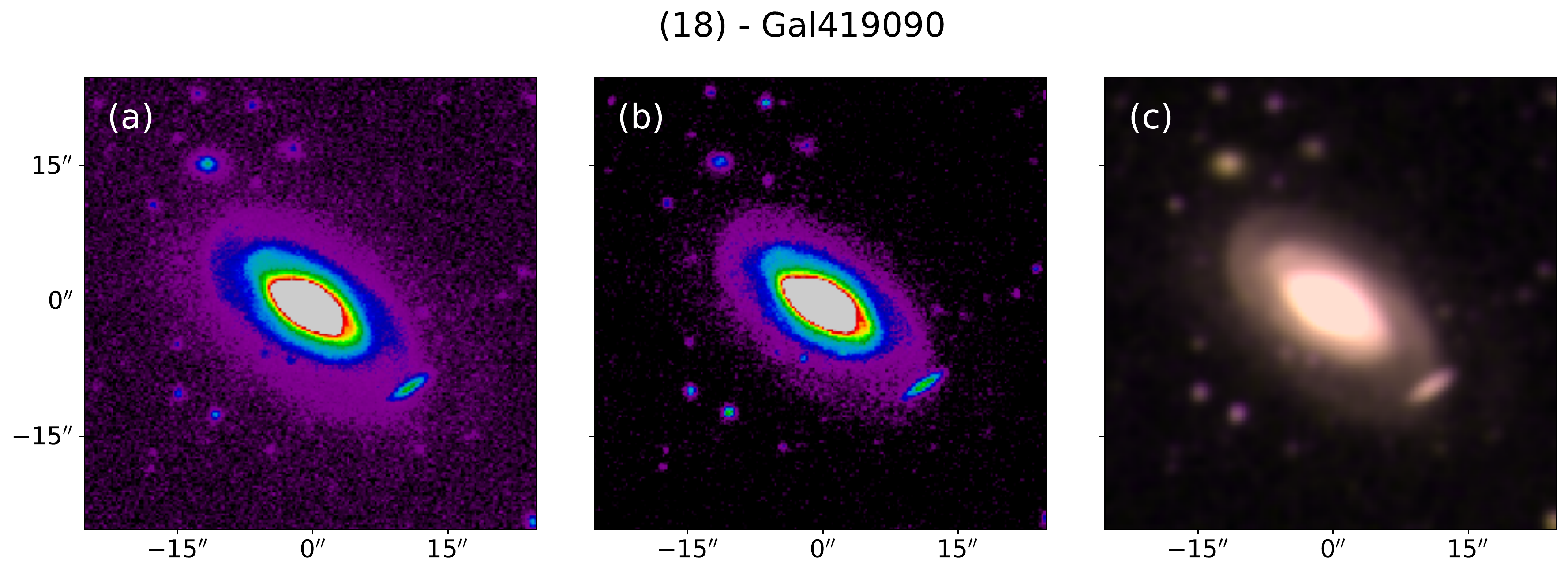}
\end{figure*}

\begin{figure*}
\centering
\includegraphics[width = 17cm]{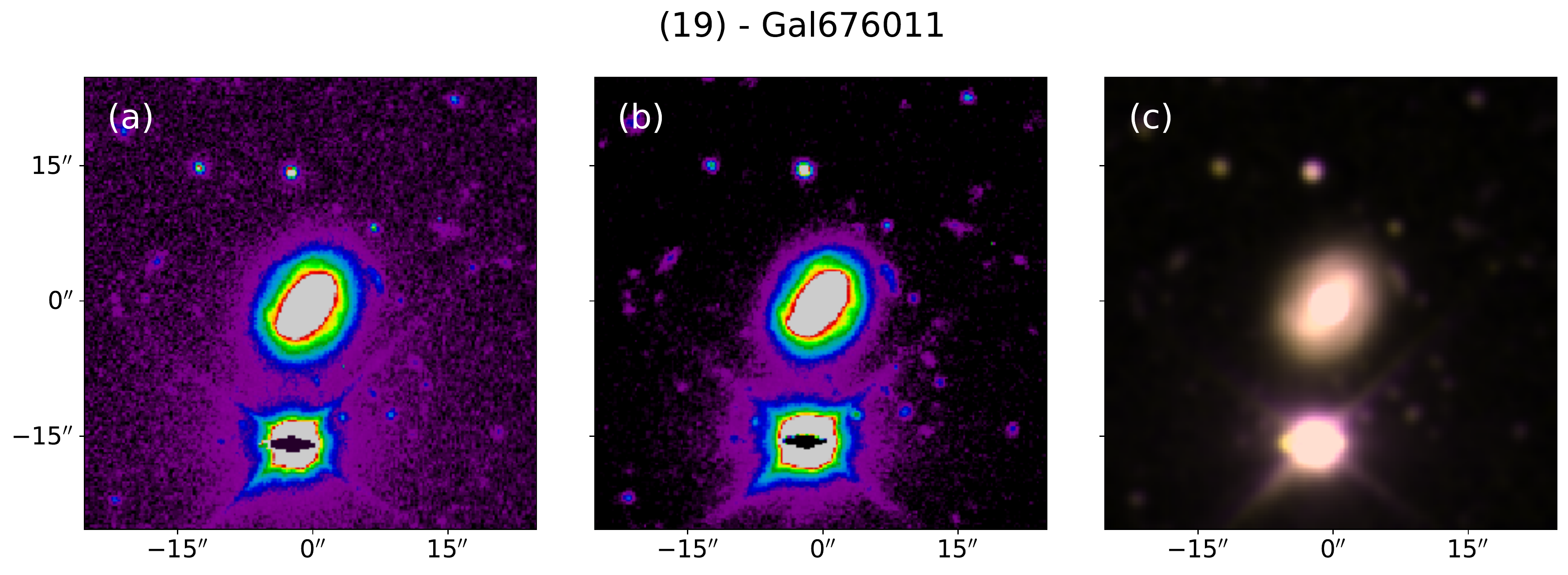}
\end{figure*}

\begin{figure*}
\centering
\includegraphics[width = 17cm]{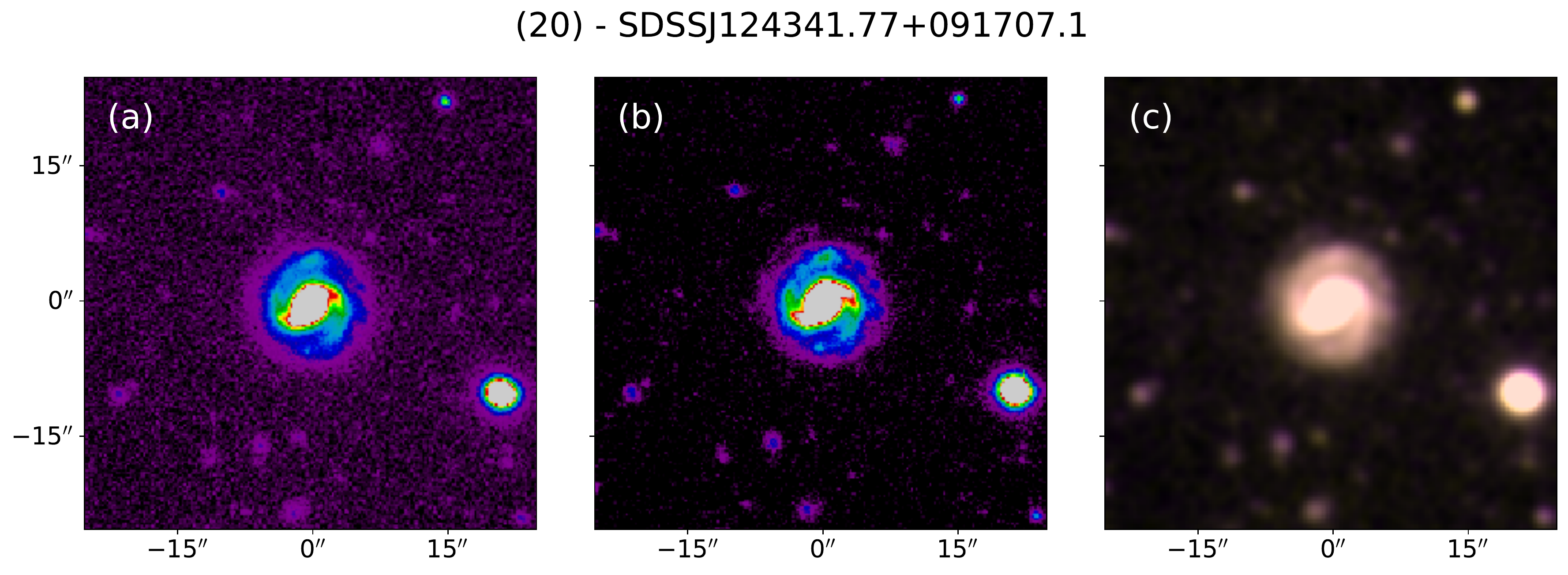}
\figurenum{8}
\caption{(Continued.)}
\end{figure*}

\clearpage

\begin{figure*}
\centering
\includegraphics[width = 17cm]{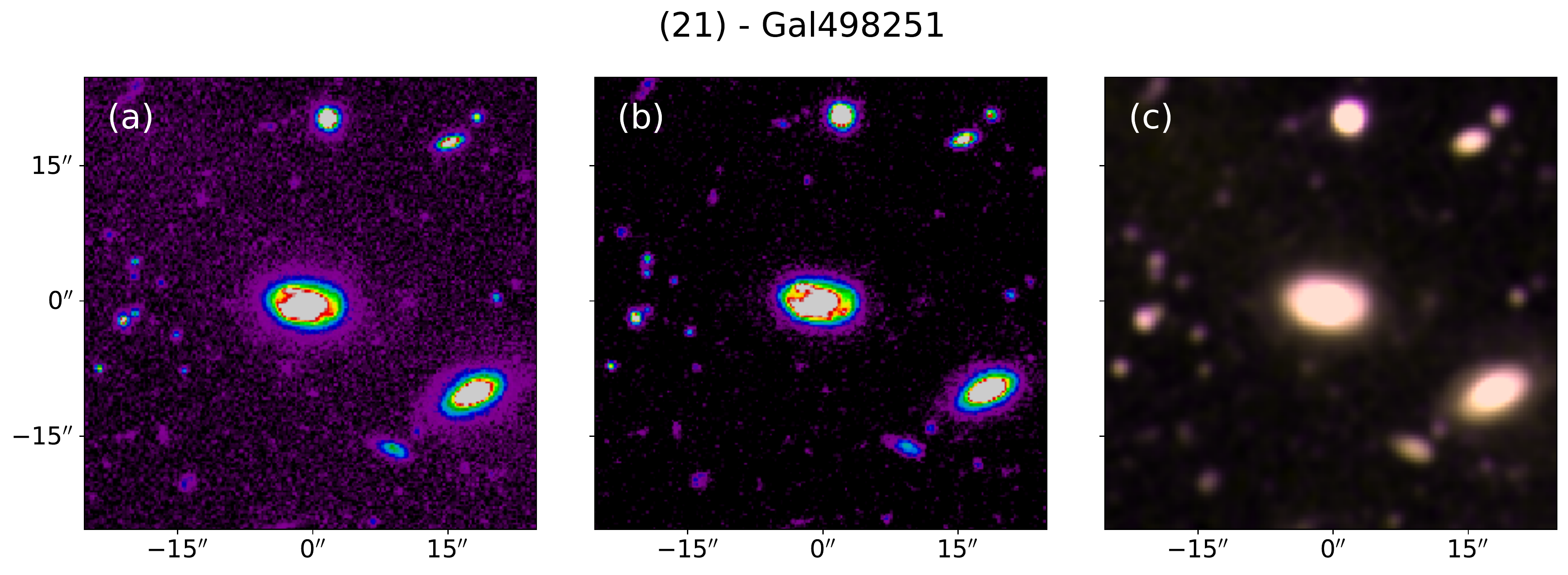}
\end{figure*}

\begin{figure*}
\centering
\includegraphics[width = 17cm]{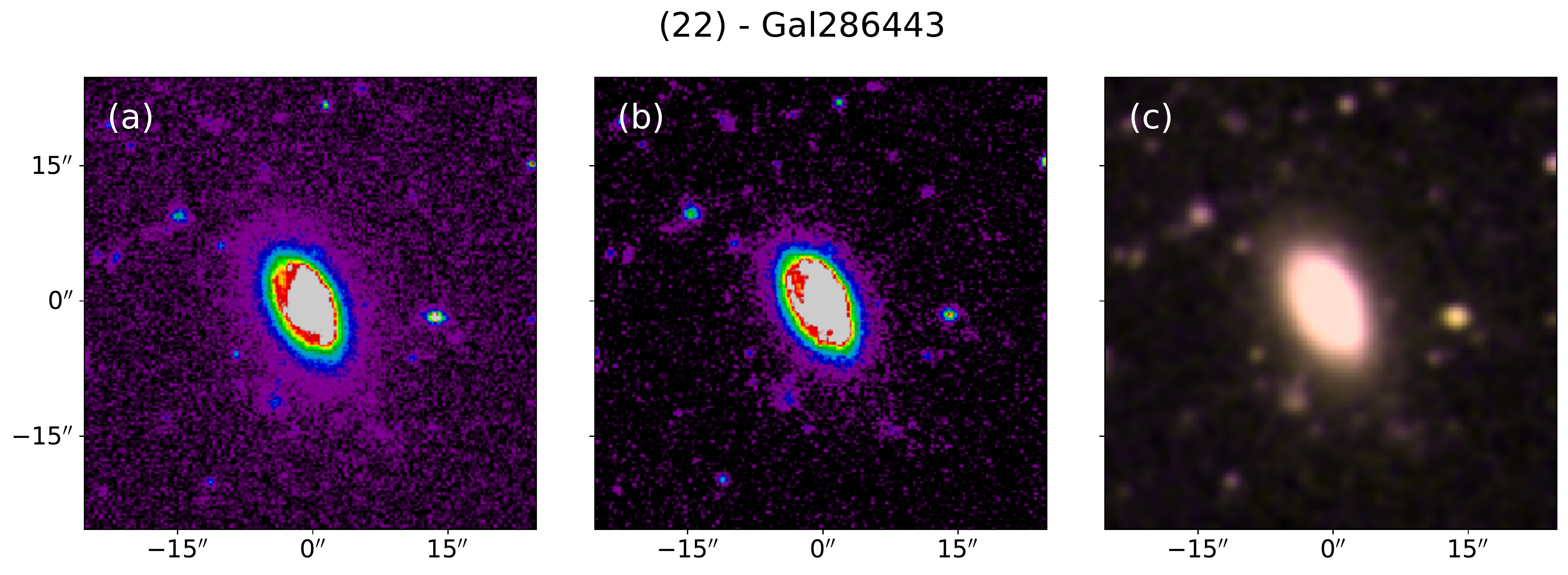}
\end{figure*}

\begin{figure*}
\centering
\includegraphics[width = 17cm]{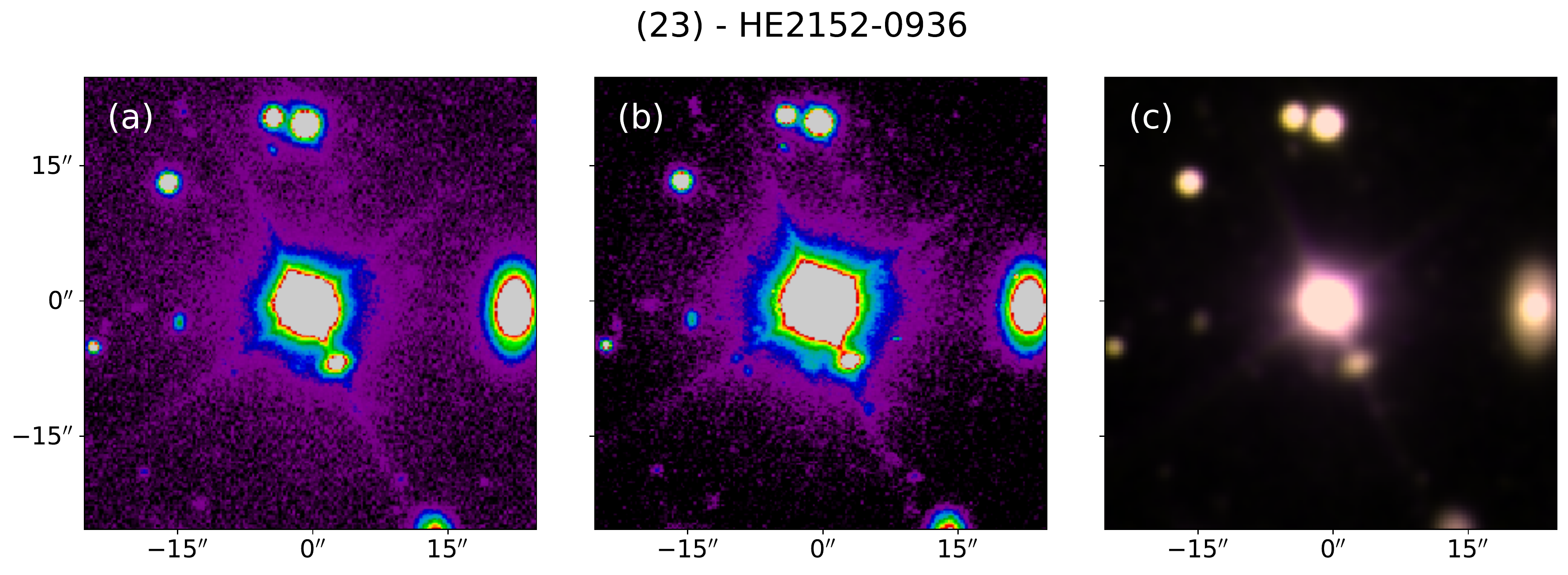}
\figurenum{8}
\caption{(Continued.)}
\end{figure*}

\clearpage

\begin{figure*}
\centering
\includegraphics[width = 17cm]{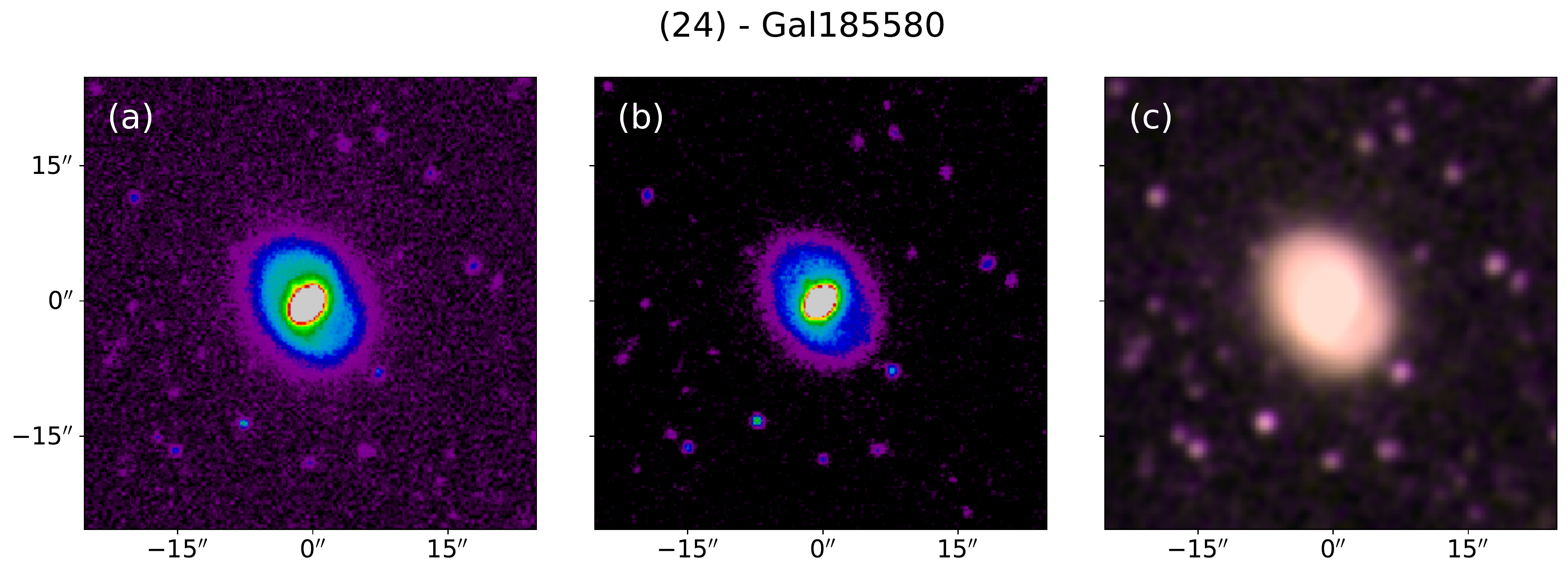}
\end{figure*}

\begin{figure*}
\centering
\includegraphics[width = 17cm]{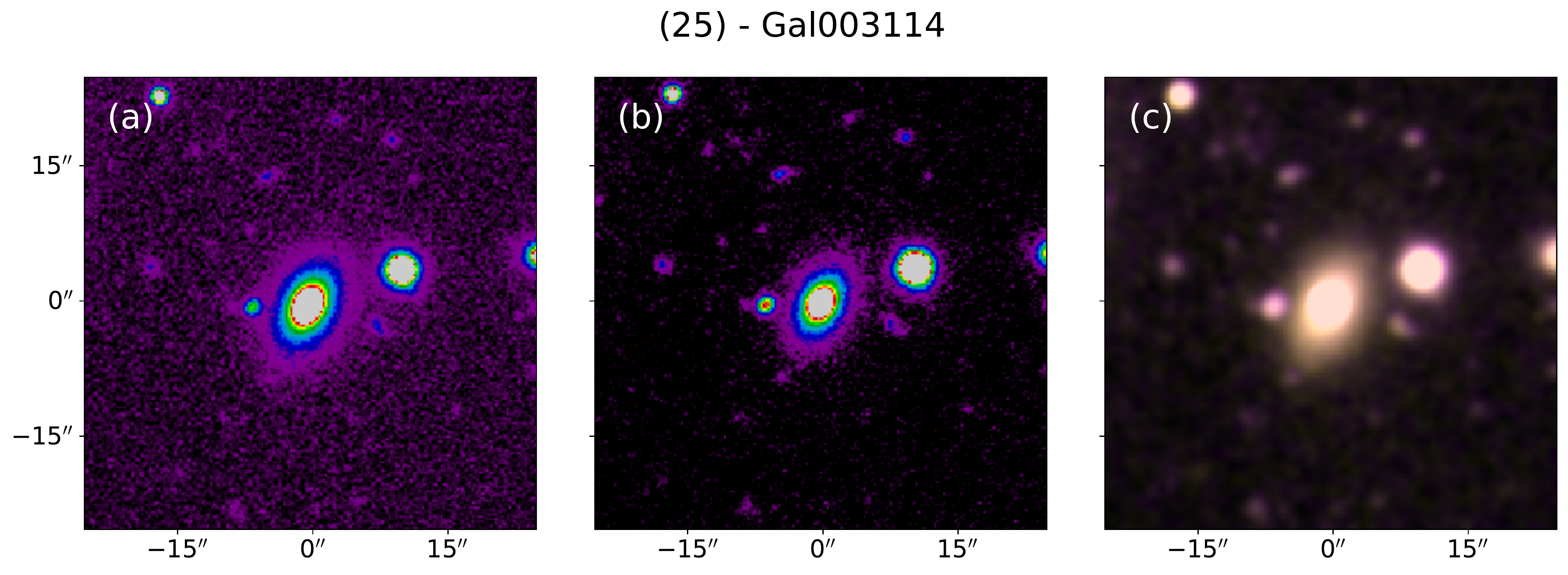}
\end{figure*}

\begin{figure*}
\centering
\includegraphics[width = 17cm]{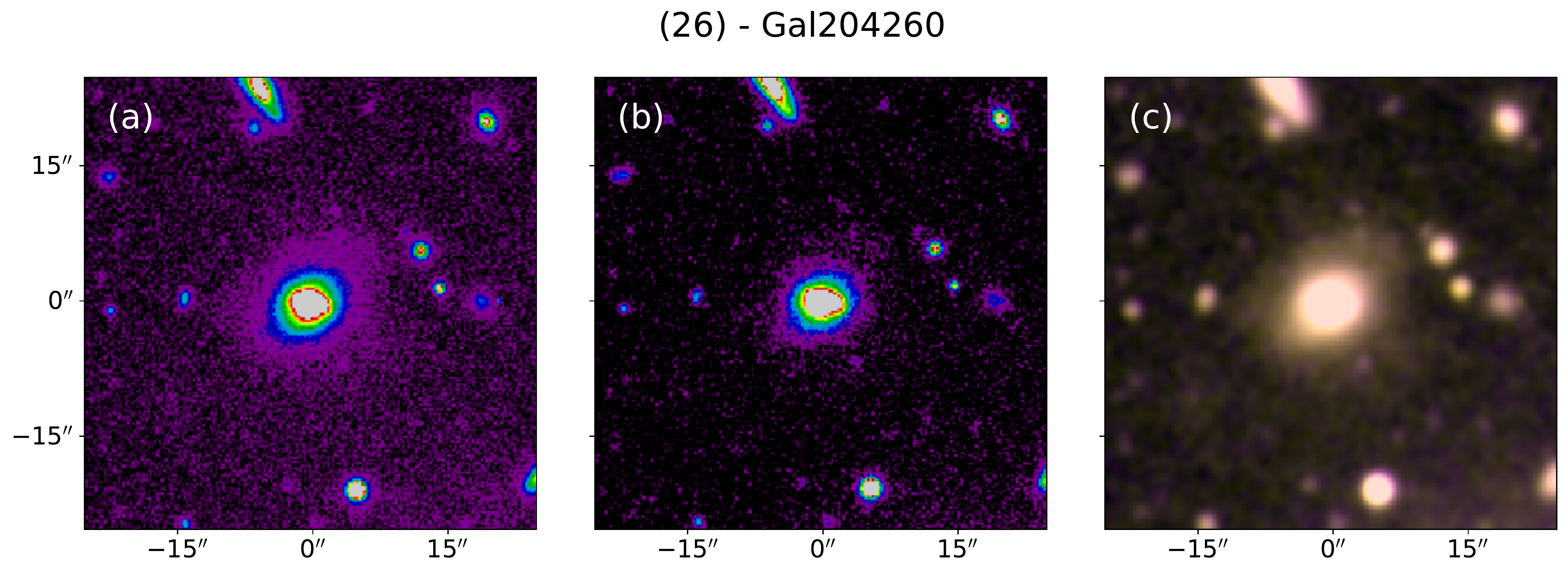}
\figurenum{8}
\caption{(Continued.)}
\end{figure*}

\clearpage

\begin{figure*}
\centering
\includegraphics[width = 17cm]{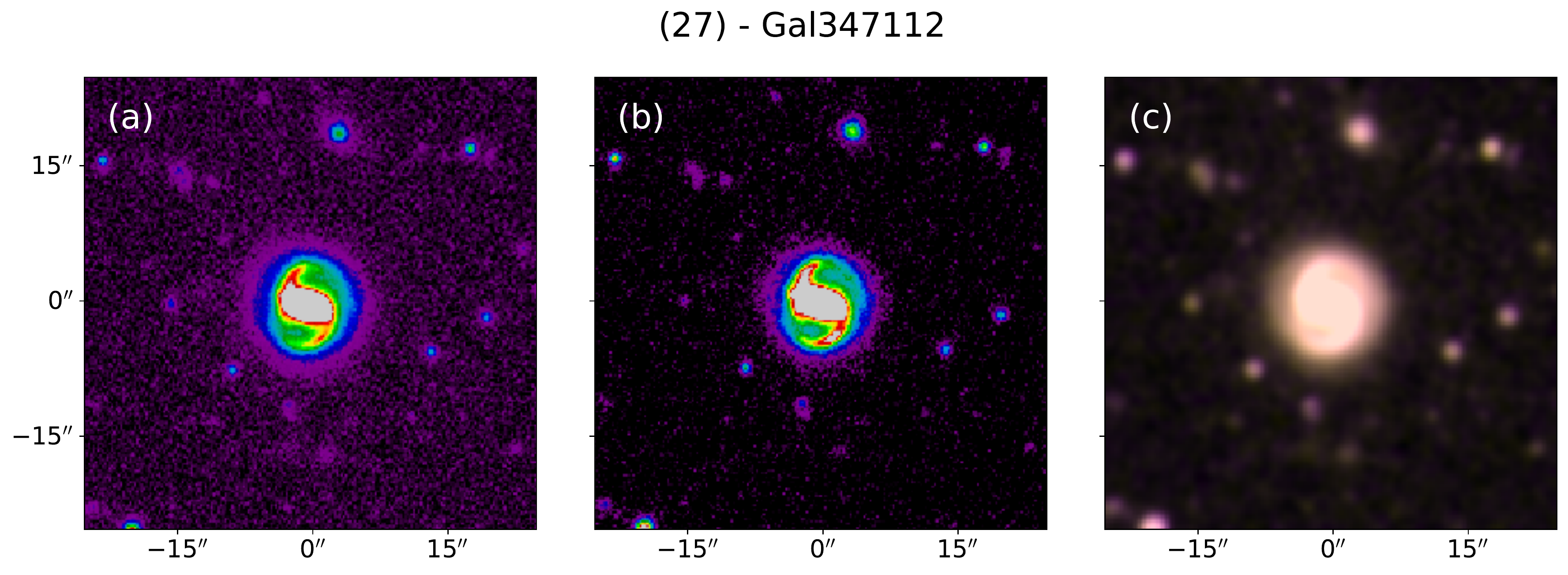}
\end{figure*}

\begin{figure*}
\centering
\includegraphics[width = 17cm]{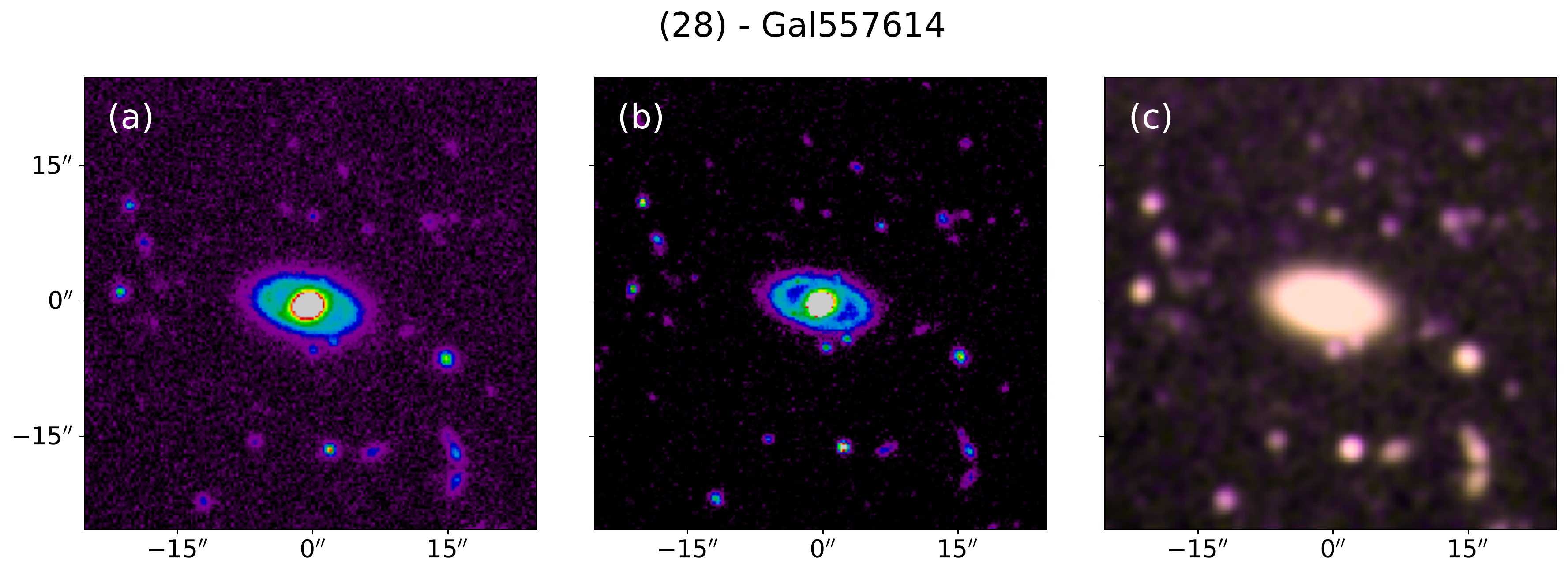}
\end{figure*}

\begin{figure*}
\centering
\includegraphics[width = 17cm]{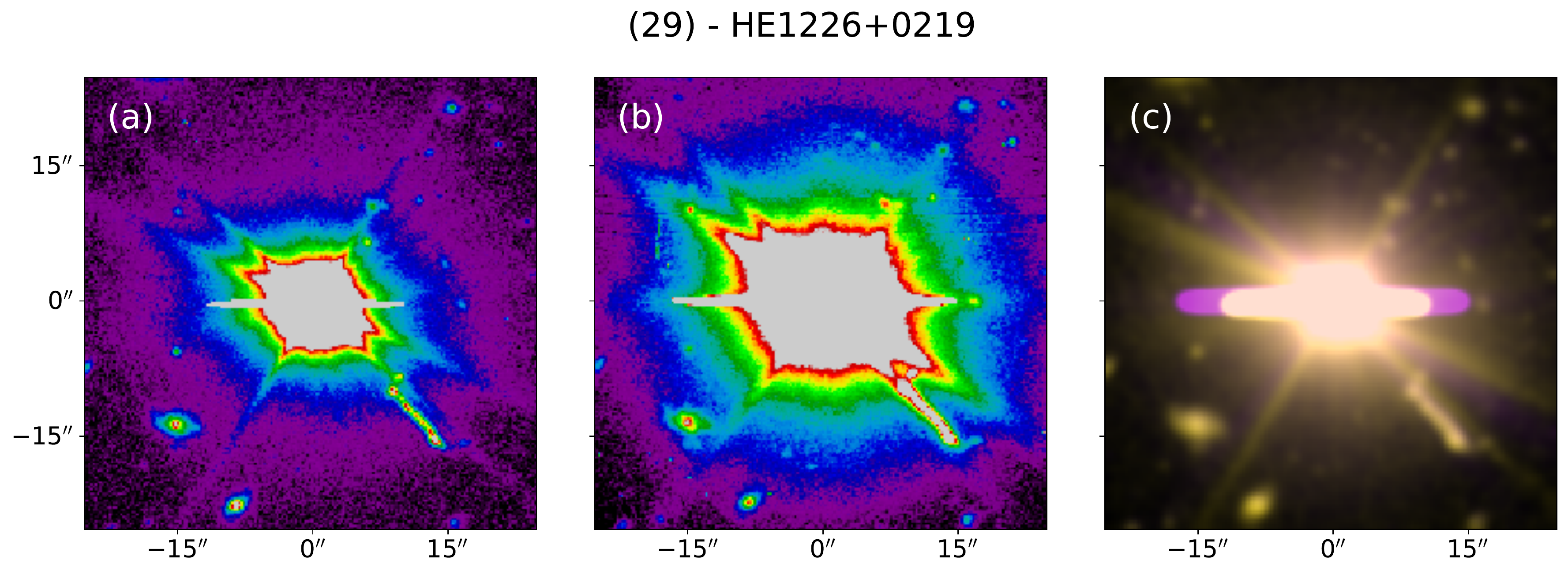}
\figurenum{8}
\caption{(Continued.)}
\end{figure*}

\clearpage

\begin{figure*}
\centering
\includegraphics[width = 17cm]{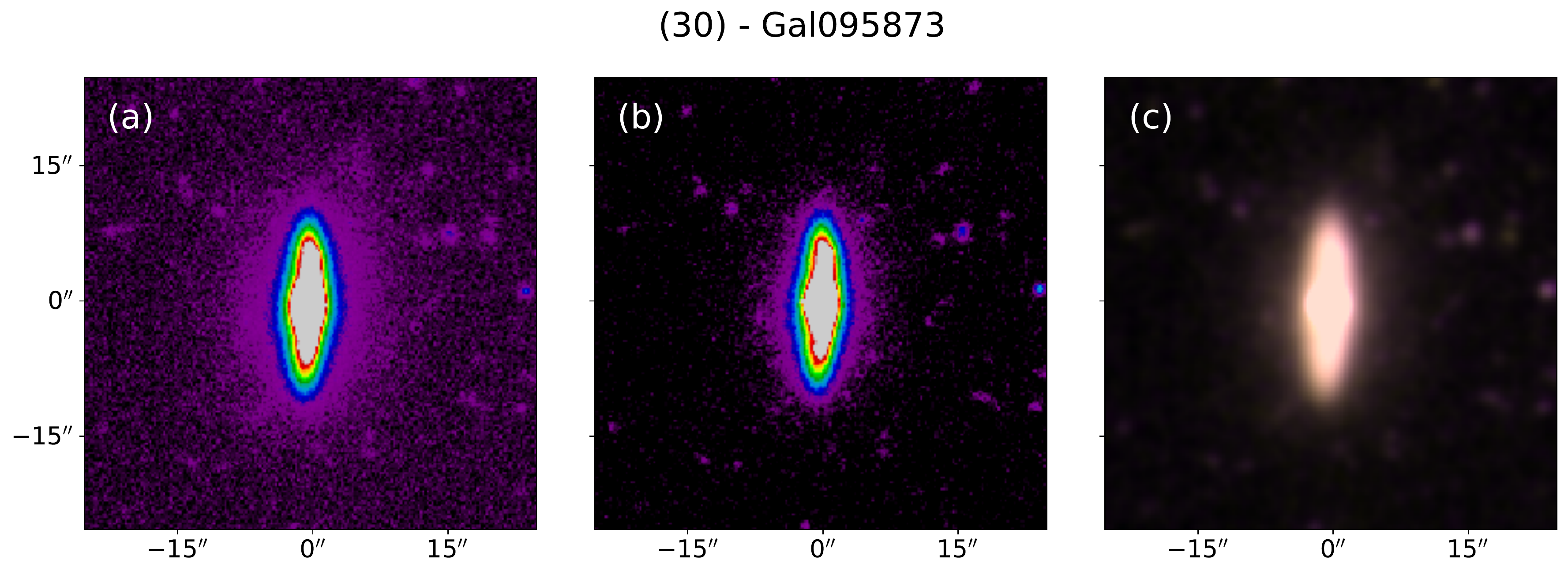}
\end{figure*}

\begin{figure*}
\centering
\includegraphics[width = 17cm]{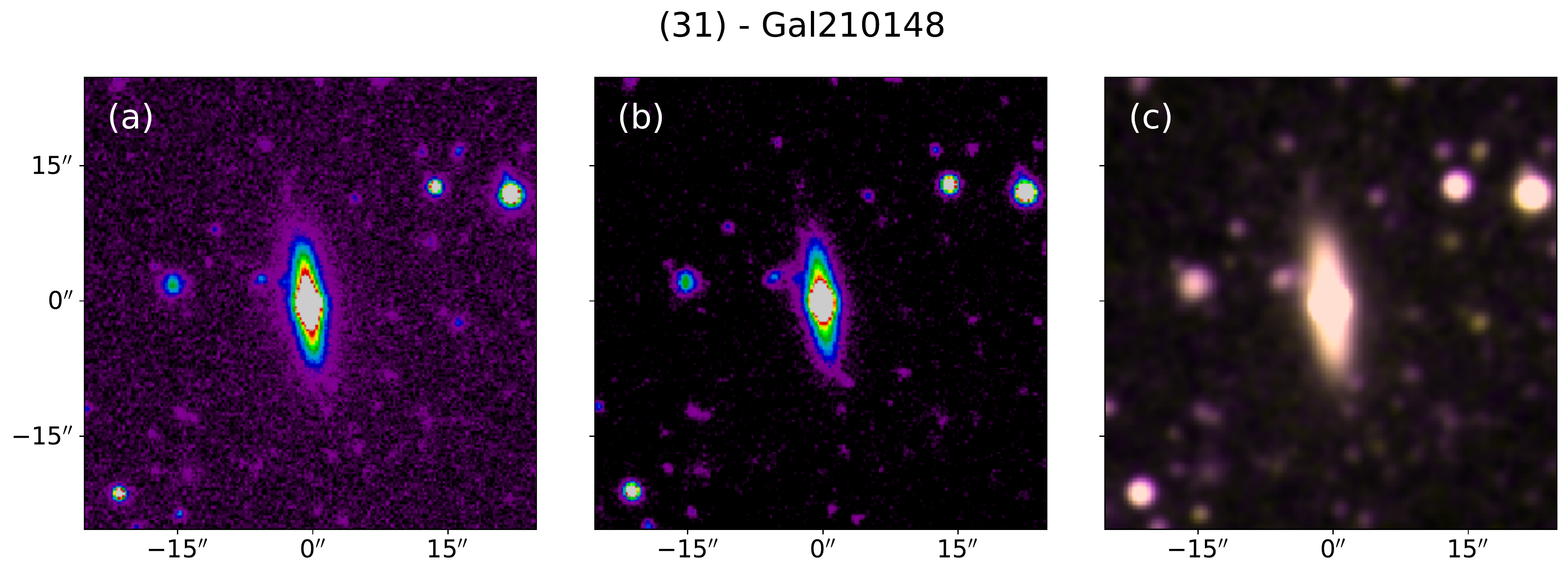}
\end{figure*}

\begin{figure*}
\centering
\includegraphics[width = 17cm]{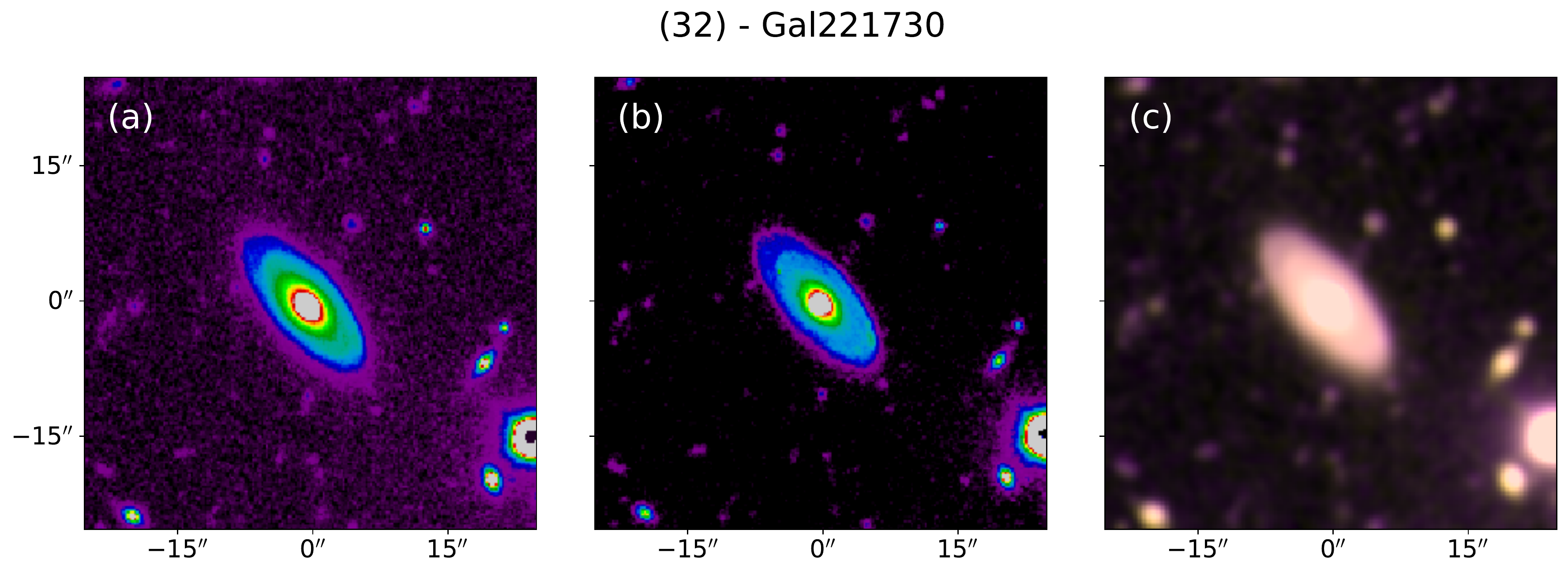}
\figurenum{8}
\caption{(Continued.)}
\end{figure*}

\clearpage

\begin{figure*}
\centering
\includegraphics[width = 17cm]{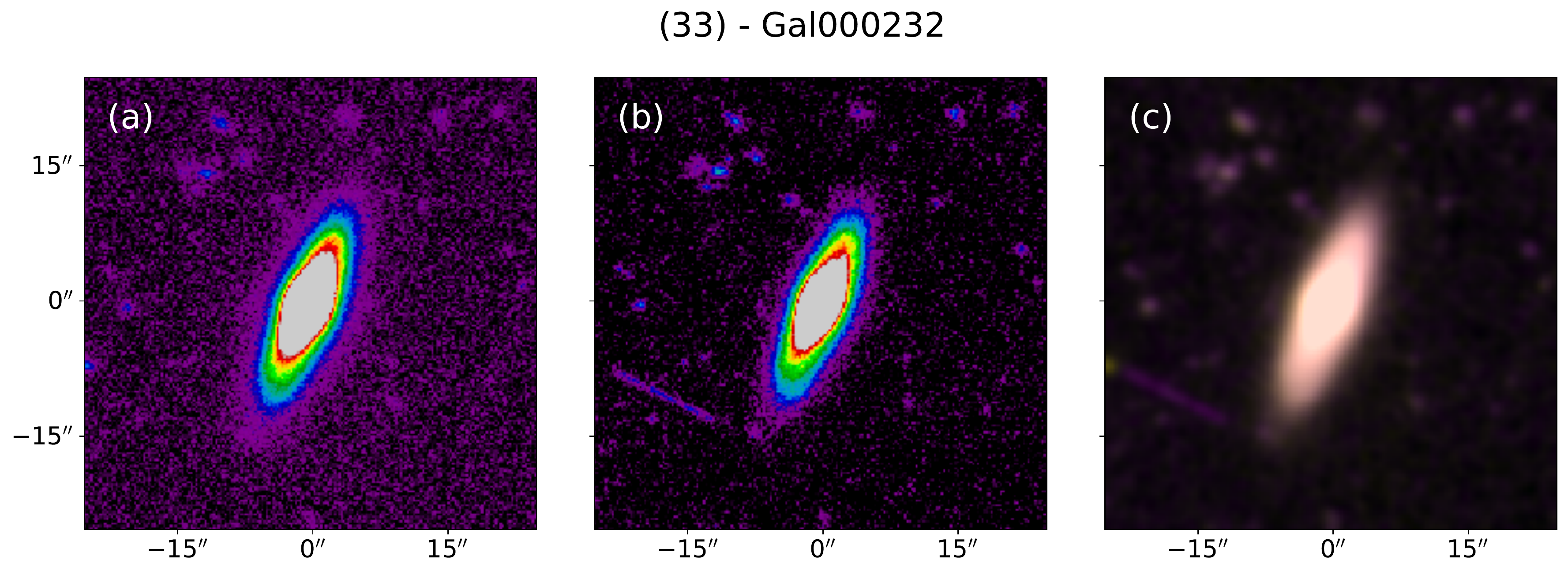}
\end{figure*}

\begin{figure*}
\centering
\includegraphics[width = 17cm]{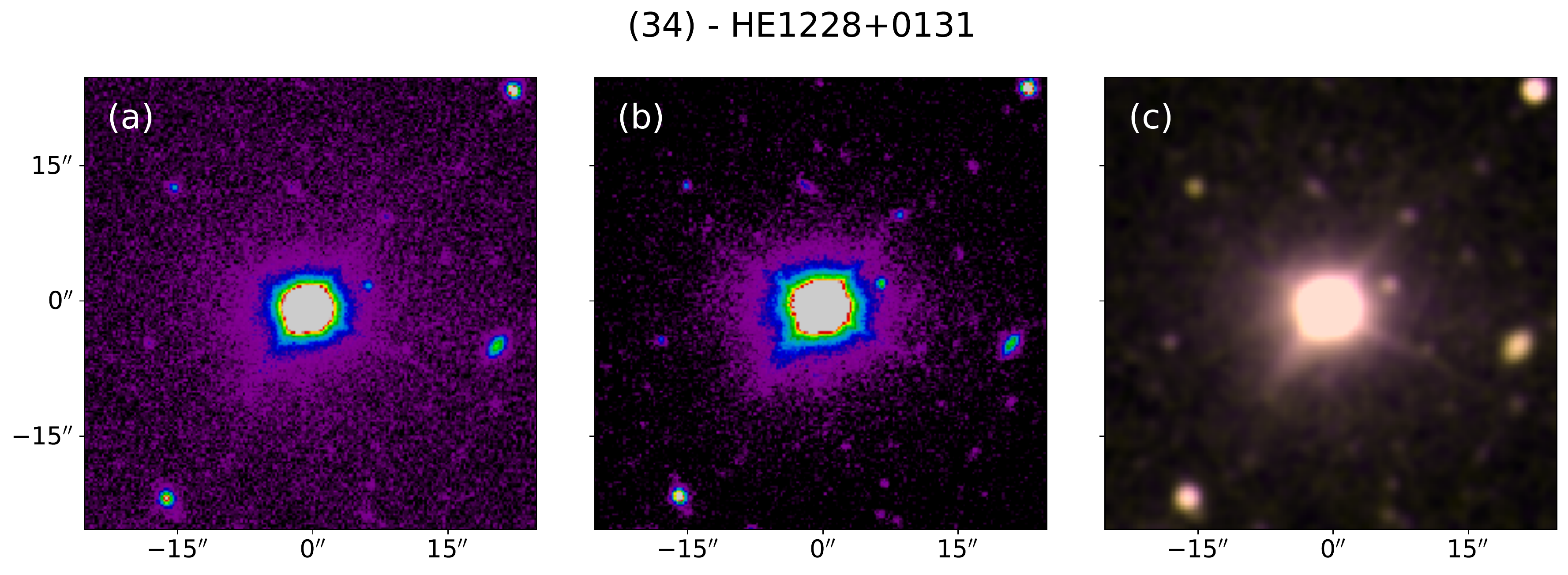}
\end{figure*}

\begin{figure*}
\centering
\includegraphics[width = 17cm]{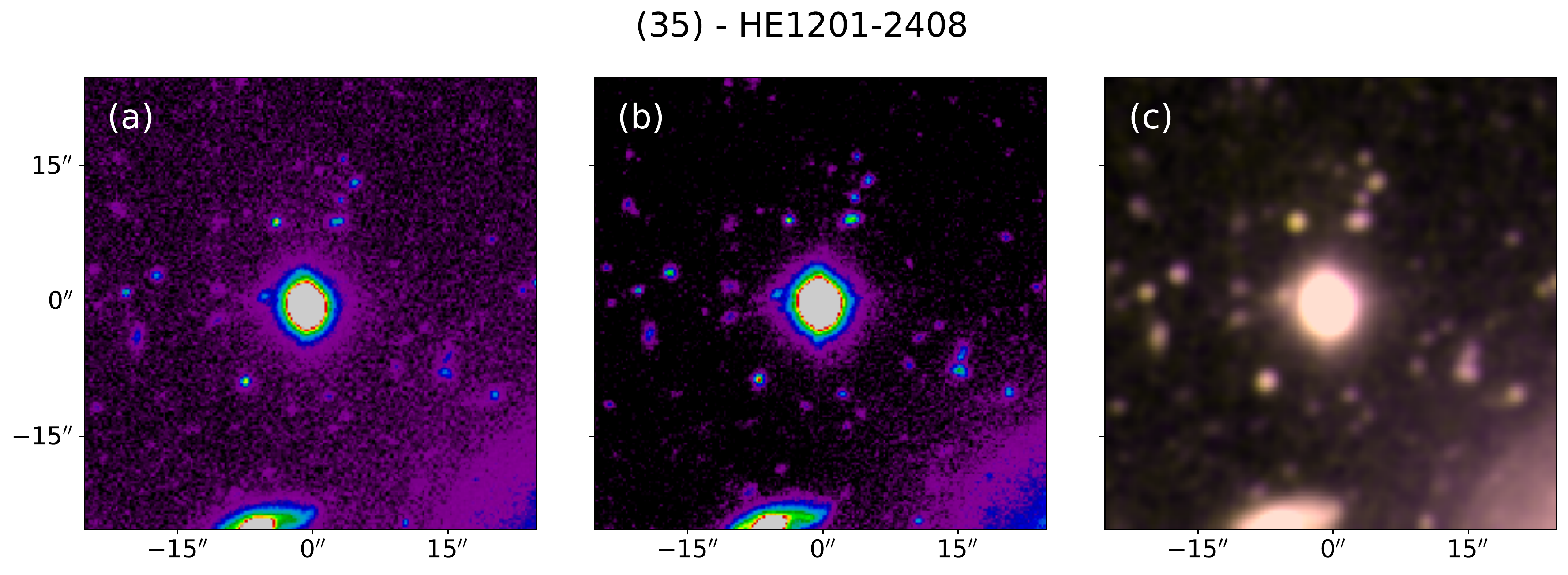}
\figurenum{8}
\caption{(Continued.)}
\end{figure*}

\clearpage

\begin{figure*}
\centering
\includegraphics[width = 17cm]{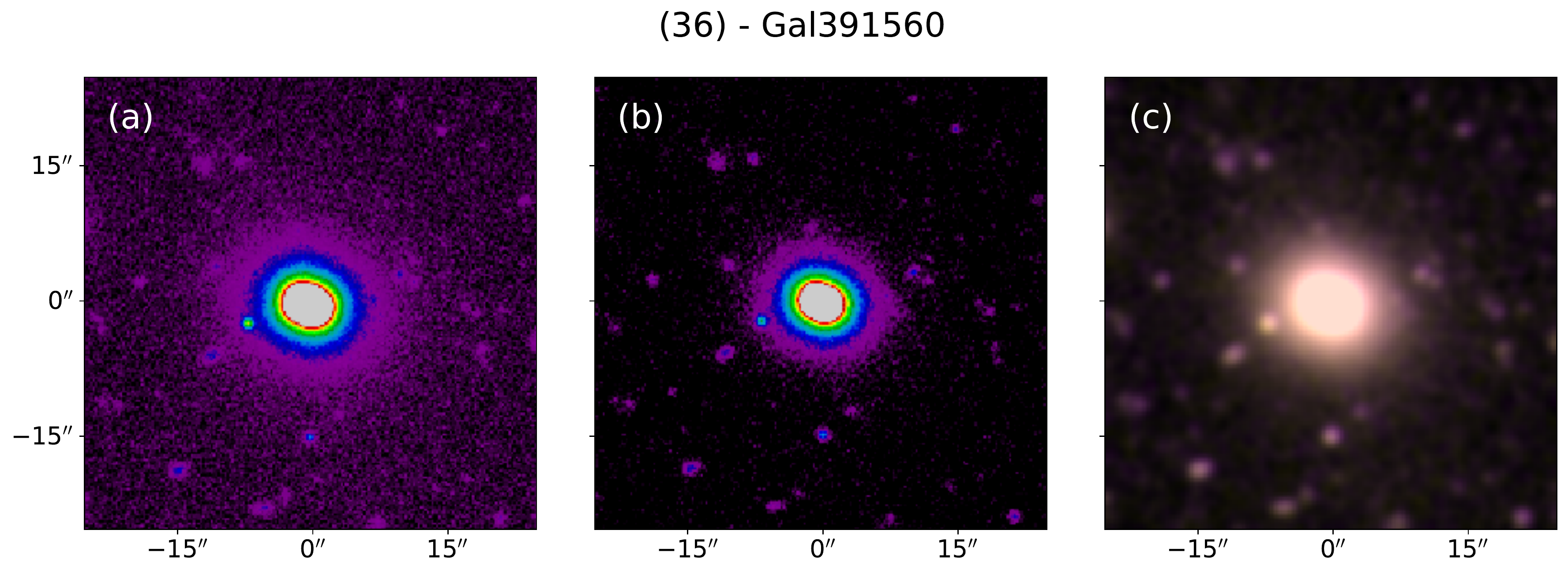}
\end{figure*}

\begin{figure*}
\centering
\includegraphics[width = 17cm]{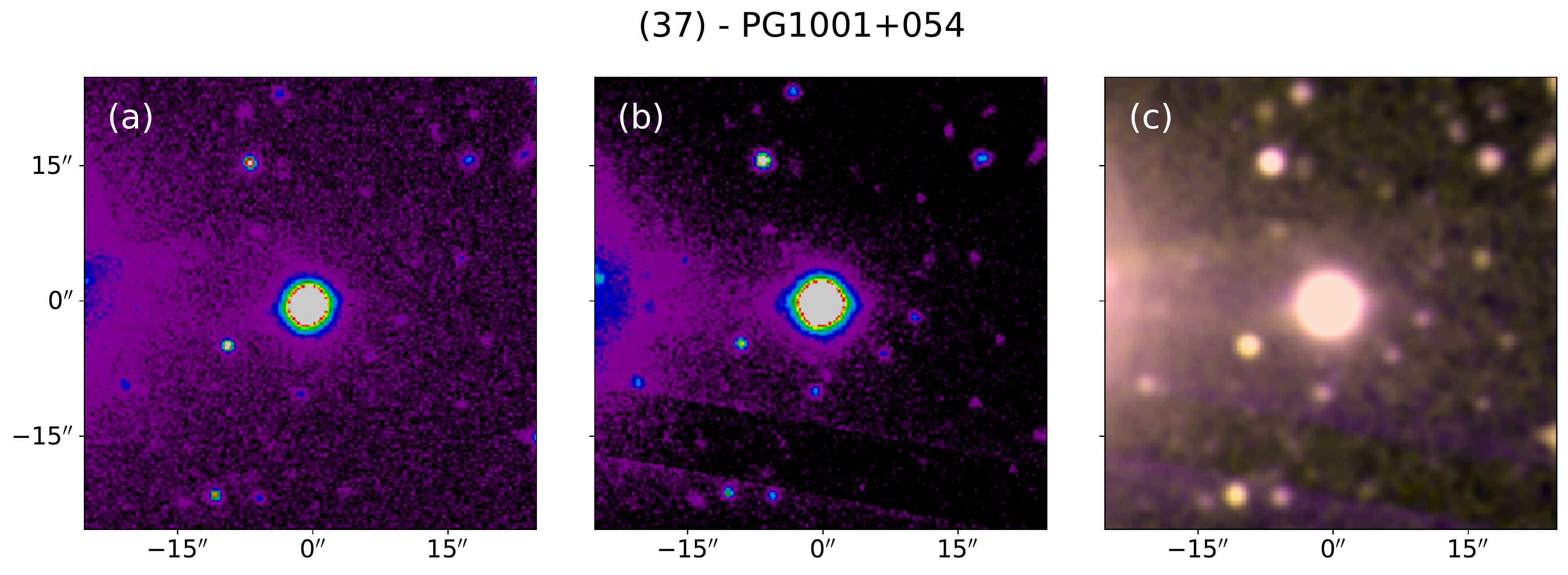}
\end{figure*}

\begin{figure*}
\centering
\includegraphics[width = 17cm]{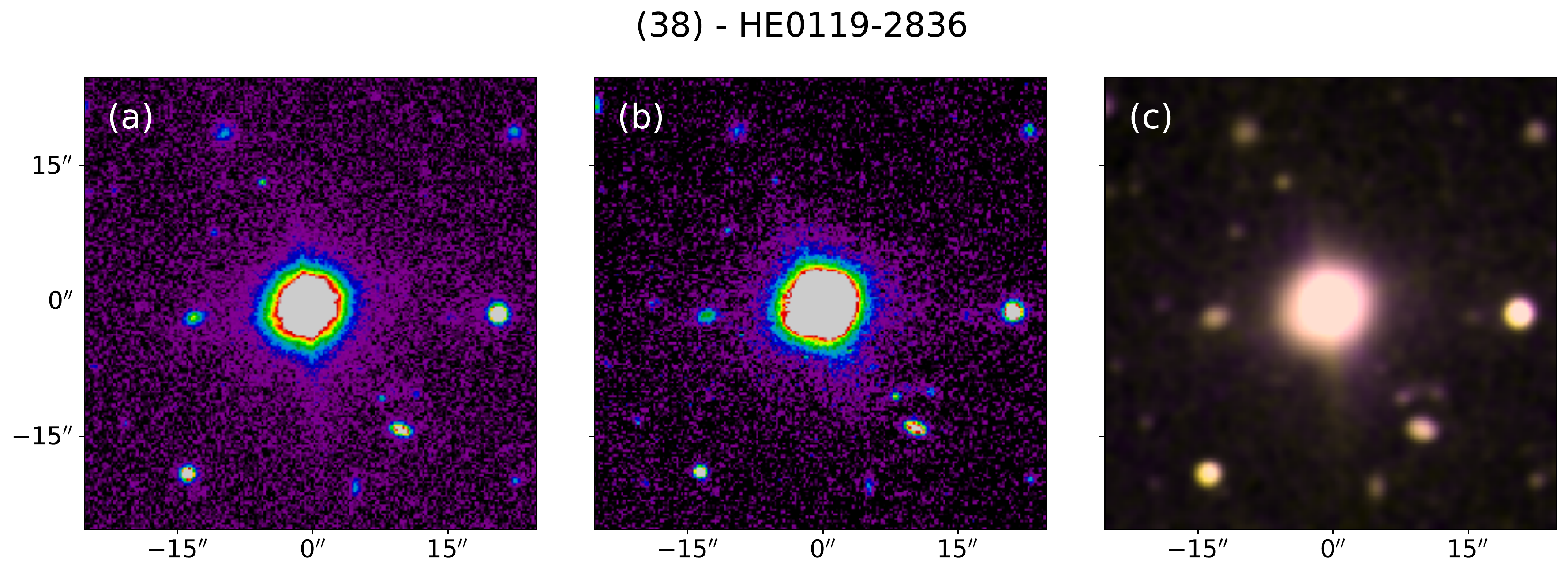}
\figurenum{8}
\caption{(Continued.)}
\end{figure*}

\clearpage

\begin{figure*}
\centering
\includegraphics[width = 17cm]{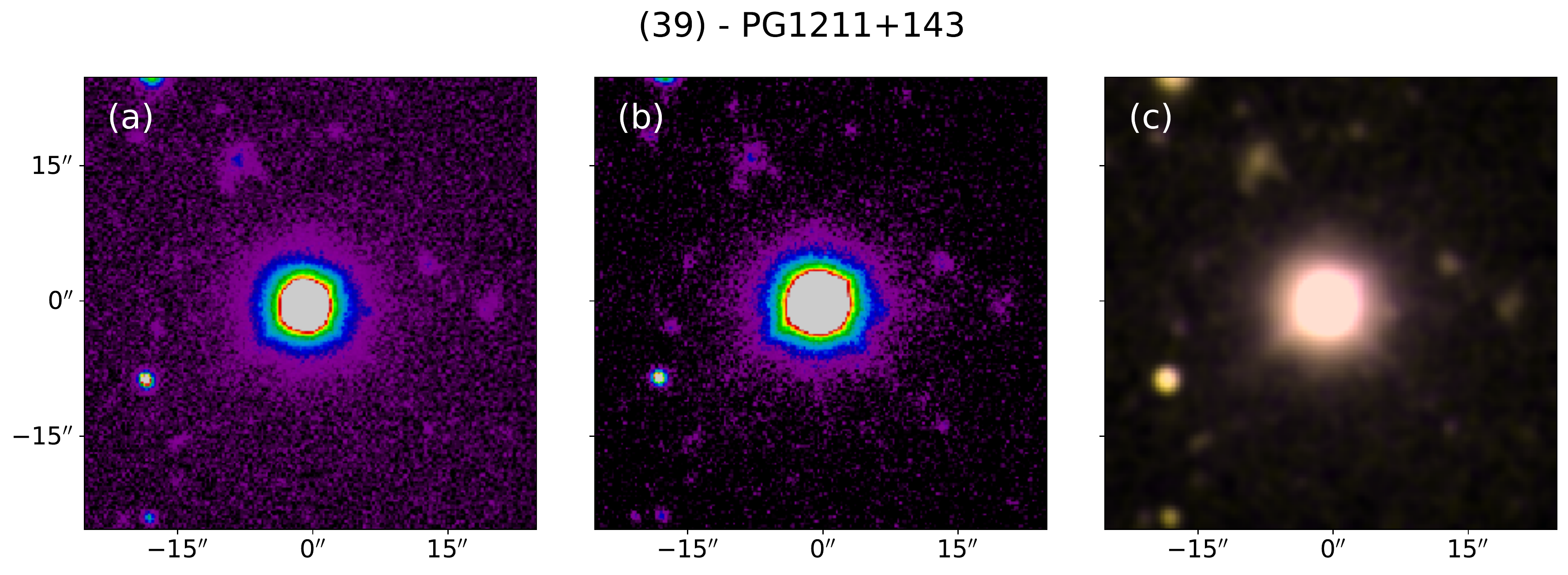}
\end{figure*}

\begin{figure*}
\centering
\includegraphics[width = 17cm]{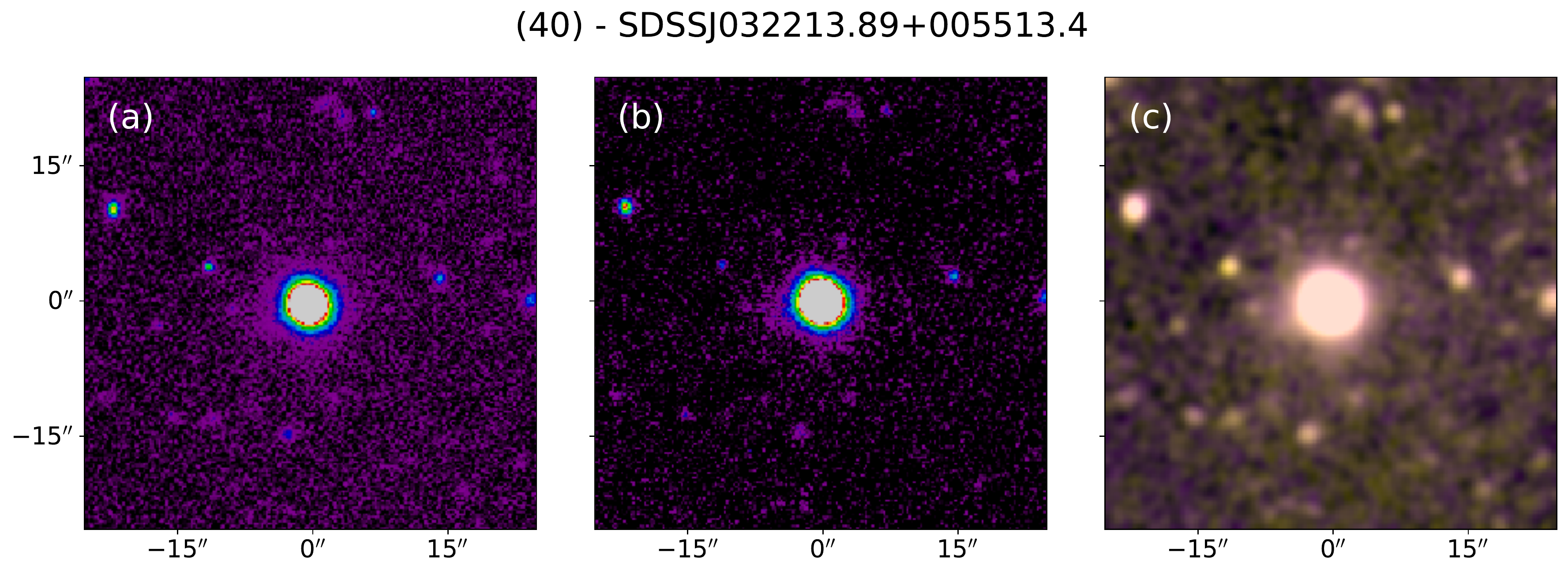}
\end{figure*}

\begin{figure*}
\centering
\includegraphics[width = 17cm]{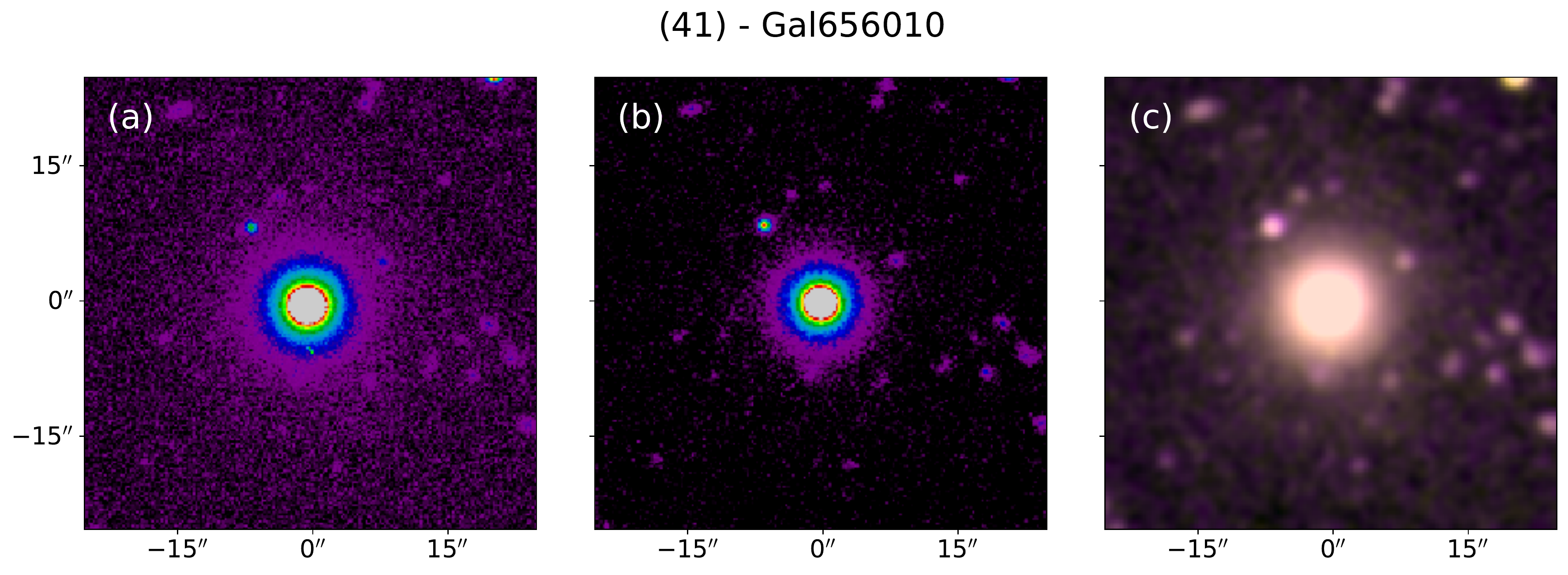}
\figurenum{8}
\caption{(Continued.)}
\end{figure*}

\clearpage

\begin{figure*}[t]
\centering
\includegraphics[width = 17cm]{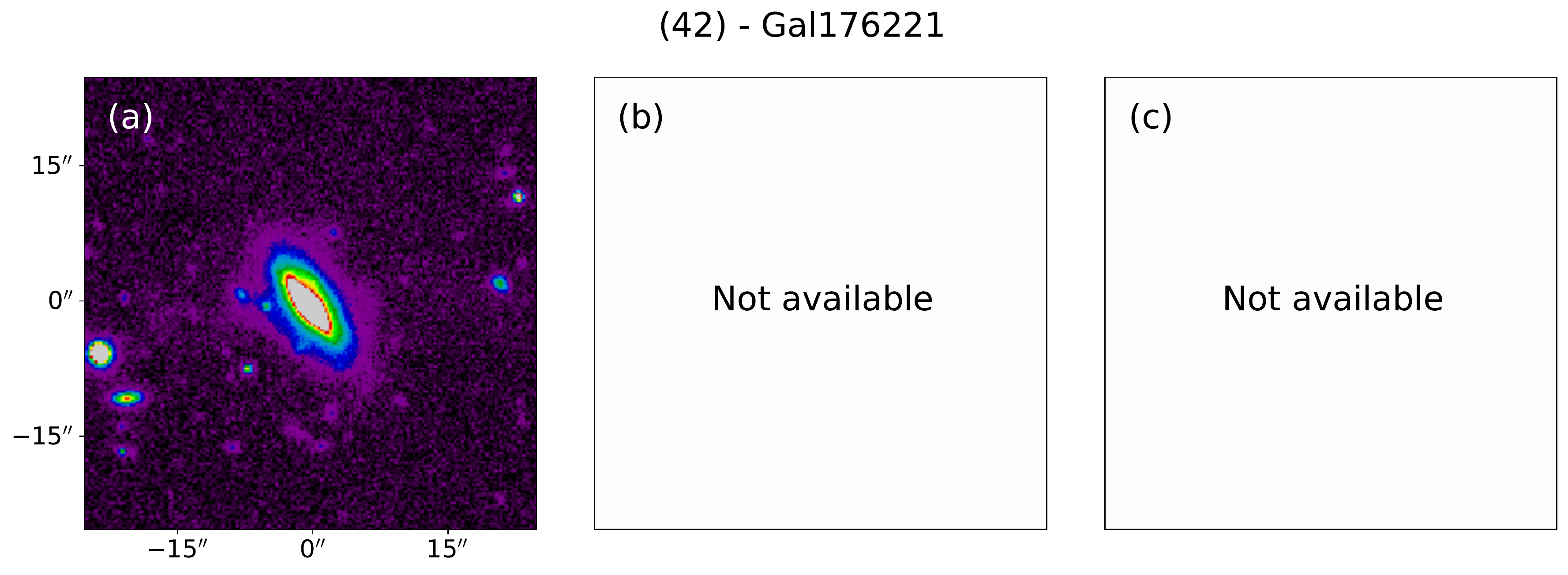}
\figurenum{8}
\caption{(Continued.)}
\end{figure*}

\section{Tabular overall consensus rankings}\label{appendix:meta_ranking_tab}

Complementary to Appendix~\ref{appendix:meta_ranking} we present in this section for referential use the consensus ranks for each target for all sets and combination methods. As in Appendix~\ref{appendix:meta_ranking} the sources are sorted by rank of the `meta' consensus ranking, i.e. the combined ranking of the nine overall rankings (see Sect.~\ref{subsec:robustness_of_fracs}).

\startlongtable
\begin{deluxetable*}{cccccccccc}
\tablenum{3}
\tablecaption{Final consensus ranks\label{tab:overall_rank_results}}
\tablewidth{0pt}
\tablehead{
\colhead{Target} &
\multicolumn3c{Borda} & \multicolumn3c{Average} & \multicolumn3c{Schulze} \\
\colhead{} & 
\colhead{V-band} & \colhead{B-band} & \colhead{Color} &
\colhead{V-band} & \colhead{B-band} & \colhead{Color} &
\colhead{V-band} & \colhead{B-band} & \colhead{Color} \\
}
\startdata
SDSS-J105007.75+113228.6	&	1	&	3	&	3	&	1	&	4	&	4	&	1	&	3	&	5	\\
Gal030481	&	4	&	6	&	1	&	2	&	5	&	1	&	2	&	5	&	1	\\
HE0157+0009	&	3	&	1	&	4	&	3	&	1	&	6	&	3	&	1	&	6	\\
HE2011-6103	&	2	&	2	&	5	&	4	&	2	&	5	&	4	&	2	&	4	\\
HE2258-5524	&	5	&	4	&	2	&	6	&	6	&	2	&	5	&	4	&	2	\\
HE0132-0441	&	7	&	7	&	6	&	5	&	7	&	3	&	6	&	7	&	3	\\
HE0558-5026	&	6	&	5	&	8	&	7	&	3	&	8	&	7	&	6	&	8	\\
PG1012+008	&	8	&	8	&	7	&	8	&	8	&	7	&	8	&	8	&	7	\\
Gal458007	&	10	&	11	&	9	&	9	&	10	&	9	&	9	&	10	&	10	\\
Gal079769	&	12	&	10	&	10	&	13	&	9	&	13	&	10	&	9	&	9	\\
Gal270096	&	11	&	12	&	12	&	10	&	11	&	12	&	11	&	11	&	11	\\
Gal698144	&	18	&	13	&	13	&	15	&	12	&	10	&	18	&	12	&	12	\\
Gal782980	&	9	&	9	&	15	&	12	&	13	&	16	&	12	&	13	&	16	\\
HE0444-3449	&	13	&	21	&	11	&	11	&	22	&	11	&	13	&	21	&	13	\\
Gal534882	&	15	&	16	&	16	&	20	&	19	&	18	&	17	&	17	&	15	\\
Gal510223	&	19	&	14	&	21	&	17	&	15	&	19	&	22	&	14	&	19	\\
Gal050873	&	20	&	17	&	17	&	19	&	20	&	17	&	19	&	18	&	17	\\
Gal419090	&	22	&	22	&	18	&	18	&	18	&	15	&	20	&	16	&	18	\\
Gal676011	&	23	&	15	&	20	&	23	&	16	&	21	&	23	&	15	&	20	\\
SDSS-J124341.77+091707.1	&	17	&	19	&	22	&	21	&	14	&	22	&	15	&	22	&	24	\\
Gal498251	&	21	&	18	&	26	&	22	&	17	&	25	&	21	&	19	&	22	\\
Gal286443	&	16	&	20	&	27	&	16	&	21	&	26	&	16	&	20	&	23	\\
HE2152-0936	&	24	&	23	&	37	&	24	&	24	&	35	&	24	&	23	&	35	\\
Gal185580	&	26	&	30	&	14	&	25	&	29	&	14	&	25	&	28	&	14	\\
Gal003114	&	28	&	24	&	29	&	28	&	23	&	27	&	27	&	24	&	26	\\
Gal204260	&	31	&	26	&	19	&	29	&	26	&	20	&	30	&	29	&	21	\\
Gal347112	&	30	&	29	&	23	&	26	&	27	&	23	&	26	&	26	&	27	\\
Gal557614	&	27	&	27	&	24	&	30	&	28	&	24	&	28	&	27	&	25	\\
HE1226+0219	&	25	&	28	&	25	&	27	&	31	&	29	&	33	&	33	&	29	\\
Gal095873	&	29	&	31	&	32	&	31	&	30	&	31	&	29	&	30	&	32	\\
Gal210148	&	33	&	25	&	35	&	33	&	25	&	33	&	31	&	25	&	33	\\
Gal221730	&	34	&	33	&	28	&	32	&	33	&	28	&	35	&	32	&	31	\\
Gal000232	&	36	&	34	&	30	&	37	&	34	&	32	&	34	&	31	&	30	\\
HE1228+0131	&	35	&	32	&	33	&	34	&	32	&	37	&	36	&	34	&	36	\\
HE1201-2408	&	37	&	38	&	34	&	35	&	36	&	34	&	32	&	36	&	34	\\
Gal391560	&	38	&	39	&	31	&	36	&	39	&	30	&	38	&	35	&	28	\\
PG1001+054	&	32	&	36	&	41	&	38	&	35	&	40	&	37	&	37	&	39	\\
HE0119-2836	&	40	&	37	&	39	&	39	&	37	&	38	&	39	&	38	&	38	\\
PG1211+143	&	39	&	35	&	38	&	40	&	38	&	39	&	40	&	39	&	41	\\
SDSS-J032213.89+005513.4	&	41	&	40	&	40	&	41	&	40	&	41	&	42	&	40	&	40	\\
Gal656010	&	42	&	41	&	36	&	42	&	41	&	36	&	41	&	41	&	37	\\
Gal176221	&	14	&	N/A	&	N/A	&	14	&	N/A	&	N/A	&	14	&	N/A	&	N/A	\\
\enddata
\tablecomments{The final ranks for each source depending on combination method (Borda, Average or Schulze) and set ($B$, $V$ or color images. The targets are sorted by a repeated use of the Schulze method on this nine overall rankings resulting in a singular consensus sequence. Since we have for Gal176221 only observations in $V$-band it is ranked last by the algorithm.}
\end{deluxetable*}

\onecolumngrid

\clearpage

\bibliography{references}{}
\bibliographystyle{aasjournal}

\end{document}